\documentclass[a4paper,11pt]{article}

\usepackage[
backend=biber,
style=numeric-comp, 
sorting=none 
]{biblatex}
\usepackage[utf8]{inputenc}
\usepackage[british]{babel}
\usepackage{csquotes}

\usepackage{manfnt}
\usepackage{braket}

\newcommand{\COMMENTOOK}[1]{}

\usepackage{graphicx}
\usepackage{subcaption} 
\usepackage{color} 

\usepackage{pict2e}
\usepackage{import}

\usepackage{authblk}
\usepackage{hyperref}
\usepackage{amsmath}
\usepackage{amssymb}
\usepackage{amsfonts}
\usepackage{enumitem}
\usepackage{breqn}

\numberwithin{equation}{section}
\usepackage{ytableau}

\usepackage{cleveref}

\usepackage{lscape} 

\ifx\GSW\undefined
  \newcommand{\mys}{s}
  \newcommand{\myt}{t}
  \newcommand{\myu}{u}  
\else
  \newcommand{\mys}{s}
  \newcommand{\myt}{u}
  \newcommand{\myu}{t}  
\fi

\newcommand{\wrt}{{w.r.t.} }

\newcommand{\lc}{{lightcone} }

\newcommand{\oh}{ \frac{1}{2} }
\newcommand{\ap}{ {\alpha'} }

\newcommand{\N}{ \mathbb{N} }

\newcommand{\R}{ \mathbb{R} }

\newcommand{\TOGGLE}[1]{}

\newcommand{\ZZ}{{Z}}

\begin{document}

\begin{center}
  {\Large \bf Bosonic string massive scalar amplitudes
  and absence of ``chaos'' 
  }
\end{center}

\vskip .6cm
\medskip

\vspace*{2.0ex}

\baselineskip=18pt

\begin{center}

{\large 
\rm  Igor Pesando }

\end{center}

\vspace*{4.0ex}
\centerline{ \it Dipartimento di Fisica, Universit\`{a} di Torino, 
  and I.N.F.N., Sezione di Torino}
  
\centerline{ \it \small 
  Via P.\ Giuria 1, I-10125 Torino}
  
\vspace*{1.0ex}
\centerline{\small E-mail: igor.pesando@to.infn.it}

\vspace*{5.0ex}

\centerline{\bf Abstract} \bigskip

We compute directly in covariant formalism various four point
massive scalar amplitudes in bosonic string
with scalars up to level 11.

The most ``difficult'' amplitude we consider is four identical scalars
at level 4, i.e. the first non trivial amplitude with four identical
massive scalars.

All other amplitudes have at least one tachyon.

All computed amplitudes are perfectly regular and show no sign of
erratic behavior.
While this could be naively expected it is not completely trivial
because of higher spins exchanged.

We give a naive argument based on a very simple model
of why this can reasonably be expected and we check on
the explicit examples that the intuition derived from the model is sensible.

Furthermore we discuss the complications due to null states (BRST exact
states) in actually computing the covariant bosonic string spectrum and
how these can be overcome by using a ``$\Pi$ gauge''
which makes manifest that the degeneration of any spin at any mass
level is independent on the spacetime dimension as predicted by
Curtright et al at the end of 90s.

\vfill

\vfill \eject
\baselineskip18pt

\tableofcontents

\section{Introduction and conclusions}

A hallmark of string theory is the presence of an infinite number of
higher-mass spin excitations in its spectrum, which fill the
higher-dimensional representations of the Poincaré group.
String theory offers a consistent description of these massive
modes. However, while the interactions of massless modes are well
understood, the study of amplitudes involving higher-mass string modes
remains relatively under explored (see however
\cite{Mitchell:1990cu,Manes:2001cs,Manes:2003mw,Manes:2004nd} for a
discussion of string form factors,
\cite{Iengo:2002tf,Iengo:2003ct,Chialva:2003hg,Chialva:2004ki,Chialva:2004xm,Iengo:2006gm,Iengo:2006if,Iengo:2006na}
for the discussion on the stability of massive states which is
relevant for the correspondence principle and \cite{Arduino:2020axy}
for an explanation of divergences in temporal string orbifolds due to
massive states).

There are compelling reasons to address this gap.
In the context of gravitational waves, and in line with the no-hair
theorem, black holes are fully characterized by their mass, charge,
and angular momentum. At large distances, their interactions can be
described by the classical limit of on-shell amplitudes, where black
holes are modeled as massive, charged, and spinning point
particles. The study of scattering amplitudes involving higher-spin
particles is an active area of research, due to its relevance in
understanding binary black hole dynamic in the inspiral
phase\cite{Bern:2020buy,Cangemi:2022abk,Cangemi:2022bew}.
		
Investigating higher-spin interactions is also of significant interest
due to the black hole/string correspondence principle
\cite{Polchinski:1995ta,Horowitz:1996nw,Horowitz:1997jc,Damour:1999aw,Veneziano:2012yj}.
One of its incarnations suggests that perturbative string states may collapse into black
holes when the closed string coupling constant reaches a critical
threshold, $g_s N^{1/4}\sim 1$, where $N$ represents the string’s
excitation level \cite{Horowitz:1996nw}.
String theory interactions are encapsulated in correlators of
BRST-invariant vertex operators. While these operators are well
understood for highly massive states on the leading Regge trajectory,
there have been few systematic attempts to formulate vertex operators
for arbitrary massive string
states\cite{Manes:1988gz,Bianchi:2010es,Markou:2023ffh}.
		
In the 70s, the Del Giudice, Di Vecchia, and Fubini operators,
commonly named DDF operators, were introduced to describe excited
massive string states in bosonic string theory
\cite{DelGiudice:1971yjh,Ademollo:1974kz}.
An important feature of these operators is that they commute with the
generators of the Virasoro algebra.
This property allows for the generation of the complete Hilbert space
of non null (non BRST exact) physical states by applying an arbitrary
combination of them to the ground state.
Generic three point amplitudes involving such states of arbitrary
string level were computed \cite{Ademollo:1974kz} and subsequently,
the framework was extended to the fermionic Neveu-Schwarz model
\cite{Hornfeck:1987wt}.
%

%
Recently tree level scattering amplitudes involving DDF operators
of excited strings have been explored
\cite{Bianchi:2019ywd,Gross:2021gsj,Rosenhaus:2021xhm,Firrotta:2022cku,Hashimoto:2022bll,Firrotta:2023wem,Savic:2024ock,Biswas:2024unn,Firrotta:2024qel,Firrotta:2024fvi,Bhattacharya:2024szw,Biswas:2024mdu}.
In particular in \cite{Biswas:2024mdu} a compact expression for the
general tree level open string amplitude involving an arbitrary number
of DDF states was given.
The same expression was then shown to hide the \lc Mandelstam maps and
be equivalent to the \lc formulation \cite{Biswas:2024epj}
(see \cite{Erler:2020beb,Erler:2024lje} for new insight from the
string field point of view).

The erratic behavior of these amplitudes has challenged our
understanding of chaotic phenomena in the quantum field and string
S-matrix \cite{Rosenhaus:2020tmv,Gross:2021gsj,Das:2023xge}.
A quantitative indicator for chaotic quantum processes has been
introduced in \cite{Bianchi:2022mhs,Bianchi:2023uby,Bianchi:2024fsi},
which draws upon the principles of Random Matrix Theory and the
$\beta$-ensemble \cite{Bianchi:2023uby}.
While DDF operators generate a complete set of physical states, their
connection with BRST-invariant vertex operators describing single
string excitations in the usual transverse and traceless gauge remains
cryptic.
Significant progress in understanding this connection has been made
recently \cite{Markou:2023ffh,Biswas:2024unn,Basile:2024uxn}.

However as noticed in \cite{Bianchi:2023uby,Bianchi:2024fsi} and more
quantitatively in \cite{Pesando:2024lqa}
``chaos'' in string amplitudes in three point
amplitudes with a massive state and two tachyons is not due to the
string itself but rather to the ``chaotic'' mixture of spins which is
in the DDF states
(at level about 20 scalars may be normalized to have integer
coefficients and some coefficients are of order
$O(10^{2000}$) \cite{Pesando:2024lqa}).
In fact on shell 3 point amplitudes with a state with a definite spin
and two scalars are
completely determined by kinematics.
It is worth stressing the ``on shell'' since off shell there are many
coefficients which multiply higher powers of momenta \cite{Pesando:2024lqa}.
These coefficients show up in the 4 point amplitudes and therefore it
is worth examining whether 4 massive scalar amplitudes are really
smooth and not erratic.
This is one of the motivations behind this work.

Another motivation is to have some explicit examples of 4 point
amplitudes with massive scalars.
In particular with 4 identical massive scalars since this amplitude
can be used as benchmark for general assertions derived from $S$ matrix theory.


%
Another point which is worth stressing is that originally ``chaos''
showed up already in 3 point DDF amplitudes with one single massive state
and two tachyons at level $N=50$.
This means roughly a dependence on angular variable $\theta$ (massive
higher spin states have preferred directions set by polarization and
therefore also a 3 point amplitudes have angular dependence)
up to  circa $cos(N \theta)$  but in the cases considered
in the previously cited paper  $cos(26 \theta)$ up to multiple zeros.
In this paper we reach up to $t^{19}\sim \cos(19\theta)$ but we plot
only the case $t^{17}\sim \cos(17\theta)$ in fig. \ref{fig:ratios_for_8640}.

The result is that for all the scalar amplitudes we have computed
 which involve massive scalars up to level $N=11$ and always at least
 one tachyon but for the 4 level $N=4$ massive scalars
 there is no ``chaos'' or erratic behavior and therefore the
 coefficients of higher momentum couplings  do not increase too fast
 (and actually amplitudes seem to be dominated by spin $0$ exchange as
 shown by the ratios between $cos(k \theta)$ and $cos(0 \theta)$).

A very rough model can hint for an explanation.
The same model hints that in all amplitudes the angular dependence
is roughly dominated by $1=cos(0 \theta)$ (and even better approximated with
$\cos(\theta)$ and  $\cos(2\theta)$ ). 
This is verified by the explicit computations
(see
fig.s \ref{fig:ratios_for_8640}, \ref{fig:S4TTT_ratios}, \ref{fig:S6zS4zTT_ratios},
\ref{fig:S6zTTS4z_ratios}, \ref{fig:S6zTS4zT_ratios} and \ref{fig:S4zS4zS4zS4z_ratios}
).

The paper is organized as follows.

In section \ref{sec:naive_model} we explain the very naive model of
why we do not expect erratic behavior in the generic case.
Essentially it is the reason why Born amplitudes are dominated by low
angular momenta.

In section \ref{sec:states} we describe the straightforward way we
have used to compute some representative of the massive states.
In the same section we discuss how the states can be obtained in a by
far simpler and instructive way in the ``$\Pi$ gauge''.
This gauge makes the computations less cumbersome and reveals that we
can avoid completely the discussion on null (BRST exact states).
In this way we make explicit the statement made
in \cite{Curtright:1986di,Curtright:1986ie,Curtright:1986gs} that the
degeneration of any spin at a given level is independent on the spacetime
dimension.
Despite their use would make computations less demanding we have
preferred to use the more ``complex'' states since this allows to do
more checks on the code we have used.
In particular we have checked that 3 point amplitudes involving at
least one null state vanish identically and that
4 point amplitudes involving at
least one null state vanish identically after integrating over the
moduli space, i.e. the position of the movable puncture.

Finally in section \ref{sec:amplitudes} we have collected some of the explicit
amplitudes we have computed.
We have chosen three different amplitudes.
The simplest one ``S4zTTT'' (one massive scalar at level $4$ and 3
tachyons).
The first amplitude which shows that amplitudes derived from the same
correlator can be quite different,
i.e. ``S6zS4zTT'' (one massive scalar at level $6$, one massive scalar
at level $4$ and 2 tachyons).
And finally the first amplitude with 4 identical non trivial scalars, i.e. not
KK from massless states, i.e ``S4zS4zS4zS4z'' (4 massive scalars at level $4$).

All the amplitudes and their associated correlators are given as
supplementary materials.
They behave in a very similar way, at least to a human eye.
This is the reason why we have made many plots in order to reveal the
diversities hidden into the mathematical expressions.

It is obvious that the results of this paper are not a proof, they are
a hint and do not imply in an obvious way anything for higher point amplitudes.
Even if there are some reasons to believe that they can hold true.
In order to improve and refine the results is our intention to switch
to the $\Pi$ gauge states and consider also states with spin.
The needed program in maxima is included in the supplementary material.

Already now it is necessary to handle expressions which have $O(10^5)$
terms if full expanded.
The reason of so many terms is the very rapid growth of the number of
cycles from Wick contraction.

There is the possibility that no erratic behavior is present in 4pts
amplitudes but only in higher amplitudes as suggested
in \cite{Rosenhaus:2020tmv}.
The point is worth investigating since
if black holes exist then the correspondence principle
requires some ``chaos'' 
%

\section{A naive model for explaining the absence of chaos}
\label{sec:naive_model}

\begin{figure}[h]
  \centering
    \begin{subfigure}[b]{0.3\textwidth}
    \centering    
    \includegraphics[width=0.8\textwidth]{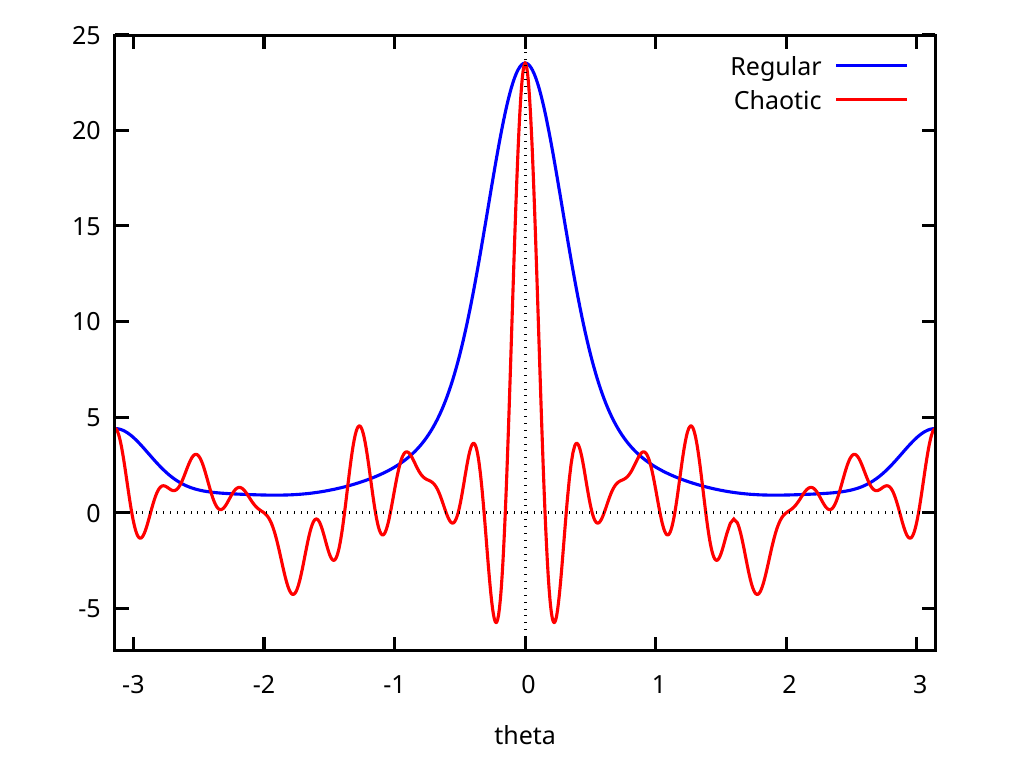}
    \end{subfigure}
    \begin{subfigure}[b]{0.3\textwidth}
    \centering    
    \includegraphics[width=0.8\textwidth]{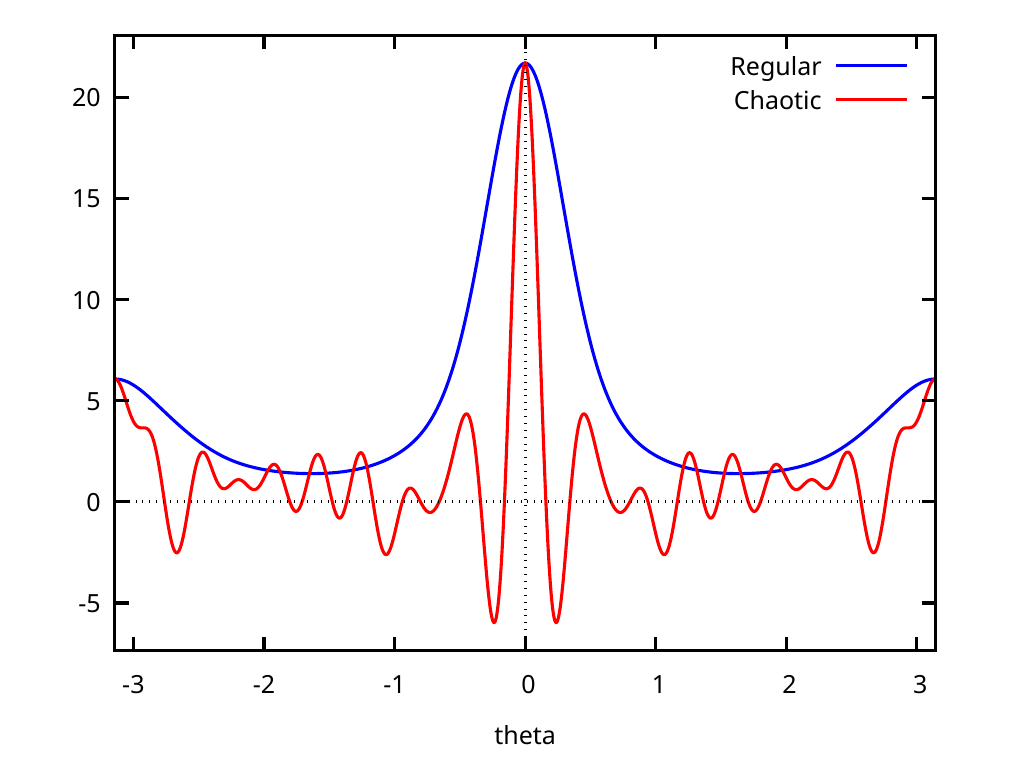}
    \end{subfigure}
    \begin{subfigure}[b]{0.3\textwidth}
    \centering    
    \includegraphics[width=0.8\textwidth]{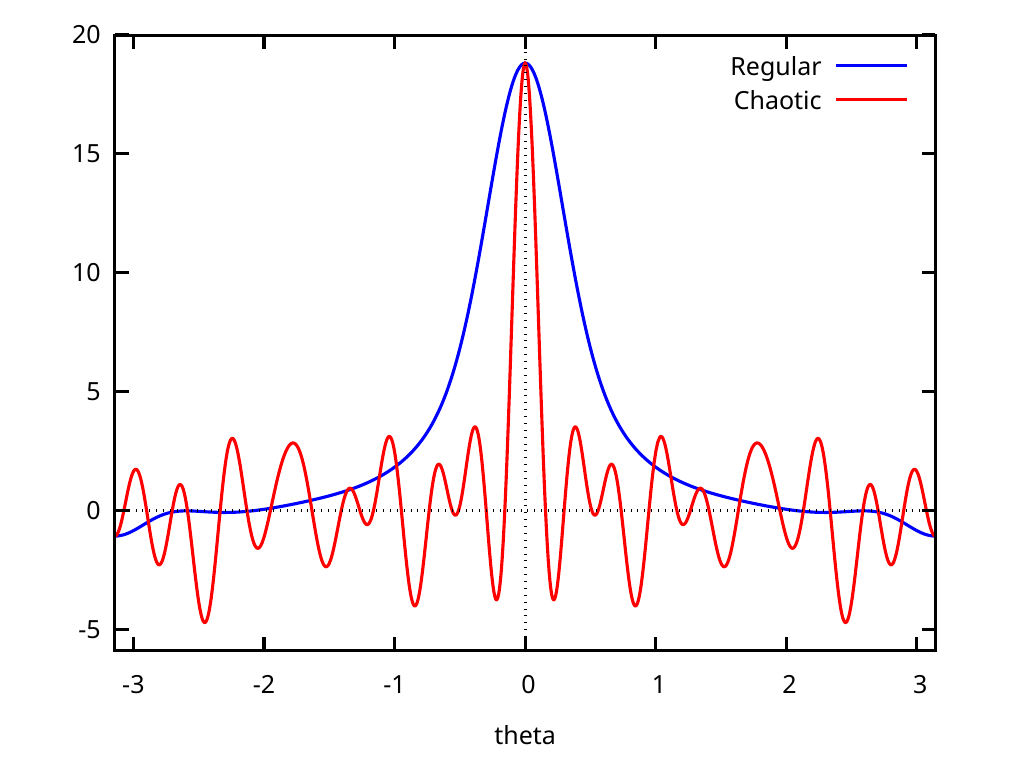}
    \end{subfigure}
\caption{Three graphs of the functions $C_N(\theta)$ and $R_N(\theta)$
with highest frequency $N=20$}    
\label{fig:N20chaos}
\end{figure}

\begin{figure}[h]
  \centering
    \begin{subfigure}[b]{0.3\textwidth}
    \centering    
    \includegraphics[width=0.8\textwidth]{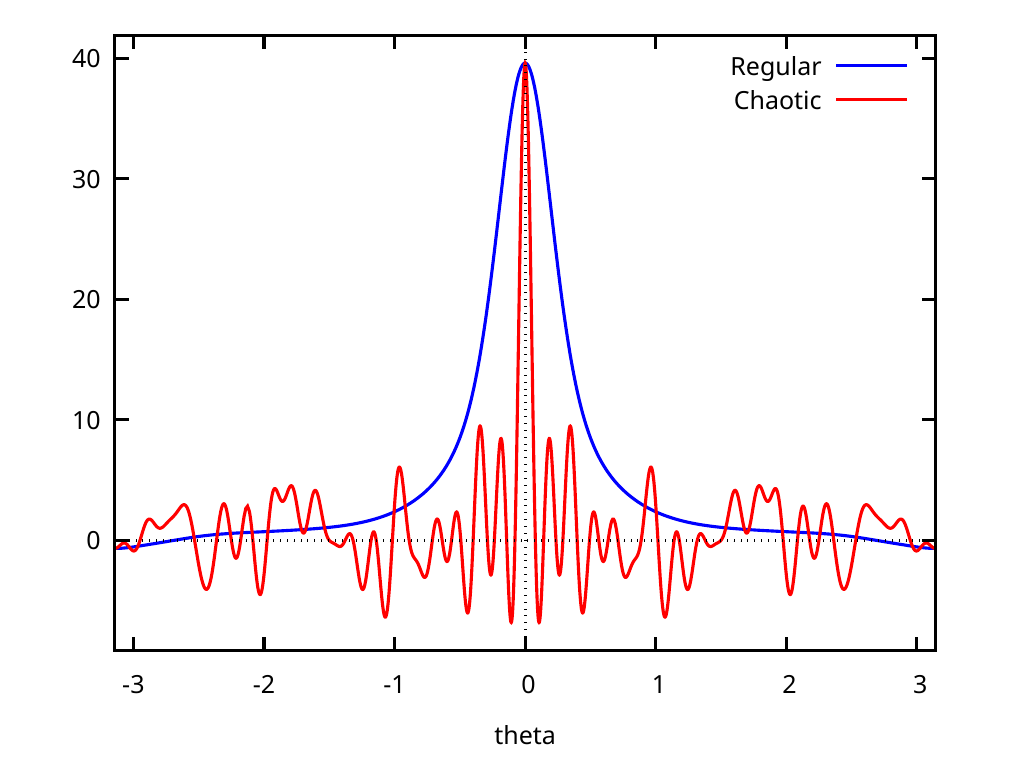}
    \end{subfigure}
    \begin{subfigure}[b]{0.3\textwidth}
    \centering    
    \includegraphics[width=0.8\textwidth]{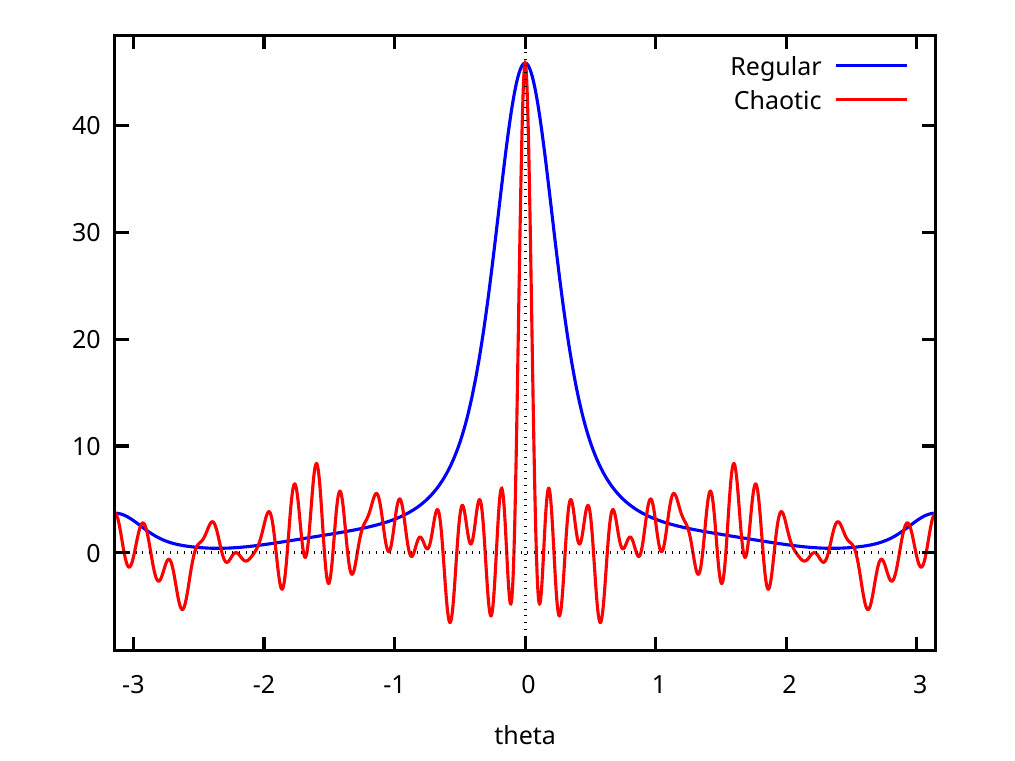}
    \end{subfigure}
    \begin{subfigure}[b]{0.3\textwidth}
    \centering    
    \includegraphics[width=0.8\textwidth]{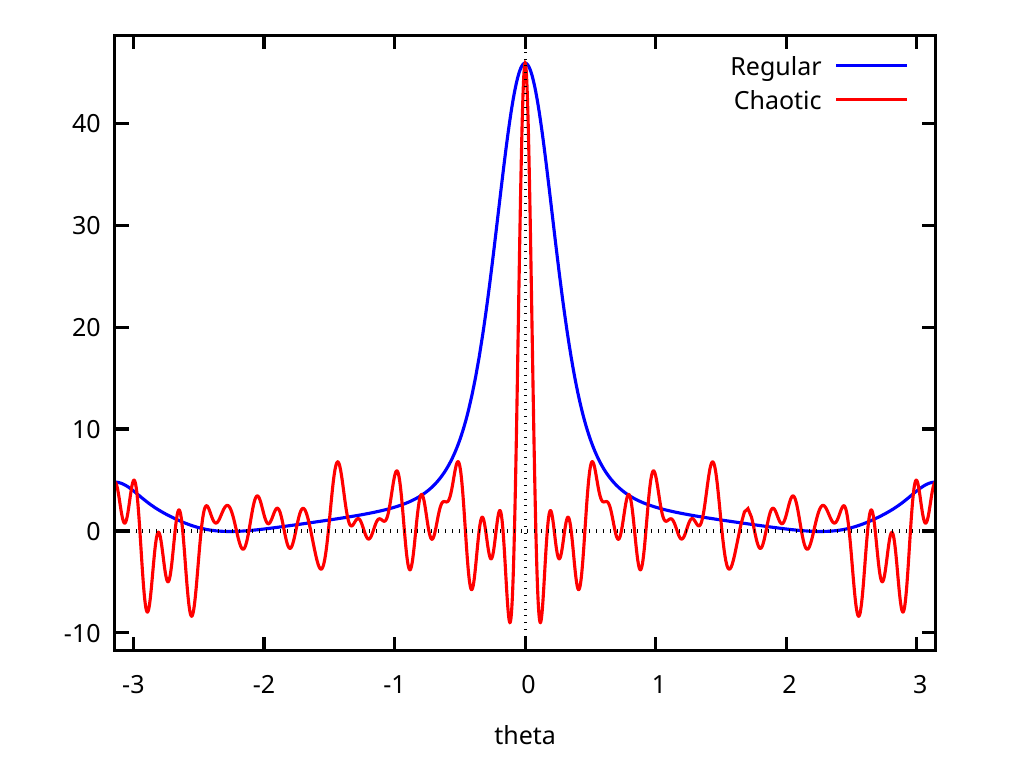}
    \end{subfigure}
\caption{Three graphs of the functions $C_N(\theta)$ and $R_N(\theta)$
with highest frequency $N=40$}    
\label{fig:N40chaos}
\end{figure}

Let us start considering the graphs in figures \ref{fig:N20chaos}
and \ref{fig:N40chaos}.
.

The two completely different behavior are associated to the same
sequence of identically and independent random uniform variable in the
interval $[-1,1]$.
In fact suppose we draw $N+1$ random numbers $c_k$ ($0\le k \le N$)
from a uniform distribution in the interval $[-1,1]$.
We can then define two different functions which at first sight should
have a similar graph
\begin{align}
C_N(\theta)
=&
\sum_{k=0}^N\, c_k \cos(k \theta)
,
\nonumber\\
R_N(\theta)
=&
\sum_{k=0}^N\, c_k  \left( \cos(\theta) \right)^k
,
\end{align}     
but in reality the first function $C_N(\theta)$ has a ``chaotic'' behavior
while the second function $R_N(\theta)$ has a regular behavior.

The reason of the very different behaviors is due to the fact that
$|\cos(\theta)|\le 1$ and therefore
the contribution from $\left( \cos(\theta) \right)^k$ for big $k$ is
suppressed \wrt  f.x. $\cos(\theta)$ and $1$.

The idea is that the ``non trivial part'' of 4pts string amplitudes is
encoded in a polynomial $P_N(s,t)$ in Mandelstam variables $s$ and $t$.
At fixed center of mass energy $s$ the angular dependence is contained
in $t\sim cos(\theta)$.
Hence almost all angular dependence besides the one in Veneziano
amplitude is in form of a polynomial
$\hat P_N(s,cos(\theta))$.
If we suppose that all coefficients in $P_N(s,t)$ are of the same
order of magnitude then almost all interesting angular dependence is
in the low $k$ terms, roughly
$\cos(2 \theta)$, $\cos(\theta)$ and $1$ because
$|\cos(\theta)|\le 1$.

We have written only $1$, $\cos(\theta)$ and $\cos(2\theta)$
because we can expect that higher $k$s dominate when all lower $k$
term vanish.
But only $\cos(2^m \theta)$ ($m\in \N$) have no common zeros therefore
generically $1$, $\cos(\theta)$ and $\cos(2\theta)$ will determine the
behavior since their coefficients are also bigger for $R_N$ when
written as a $\cos(k \theta)$ series.

We can make this statement more rigorous.
If we rewrite the regular  function $R_N(\theta)$ using $\cos(k\theta)$ we get
\begin{align}
\hat c_{2 m}
=&
\sum_{l=0}^{[N/2]-m} { 2 l + 2 m \choose l} \frac{1}{2^{k+l} } 2 c_{2l+2m}
,
\nonumber\\
\hat c_{2 m+1}
=&
\sum_{l=0}^{[N/2]-m} { 2 l + 2 m + 1\choose l} \frac{1}{2^{2l+2m} } c_{2l+2m+1}
.
\end{align}
Then their expectation values under the hypothesis that $c_k$
are independent and identically distributed random variables
is zero.
We can however compute the expectation values of their squares
under the same hypothesis that $c_k$
are independent and identically distributed random variables.
We easily get
\begin{align}
E[\hat c_{2 m}^2]
=&
2 \sigma^2
\sum_{l=0}^{[N/2]-m} \left( { 2 l + 2 m \choose l} \frac{1}{2^{k+l} } \right)^2
,
\nonumber\\
E[\hat c_{2 m+1}^2]
=& \sigma^2
\sum_{l=0}^{[N/2]-m} \left( { 2 l + 2 m + 1\choose
l} \frac{1}{2^{2l+2m} } \right)^2
,
\end{align}
where
$\sigma^2=E[c_k^2]$.
We can evaluate some cases and get
\begin{align}
E[\hat c_0^2 ] \sim 4/3 \sigma^2,
E[\hat c_N^2 ] = E[\hat c_{N-1}^2 ] = \frac{1}{2^{2 N}} \sigma^2,
,
\end{align}     
so we can expect
\begin{align}
\sqrt{ \frac{E[\hat c_N^2 ]}{ E[\hat c_0^2 ]} }
\sim
\frac{1}{2^{N}}
\end{align}     

In section \ref{sec:amplitudes} on amplitudes we plot
$\Re \log c_N(s) / c_0(s)$ for a range of $s$ and we show that the
scattering is dominated by spin $0$.

Here we plot in figure \ref{fig:ratios_for_8640}
the same quantities $\Re \log c_N(s) / c_0(s)$ for amplitudes we
have not reported in the main text in order to give an immediate
visual confirmation of what asserted.

\begin{figure}[!h]
  \centering
    \begin{subfigure}{0.49\textwidth}
    \centering    
   \resizebox{1.0\textwidth}{!}{\input{S8z1S6zS4zT.log.ratios.s_11.91607978309962_119.1607978309962_steps_1000_tex.tex}}
    \end{subfigure}
    \begin{subfigure}{0.49\textwidth}
    \centering    
   \resizebox{1.0\textwidth}{!}{\input{S8z1TS6zS4z.log.ratios.s_0.0_64.1109_steps_1000_tex.tex }}
    \end{subfigure}
\caption{
Some ratios $\Re \log c_N(s) / c_0(s)$ for two amplitudes
``S8z1S6zS4zT'' and ``S8z1TS6zS4z'' which can be derived from the same
correlator but are quite different.
In particular these amplitudes have contribution up to ``spin''  $16$
and $17$ and are defined for different $s$ ranges.
}
\label{fig:ratios_for_8640}
\end{figure}
%
%

\section{Massive scalar states and $\Pi$ gauge}
\label{sec:states}

In this section we would like to describe how we build the massive
scalars we are going to use and the ambiguities related to the
existence of null (BRST exact) states.
The explicit computation shows explicitly what is known from theory:
the degeneration of scalars at a given arbitrary level is
independent on the space time dimension but the number of null states
does depend on it.

We then describe how all is simplified in the $\Pi$ gauge.
The gauge is also all very well suited to show explicitly that the
degeneration of a given arbitrary spin at a given arbitrary level is
independent on the spacetime dimension and on the existence of null states.
What depends on spacetime dimension is the dimension of the irrep.

\subsection{Building explicit representatives of scalar states}
In order to describe the straightforward and simplest approach we use
we start with the lightest massive scalar, the one at level $N=4$.

We first compute the scalar basis at level $N=4$.  Differently from
the \lc we need to include also the zero modes $\alpha^\mu_0$ this
makes the dimension of the vector space bigger.
The explicit basis reads 
\begin{align}
\Big\{
&
\left(1 , 0\right)^4 , \left(1 , 0\right)^2\,\left(1 , 1\right)
,
\left(1 , 1\right)\,\left(2 , 0\right) , \left(2 , 0\right)^2,
\nonumber\\
&
\left(1 , 0\right)\,\left(2 , 1\right) , \left(2 , 2\right),
\left(1 , 0\right)\,\left(3 , 0\right) , \left(3 , 1\right),
\left(4 , 0\right)
\Big\}
,
\end{align}
where we use the short hand notation
\begin{equation}
(m,n) \equiv \alpha^\mu_{-m} \alpha^\mu_{-n} 
,
\end{equation}
and we do not explicitly write the momentum
$| k_\nu \rangle $.
This means f.x. that
\begin{align}
\left(1 , 1\right)\,\left(2 , 0\right)
\equiv
\alpha_{-1}^\mu \alpha_{-1}^\mu\,
\alpha_{-2}^\nu \alpha_{0}^\nu\,
|
.
\end{align}     
This basis has $11$ elements while the \lc and $\Pi$ gauge basis have
only $3$ elements, i.e. the previous elements without any $\alpha_0$.
Explicitly
\begin{align}
 \Big\{
 \left(1 , 1\right)^2 , \left(2 , 2\right) ,
 \left(3 ,1\right)
 \Big\}
 .
\end{align}

Conceptually this is not a big difference but computationally it is
since the computation of amplitudes requires by far more terms.
For example the 3 point amplitude requires in principle without
considering permutation symmetries $11^3$
sub-amplitudes in the straightforward computation versus $3^3$
sub-amplitudes in \lc or $\Pi$ gauge.

The difference becomes quickly very big as the following table shows
\begin{equation}
\begin{matrix}

 \\
 \mbox{cov}
 & 1 & 3 & 5 & 11 & 18 & 35 & 57 & 102 & 165 & 279 & 444 & 726
 \\
 \Pi \mbox{ gauge}

\end{matrix}
.
\end{equation}

After we have construct the basis we can consider the most general scalar
state as a linear combination of the basis elements
\renewcommand{\ZZ}{(2\ap k^2)}
\begin{align}
|SC_4\rangle
=&
\left(4 , 0\right)\,s_{11}
+\left(3 , 1\right)\,s_{10}
+\left(1 , 0\right)\,\left(3 , 0\right)\,s_{9}
+\left(2 , 2\right)\,s_{8}
\nonumber\\
&
+\left(1 , 0\right)\,\left(2 , 1\right)\,s_{7}
+\left(2 , 0\right)^2\,s_{6}
+\left(1 , 1\right)\,\left(2 , 0\right)\,s_{5}
\nonumber\\
&
+\left(1 , 0\right)^2\,\left(2 , 0\right)\,s_{4}
+\left(1 , 1\right)^2\,s_{3}
+\left(1 , 0\right)^2\,\left(1 , 1\right)\,s_{2}
+\left(1 , 0\right)^4\,s_{1}
,
\end{align}
and then require it a primary field by imposing
\begin{align}
L_{1} |SC_4\rangle
=
L_{2} |SC_4\rangle
=
0.
\end{align}
This gives $3$ independent (but not orthogonal) solutions, i.e
\begin{align}
|s_{4\, 1}\rangle
=&
-{{2\,\ZZ^2\,\left(D+1\right)}\over{3}}
\,\left(3 ,1\right)
+
{{2\,\ZZ\,\left(D+1\right)}\over{3}}
\,\left(1 , 0\right)\,\left(3 , 0\right)
+
{{\ZZ^2\,\left(D+1\right)}\over{2}}
\,\left(2 ,2\right)
\nonumber\\
&
-
{{\ZZ\,\left(D+1\right)}\over{2}}
\,\left(2 ,0\right)^2
+
\ZZ^2\,\left(1 , 1\right)^2
-
2\,\left(0 ,0\right)\,\left(1 , 0\right)^2\,\left(1 , 1\right)
+
\left(1 , 0\right)^4
\end{align}
\begin{align}
|s_{4\, 2}\rangle
=& 
-{{\left(4\,D^2-2\,\ZZ \,D+21\,D-2\,\ZZ^2
+\ZZ + 20\right)}
\over
{4\,\ZZ\,\left(2\,\ZZ-9\right)}
}
\left(2 , 0\right)^2
\nonumber\\
&
+
{{\left(4\,D^2-4\,\ZZ\,D+18\,D-2\,\ZZ^2+3\,\ZZ +23\right)}
\over
{4\,\left(2\,\ZZ-9\right)}
}
\left(2 , 2\right)
\nonumber\\
&
-
{{\left(4\,D^2-4\,\ZZ\,D+18\,D-6\,\ZZ +23\right)}
\over
{3\,\left(2\,\ZZ-9\right)}}
\left(3 , 1\right)
\nonumber\\
&
+
{{\left(4\,D^2-10\,\ZZ\,D+21\,D+20\right)}
\over
{3\,\ZZ\,\left(2\,\ZZ-9\right)}}
\left(1 , 0\right)\,\left(3 , 0\right)
+
{{\left(4\,D-2\,\ZZ+5\right)}
\over
{\ZZ\,\left(2\,\ZZ-9\right)}
}
\left(1 , 0\right)^2\,\left(2 , 0\right)
\nonumber\\
&
-
{{\left(4\,D-2\,\ZZ+5\right)}
\over
{\ZZ\,\left(2\,\ZZ -9\right)}}
\left(1 ,0\right)^2\,\left(1 , 1\right)
+
{{\left(2\,D-2\,\ZZ+7\right)}
\over
{2\,\ZZ-9}}
\left(1 , 1\right)^2\,
\nonumber\\
&
+
{{\left(2\,\ZZ-1\right)\,\left(D-1\right)}
\over{4\,\left(2\,\ZZ-9\right)}}
\,\left(4 , 0\right)
+
\left(1 , 0\right)\,\left(2 , 1\right) 
\end{align}
\begin{align}
|s_{4\, 3}\rangle
=& 
-{{\left(D^2-2\,\ZZ\,D+12\,D
+4\,\ZZ^2-26\,\ZZ+32\right)}
\over
{4\,\ZZ\,\left(2\,\ZZ-9\right)}}
\left(2 , 0\right)^2\,
\nonumber\\
&
-
{{\left(D^2-4\,\ZZ\,D+18\,D+4\,\ZZ^2
-24\,\ZZ+26\right)}
\over
{3\,\left(2\,\ZZ-9\right)}}
\left(3 , 1\right)\,
\nonumber\\
&
+
{{\left(D^2-4\,\ZZ\,D+18\,D+4\,\ZZ^2
-24\,\ZZ+26\right)}
\over
{4\,\left(2\,\ZZ-9\right)}}
\left(2 , 2\right)\,
\nonumber\\
&
+
{{\left(D-2\,\ZZ+4\right)\,
\left(D-2\,\ZZ+8\right)}
\over
{3\,\ZZ\,\left(2\,\ZZ-9\right)}}
\left(1 , 0\right)\,\left(3 , 0\right)\,
+
{{\left(D-2\,\ZZ+8\right)}
\over
{\ZZ\,\left(2\,\ZZ-9\right)}}
\left(1 , 0\right)^2\,\left(2 , 0\right)\,
\nonumber\\
&
-
{{\left(D-2\,\ZZ+8\right)}
\over
{\ZZ\,\left(2\,\ZZ-9\right)} }
\left(1 , 0\right)^2\,\left(1 , 1\right)\,
+
{{\left(D-4\,\ZZ+17\right)}
\over
{2\,\left(2\,\ZZ-9\right)}}
\left(1 , 1\right)^2\,
\nonumber\\
&
+
{{\left(D-1\right)}\over{2\,\left(2\,\ZZ-9\right)}}
\left(4 , 0\right)\,
+
\left(1 , 1\right)\,\left(2 , 0\right)
.
\end{align}
In the previous expression $D$ is the spacetime dimension.
The number of solutions does not depend of $D$ or the mass shell
condition.

On the other side we know that there is only one scalar at level $N=4$
therefore all the previous states must be describe the same physical
scalar.
There is also the possibility that some of them are null (BRST exact)
states.

To exactly count the physical scalars at this level we have to find
the null states.
This requires to find a basis for scalar states at level $N-1=3$ 
\begin{align}
\left\{
\left(1 , 0\right)^3 , \left(1 , 0\right)\,\left(1 , 1\right) ,
\left(1 , 0\right)\,\left(2 , 0\right) , \left(2 , 1\right) ,
\left(3 , 0\right) \right\}
,
\end{align}
and at level $N-2=2$
\begin{align}
\left\{ \left(1 , 0\right)^2 ,
\left(1 , 1\right) , \left(2 , 0\right)
\right\}
,
\end{align}
then to make the general linear superposition
and then build their images under $L_{-1}$ and $L_{-2}$ respectively.
The result of the previous steps is a scalar state at level $N=4$
which is not necessarily associated with a primary field, i.e.  physical.
We get
\begin{align}
=
&
2\,\left(4 , 0\right)\,n_{8}
+
\left(2 , 0\right)^2\,n_{8}
+
{{\left(1 , 1\right)\,\left(2 , 0\right)\,n_{8}}\over{2}}
\nonumber\\
&
+
2\,\left(3 , 1\right)\,n_{7}
+
\left(1 , 1\right)\,\left(2 , 0\right)\,n_{7}
+
{{\left(1 , 1\right)^2\,n_{7}}\over{2}}
\nonumber\\
&
+
2\,\left(1 , 0\right)\,\left(3 , 0\right)\,n_{6}
+
\left(1 , 0\right)^2\,\left(2 , 0\right)\,n_{6}
+
{{\left(1 , 0\right)^2\,\left(1 , 1\right)\,n_{6}}\over{2}}
\nonumber\\
&
+
3\,\left(4 , 0\right)\,n_{5}
+
\left(1 , 0\right)\,\left(3 , 0\right)\,n_{5}
\nonumber\\
&
+
2\,\left(3 , 1\right)\,n_{4}
+
\left(2 , 2\right)\,n_{4}
+
\left(1 , 0\right)\,\left(2 , 1\right)\,n_{4}
\nonumber\\
&
+
2\,\left(1 , 0\right)\,n_{3}\,\left(3 , 0\right)
+
\left(2 , 0\right)^2\,n_{3}
+
\left(1 , 0\right)^2\,\left(2 , 0\right)\,n_{3}
\nonumber\\
&
+
2\,\left(1 , 0\right)\,n_{2}\,\left(2 , 1\right)
+
\left(1 , 1\right)\,n_{2}\,\left(2 , 0\right)
+
\left(1 , 0\right)^2\,\left(1 , 1\right)\,n_{2}
\nonumber\\
&
+
3\,n_{1}\,\left(1 , 0\right)^2\,\left(2 , 0\right)
+
n_{1}\,\left(1 , 0\right)^4
.
\end{align}
This expression has $8=5+3$ undetermined coefficients as many as the
dimension of the scalar basis used to build it.

In critical dimension $D=26$ and on mass shell
$(0,0)=\ZZ= -2 (4-1)=-6$
{\sl only}
we get $2$ null states.

These null states may be used to simplify the physical level $4$
scalar
$r_1 |s_{4\, 1}\rangle + r_2 |s_{4\, 2}\rangle + r_3 |s_{4\,3}\rangle$.
There is no unique way.
One possible gauge is to try to set to zero the terms with the highest
number of zero modes $\alpha^\mu_0$.
The result is reported in section \ref{sec:states_used}
which contains the states we actually used to perform the computations.

There are also other possibilities.
For example to try to set to zero the coefficients of the terms
with the highest excitation $m$  $\alpha^\mu_{-m}$.

Actually there is a very efficient gauge which we describe now.

\subsection{Scalar and massive states in $\Pi$ gauge}

In the previous subsection we have seen that building explicitly scalar
(and massive) states in covariant formalism may be computationally
more demanding than build them in \lc as done in \cite{Pesando:2024lqa}.
This happens because the covariant basis are bigger.
We want now to show that there exists a very effective and
conceptually illuminating gauge which we dub $\Pi$ gauge.

The idea is very simple.
In \cite{Manes:1988gz} it was noticed that replacing
the usual polarizations  with the projected polarizations as
\begin{align}
\epsilon_\mu
~~\rightarrow~~
(\Pi \epsilon)_\mu
=
\left( \delta_\mu^\nu - \frac{k_\mu k^\nu}{ k^2} \right) \epsilon_\nu
,
\end{align}
led to a simplification in the conditions imposed by Virasoro
constraints.
Now we further generalize this approach and we build the scalar basis
(and all the other states) using the projector
$\Pi_\mu^{.\, \nu}(\alpha_0)
=
\left( \delta_\mu^\nu - \frac{\alpha_{0\, \mu} \alpha_0^\nu}{
(\alpha_0)^2 } \right)$
using
\begin{equation}
[m, n]=
(m,n)_\Pi
\equiv
\alpha^\mu_{-m} \, \Pi_{\mu \nu}(\alpha_0)\, \alpha^\nu_{-n} 
.
\end{equation}
This immediately leads to the collapse of the size of the covariant
basis to the same size of the \lc one, i.e for $N=4$ scalars
\begin{align}
 \Big\{
 \left[1 , 1\right]^2 , \left[2 , 2\right] ,
 \left[3 ,1\right]
 \Big\}
 .
\end{align}

The obvious and natural question is whether this is not too restrictive
and whether we can describe all physical states in this way.
The answer is that it is possible and this a very efficient way
since it eliminates completely the necessity of considering null states.
The reason is very simple.
If we describe the massive state in the rest frame, also when it is off
shell as long as $E\ne0$,
then the projector eliminates all the objects with a temporal index
$\mu=0$
since
$\Pi_{\mu=0}^{.\, \nu}\Big|_{rest}= \Pi_\mu^{.\, \nu=0}\Big|_{rest}=0$.
This means that all states build using the previous basis have
positive norm.
This is the crucial point since any null state has by definition a
null norm hence it must contain in some way some $\alpha^{\mu=0}$.
It follows that we can forget about null states since their action
moves any state out of the $\Pi$ gauge.
This in turn means that the ``gauge orbit'' of a physical state
generated by the null states must have a representative in the $\Pi$
gauge.
Even more, because of the discussion on the previous section which
rephrases the classical knowledge that only the number of null states
does depend on the dimension of spacetime and on shell condition
we see that the degeneration of a physical (also off shell) states for
a
dimension.
This construction makes very concrete the same assertion made in
\cite{Curtright:1986di,Curtright:1986ie,Curtright:1986gs}
which was based on computing the degenerations by using characters.

The level $N=4$ scalar state reads in this gauge
\newcommand{\Deff}{D_{eff}}
\begin{align}
-{{2\,\left(\Deff+2\right)}\over{3}}
\left[3 , 1\right]\,
+
{{\left(\Deff+2\right)}\over{2}}
\left[2 , 2\right]\,
+
\left[1 , 1\right]^2
,
\end{align}
where
$\Deff=D-1$.


\subsection{The states used}
\label{sec:states_used}
The states we used in the actual computations are the following.

The level $4$ scalar with the maximum number of terms the highest
occurrences of zero modes  eliminated using the null states which has $8$ terms
out of $11$ possible reads
{\footnotesize
\begin{align}
|S4z\rangle
=&
1500\,\left(4 , 0\right)
-7104\,\left(3 , 1\right)
-184\,\left(1 , 0\right)\,\left(3 , 0\right)
+5742\,\left(2 , 2\right)
+276\,\left(1 , 0\right)\,\left(2 , 1\right)
\nonumber\\
&
+432\,\left(2 , 0\right)^2
-726\,\left(1 , 1\right)\,\left(2 , 0\right)
+225\,\left(1 , 1\right)^2
,
\end{align}
}%


%
%
%
The level $6$ scalar with
the maximum number of terms the highest
occurrences of zero modes  eliminated using the null states which has $22$ terms
out of $35$ possible reads
{\footnotesize
\begin{align}
|S6z\rangle
=&
-226500\,\left(6 , 0\right)
+1177320\,\left(5 , 1\right)
-18168\,\left(1 , 0\right)\,\left(5 , 0\right)
\nonumber\\
&
-1872750\,\left(4 , 2\right)
+214110\,\left(1 , 0\right)\,\left(4 , 1\right)
-141150\,\left(2 , 0\right)\,\left(4 , 0\right)
\nonumber\\
&
+79515\,\left(1 , 1\right)\,\left(4 , 0\right)
+929500\,\left(3 , 3\right)
-191400\,\left(1 , 0\right)\,\left(3 , 2\right)
\nonumber\\
&
+100680\,\left(2 , 0\right)\,\left(3 , 1\right)
-337500\,\left(1 , 1\right)\,\left(3 , 1\right)
+75700\,\left(3 , 0\right)^2
\nonumber\\
&
+9720\,\left(2 , 1\right)\,\left(3 , 0\right)
-43875\,\left(2 , 0\right)\,\left(2 , 2\right)
+251225\,\left(1 , 1\right)\,\left(2 , 2\right)
\nonumber\\
&
-7975\,\left(1 , 0\right)^2\,\left(2 , 2\right)
+1900\,\left(2 , 1\right)^2
+15950\,\left(1 , 0\right)\,\left(2 , 0\right)\,\left(2 , 1\right)
\nonumber\\
&
-6750\,\left(2 , 0\right)^3
+12275\,\left(1 , 1\right)\,\left(2 , 0\right)^2
-20250\,\left(1 , 1\right)^2\,\left(2 , 0\right)
+6750\,\left(1 , 1\right)^3
,
\end{align}
}%


%
%
%
The first level $8$ scalar with
the maximum number of terms the highest
occurrences of zero modes

out of $102$ possible and reads
{\footnotesize
\begin{align}
|S8z1\rangle
=&
-54920250\,\left(8 , 0\right)
+419373360\,\left(7 , 1\right)
-1427760\,\left(1 , 0\right)\,\left(7 , 0\right)
\nonumber\\
&
-2262735720\,\left(6 , 2\right)
-113561280\,\left(1 , 0\right)\,\left(6 , 1\right)
-64299480\,\left(2 , 0\right)\,\left(6 , 0\right)
\nonumber\\
&
+126077280\,\left(1 , 1\right)\,\left(6 , 0\right)
+6188984592\,\left(5 , 3\right)
+356467104\,\left(1 , 0\right)\,\left(5 , 2\right)
\nonumber\\
&
+384247584\,\left(2 , 0\right)\,\left(5 , 1\right)
-383096448\,\left(1 , 1\right)\,\left(5 , 1\right)
+1209600\,\left(1 , 0\right)^2\,\left(5 , 1\right)
\nonumber\\
&
+135449328\,\left(3 , 0\right)\,\left(5 , 0\right)
-262211040\,\left(2 , 1\right)\,\left(5 , 0\right)
+10184832\,\left(1 , 0\right)\,\left(2 , 0\right)\,\left(5 , 0\right)
\nonumber\\
&
-10789632\,\left(1 , 0\right)\,\left(1 , 1\right)\,\left(5 , 0\right)
-4300632000\,\left(4 , 4\right)
-247152360\,\left(1 , 0\right)\,\left(4 , 3\right)
\nonumber\\
&
-1212819300\,\left(2 , 0\right)\,\left(4 , 2\right)
+512024310\,\left(1 , 1\right)\,\left(4 , 2\right)
-12882555\,\left(1 , 0\right)^2\,\left(4 , 2\right)
\nonumber\\
&
-18257400\,\left(3 , 0\right)\,\left(4 , 1\right)
+529542090\,\left(2 , 1\right)\,\left(4 , 1\right)
-2116485\,\left(1 , 0\right)\,\left(2 , 0\right)\,\left(4 , 1\right)
\nonumber\\
&
+11975040\,\left(1 , 0\right)\,\left(1 , 1\right)\,\left(4 , 1\right)
-73024875\,\left(4 , 0\right)^2
-222613440\,\left(3 , 1\right)\,\left(4 , 0\right)
\nonumber\\
&
-10558800\,\left(1 , 0\right)\,\left(3 , 0\right)\,\left(4 , 0\right)
+462861000\,\left(2 , 2\right)\,\left(4 , 0\right)
+28720755\,\left(1 , 0\right)\,\left(2 , 1\right)\,\left(4 , 0\right)
\nonumber\\
&
-19570950\,\left(2 , 0\right)^2\,\left(4 , 0\right)
+16215885\,\left(1 , 1\right)\,\left(2 , 0\right)\,\left(4 , 0\right)
-11975040\,\left(1 , 1\right)^2\,\left(4 , 0\right)
\nonumber\\
&
+910342720\,\left(2 , 0\right)\,\left(3 , 3\right)
-223149500\,\left(1 , 1\right)\,\left(3 , 3\right)
+13777750\,\left(1 , 0\right)^2\,\left(3 , 3\right)
\nonumber\\
&
-156750720\,\left(3 , 0\right)\,\left(3 , 2\right)
-360009720\,\left(2 , 1\right)\,\left(3 , 2\right)
-15677340\,\left(1 , 0\right)\,\left(2 , 0\right)\,\left(3 , 2\right)
\nonumber\\
&
-772800\,\left(1 , 0\right)\,\left(1 , 1\right)\,\left(3 , 2\right)
-42000\,\left(1 , 0\right)^3\,\left(3 , 2\right)
+539448700\,\left(3 , 1\right)^2
\nonumber\\
&
-14149100\,\left(1 , 0\right)\,\left(3 , 0\right)\,\left(3 , 1\right)
-1237619880\,\left(2 , 2\right)\,\left(3 , 1\right)
-22545600\,\left(1 , 0\right)\,\left(2 , 1\right)\,\left(3 , 1\right)
\nonumber\\
&
+26813220\,\left(2 , 0\right)^2\,\left(3 , 1\right)
-631680\,\left(1 , 1\right)\,\left(2 , 0\right)\,\left(3 , 1\right)
+42000\,\left(1 , 0\right)^2\,\left(2 , 0\right)\,\left(3 , 1\right)
\nonumber\\
&
+20048000\,\left(2 , 0\right)\,\left(3 , 0\right)^2
+8144150\,\left(1 , 1\right)\,\left(3 , 0\right)^2
+10255140\,\left(1 , 0\right)\,\left(2 , 2\right)\,\left(3 , 0\right)
\nonumber\\
&
-68817420\,\left(2 , 0\right)\,\left(2 , 1\right)\,\left(3 , 0\right)
+23950080\,\left(1 , 1\right)\,\left(2 , 1\right)\,\left(3 , 0\right)
+42000\,\left(1 , 0\right)^2\,\left(2 , 1\right)\,\left(3 , 0\right)
\nonumber\\
&
-42000\,\left(1 , 0\right)\,\left(1 , 1\right)\,\left(2 , 0\right)\,\left(3 , 0\right)
+618823800\,\left(2 , 2\right)^2
+11264400\,\left(1 , 0\right)\,\left(2 , 1\right)\,\left(2 , 2\right)
\nonumber\\
&
+19192950\,\left(2 , 0\right)^2\,\left(2 , 2\right)
-35002800\,\left(1 , 1\right)\,\left(2 , 0\right)\,\left(2 , 2\right)
+63000\,\left(1 , 0\right)^2\,\left(2 , 0\right)\,\left(2 , 2\right)
\nonumber\\
&
+12448800\,\left(1 , 1\right)^2\,\left(2 , 2\right)
+23738400\,\left(2 , 0\right)\,\left(2 , 1\right)^2
-12448800\,\left(1 , 1\right)\,\left(2 , 1\right)^2
\nonumber\\
&
-126000\,\left(1 , 0\right)\,\left(2 , 0\right)^2\,\left(2 , 1\right)
+63000\,\left(1 , 1\right)\,\left(2 , 0\right)^3
,
\end{align}
}%


%
%
%
The second level $8$ scalar with
the maximum number of terms the highest
occurrences of zero modes
has again $65$ terms
out of $102$ possible and reads
{\footnotesize
\begin{align}
|S8z2\rangle
=&
-46478250\,\left(8 , 0\right)
+359941680\,\left(7 , 1\right)
-848880\,\left(1 , 0\right)\,\left(7 , 0\right) 
\nonumber\\ 
&
-1931385960\,\left(6 , 2\right)
-95941440\,\left(1 , 0\right)\,\left(6 , 1\right)
-55667640\,\left(2 , 0\right)\,\left(6 , 0\right) 
\nonumber\\ 
&
106609440\,\left(1 , 1\right)\,\left(6 , 0\right)
+5264441616\,\left(5 , 3\right)
+300357792\,\left(1 , 0\right)\,\left(5 , 2\right) 
\nonumber\\ 
&
+324832032\,\left(2 , 0\right)\,\left(5 , 1\right)
-328898304\,\left(1 , 1\right)\,\left(5 , 1\right)
+604800\,\left(1 , 0\right)^2\,\left(5 , 1\right) 
\nonumber\\ 
&
+118298544\,\left(3 , 0\right)\,\left(5 , 0\right)
-220237920\,\left(2 , 1\right)\,\left(5 , 0\right)
+8963136\,\left(1 , 0\right)\,\left(2 , 0\right)\,\left(5 , 0\right) 
\nonumber\\ 
&
-9265536\,\left(1 , 0\right)\,\left(1 , 1\right)\,\left(5 , 0\right)
-3654567000\,\left(4 , 4\right)
-208166280\,\left(1 , 0\right)\,\left(4 , 3\right) 
\nonumber\\ 
&
-1028682900\,\left(2 , 0\right)\,\left(4 , 2\right)
+439123230\,\left(1 , 1\right)\,\left(4 , 2\right)
-10779615\,\left(1 , 0\right)^2\,\left(4 , 2\right) 
\nonumber\\ 
&
-10546200\,\left(3 , 0\right)\,\left(4 , 1\right)
+453611970\,\left(2 , 1\right)\,\left(4 , 1\right)
-802305\,\left(1 , 0\right)\,\left(2 , 0\right)\,\left(4 , 1\right) 
\nonumber\\ 
&
10069920\,\left(1 , 0\right)\,\left(1 , 1\right)\,\left(4 , 1\right)
-64519875\,\left(4 , 0\right)^2
-194265120\,\left(3 , 1\right)\,\left(4 , 0\right) 
\nonumber\\ 
&
-9248400\,\left(1 , 0\right)\,\left(3 , 0\right)\,\left(4 , 0\right)
+393876000\,\left(2 , 2\right)\,\left(4 , 0\right)
+24652215\,\left(1 , 0\right)\,\left(2 , 1\right)\,\left(4 , 0\right) 
\nonumber\\ 
&
-17038350\,\left(2 , 0\right)^2\,\left(4 , 0\right)
+13616505\,\left(1 , 1\right)\,\left(2 , 0\right)\,\left(4 , 0\right)
-10069920\,\left(1 , 1\right)^2\,\left(4 , 0\right) 
\nonumber\\ 
&
772506560\,\left(2 , 0\right)\,\left(3 , 3\right)
-191999500\,\left(1 , 1\right)\,\left(3 , 3\right)
+11726750\,\left(1 , 0\right)^2\,\left(3 , 3\right) 
\nonumber\\ 
&
-134722560\,\left(3 , 0\right)\,\left(3 , 2\right)
-305834760\,\left(2 , 1\right)\,\left(3 , 2\right)
-13549620\,\left(1 , 0\right)\,\left(2 , 0\right)\,\left(3 , 2\right) 
\nonumber\\ 
&
-512400\,\left(1 , 0\right)\,\left(1 , 1\right)\,\left(3 , 2\right)
-21000\,\left(1 , 0\right)^3\,\left(3 , 2\right)
+464137100\,\left(3 , 1\right)^2 
\nonumber\\ 
&
-11458300\,\left(1 , 0\right)\,\left(3 , 0\right)\,\left(3 , 1\right)
-1064138040\,\left(2 , 2\right)\,\left(3 , 1\right)
-19210800\,\left(1 , 0\right)\,\left(2 , 1\right)\,\left(3 , 1\right) 
\nonumber\\ 
&
+21328860\,\left(2 , 0\right)^2\,\left(3 , 1\right)
-416640\,\left(1 , 1\right)\,\left(2 , 0\right)\,\left(3 , 1\right)
+21000\,\left(1 , 0\right)^2\,\left(2 , 0\right)\,\left(3 , 1\right) 
\nonumber\\ 
&
+17416000\,\left(2 , 0\right)\,\left(3 , 0\right)^2
+6809950\,\left(1 , 1\right)\,\left(3 , 0\right)^2
+7657020\,\left(1 , 0\right)\,\left(2 , 2\right)\,\left(3 , 0\right) 
\nonumber\\ 
&
-57671460\,\left(2 , 0\right)\,\left(2 , 1\right)\,\left(3 , 0\right)
+20139840\,\left(1 , 1\right)\,\left(2 , 1\right)\,\left(3 , 0\right)
+21000\,\left(1 , 0\right)^2\,\left(2 , 1\right)\,\left(3 , 0\right) 
\nonumber\\ 
&
-21000\,\left(1 , 0\right)\,\left(1 , 1\right)\,\left(2 ,0\right)\,\left(3 , 0\right)
+530541900\,\left(2 , 2\right)^2
+9601200\,\left(1 , 0\right)\,\left(2 , 1\right)\,\left(2 , 2\right) 
\nonumber\\ 
&
+17038350\,\left(2 , 0\right)^2\,\left(2 , 2\right)
-30038400\,\left(1 , 1\right)\,\left(2 , 0\right)\,\left(2 , 2\right)
+10823400\,\left(1 , 1\right)^2\,\left(2 , 2\right)
\nonumber\\
&
+31500\,\left(1 , 0\right)^2\,\left(1 , 1\right)\,\left(2 , 2\right)
+20437200\,\left(2 , 0\right)\,\left(2 , 1\right)^2
-10823400\,\left(1 , 1\right)\,\left(2 , 1\right)^2
\nonumber\\
&
-63000\,\left(1 , 0\right)\,\left(1 , 1\right)\,\left(2 ,0\right)\,\left(2 , 1\right)
+31500\,\left(1 , 1\right)^2\,\left(2 , 0\right)^2
,
\end{align}
}%

The states at level $10$ used are given in
appendix \ref{app:level10states} since they are quite uninspiring.
The state at level $11$ is only in the supplementary material.

\section{Amplitudes}
\label{sec:amplitudes}

We now discuss some examples of the computed amplitudes.
They are all qualitatively similar.
Whenever we do not write $\ap$ we set
\begin{equation}
\ap=2
.
\end{equation}

The amplitudes we have computed are the following:
\begin{itemize}
\item
with 3 tachyons:
\begin{itemize}
\item ``S4zTTT''
\item ``S6zTTT''
\item ``S81TTT''
\item ``S8z2TTT''
\item ``S10z1TTT''
\item ``S10z2TTT''
\item ``S11zTTT''
\end{itemize}

\item
with 2 tachyons:
\begin{itemize}
\item ``S4zS4zTT''
\item ``S6zS4zTT''
\item ``S6zS6zTT''
\item ``S8z1S4zTT''
\item ``S8z1S6zTT''
\item ``S8z2S4zTT''
\item ``S8z1S8z2TT''
\item ``S10z1S10z1zTT''
\item ``S11zS4zTT''
\item ``S11zS6zTT''
\end{itemize}

\item
with 1 tachyons:
\begin{itemize}
\item ``S4zS4zS4zT''
\item ``S6zS4zS4zT''
\item ``S8z1S6zS4zT''
\end{itemize}

\item
with no tachyons:
\begin{itemize}
\item ``S4zS4zS4zS4z''
\end{itemize}

\end{itemize}

Since many of these are similar to the human eye we report in the main
text only some of them.
All the others are in the supplementary material along with
instructions on how to get the tex files and the plots.
In particular we choose the following amplitudes
\begin{itemize}
\item
``S4zTTT'' since it is the simplest amplitude with one massive
scalar which is not a tachyon and can be easily inspected;

\item
``S6zS4zTT'' since it is the simplest amplitude which can be derived
from the same correlator but differs from the amplitude ``S6z T S4z
T'' in a not predictable way

\item
``S4zS4zS4zS4z'' since it is the simplest amplitude with four equal
massive scalars and can be used as benchmark for scalar massive
amplitudes in the theory of analytical S matrix.
It is quite big but not as big as the intermediate steps which involve
about $O(10^6)$ terms.

\end{itemize}

We start from the Veneziano amplitude to set the conventions and show
the effect of changing the energy flowing into the $s$ channel.
We show also the effect of changing slightly the $s$ from a non
resonant one to a resonant one.

\subsection{Veneziano amplitude and notations}
In the following we use the function
\begin{align}
A_{\mbox{ Veneziano } 1234}(s,u)
&=
\frac{ \Gamma(-\ap s-1)\, \Gamma(-\ap u-1)}{\Gamma(-\ap s-\ap u-2)}
,
\label{eq:Ven1234}
\end{align}
which is the partial color ordered Veneziano amplitude.
In the function  \eqref{eq:Ven1234} we do not use the relation among Mandelstam
variables
\begin{align}
s+t+u =& \sum_{r=1}^4\, m_{r}^2,
\end{align}
where we have defined\footnote{
These are the definitions used in Polchinski's
book \cite{Polchinski:1998rq} which differ from
the ones used in \cite{Green:1987sp} where
$
$
}
\begin{align}
s=-(k_1+k_2)^2,&&~~
t=&-(k_1+k_3)^2,&~~
u=&-(k_1+k_4)^2
.
\end{align}
Notice that with the previous definition of Mandelstam variables the 
partial color ordered Veneziano amplitude is naturally function of $s$
and $u$ and this is the reason of the definition in eq.  \eqref{eq:Ven1234}.

Upon the use of the relation among Mandelstam variables for the
tachyons
\begin{equation}
s+t+u = -\frac{4}{\ap}
,
\end{equation}
the complete $U(1)$ Veneziano amplitude reads then
\begin{align}
A& {}_{\mbox{ Veneziano }}(s,u,t)
=
\nonumber\\
&=
A_{\mbox{ Veneziano } 1234}(s,u)
+A_{\mbox{ Veneziano } 1234}(u,t)
+A_{\mbox{ Veneziano } 1234}(t,s)
\nonumber\\
&=
\frac{1}{\pi}\,\Gamma(-\ap s-1)\, \Gamma(-\ap t-1)
\,\Gamma(-\ap u-1)\,
\left(
\sin(\pi \ap s) + \sin(\pi \ap t) + \sin(\pi \ap u) 
\right)
\nonumber\\
&=
\frac{1}{\pi}\,\Gamma(-\ap s-1)\, \Gamma(-\ap t-1)
\,\Gamma(\ap s+\ap t+3)\,
\left(
\sin(\pi \ap s) + \sin(\pi \ap t) - \sin(\pi \ap (s+t)) 
\right)
,
\end{align}     
which has poles at $\ap s= 2 n +1$ but not at $\ap s= 2 n$
while the partial amplitude has poles both at $\ap s= 2 n +1$ and $\ap s= 2 n$.

In the center of mass of tachyons $1$ ad $2$ we can write
\begin{align}
(k_{1\,\mu}) =& \begin{pmatrix} E \\ p \\ 0 \\ \vec 0 \end{pmatrix},
&
(k_{2\,\mu}) =& \begin{pmatrix} E \\ -p \\ 0 \\ \vec 0 \end{pmatrix},
&
(k_{3\,\mu})
=& \begin{pmatrix}
-E \\
-p\cos\theta \end{pmatrix},
&
(k_{4\,\mu})
=& \begin{pmatrix}
-E \\
+p\cos\theta \end{pmatrix},
\nonumber\\
&
&
E^2-p^2&=-\frac{1}{\ap}.
&
\end{align}
The Mandelstam variables then read
\begin{align}
s=& 4 E^2>0,
&
t=&-4\,p^2\,\sin^2 \frac{\theta}{2}<0,
&
u=&-4\,p^2\,\cos^2 \frac{\theta}{2}<0,
&
\end{align}
We can then plot $A_{\mbox{ Veneziano } 1234}(s,u(s, \theta))$ and
$A_{\mbox{ Veneziano }}(s,\theta)$.
The plots are in figures
\ref{par_fig:veneziano_partial_vs_full_even}
,\ref{par_fig:veneziano_partial_vs_full_odd}
and \ref{par_fig:veneziano_partial_vs_full_no_pole}.

\begin{figure}[h]
\input{Partial_vs_Full_Veneziano_s_9.00_tex.tex}
\caption{
Example of
partial color ordered Veneziano amplitude $\Re \log A_{Veneziano\, 1234}(s,\theta)$
and
complete $U(1)$ Veneziano amplitude $\Re \log
A_{Veneziano}(s, \theta)$
when only the partial amplitude has a pole.
In particular when $\ap s= 2 n$ only the color ordered partial Veneziano
amplitude has a pole and this is seen by the fact that its value is
bigger and resonances are smoothed out \wrt the complete $U(1)$
Veneziano amplitude.
We have plotted the graphs with two different but small $\Im s$
in order to avoid poles and show the sensibility on the imaginary
part.
}
\label{par_fig:veneziano_partial_vs_full_even}
\end{figure}

\begin{figure}[h]
\input{Partial_vs_Full_Veneziano_s_9.50_tex.tex}
\caption{
When when $\ap s= 2 n+1$ both amplitudes have poles and their plots
are very similar.
}
\label{par_fig:veneziano_partial_vs_full_odd}  
\end{figure}
\begin{figure}[h]
\input{Partial_vs_Full_Veneziano_s_9.25_tex.tex}
\caption{
Finally in all other case, such f.x. when $\ap s= 2 n+ \oh $
neither amplitudes have a pole.
}
\label{par_fig:veneziano_partial_vs_full_no_pole}  
\end{figure}

Actually for ingoing tachyons there is also the more exotic
configuration where the center of mass is tachyonic and we cannot
observe the scattering in the center of mass.
The usual case has $s>0$ this exotic one has $s<0$.
This case can be thought of two tachyons moving in the same direction
$x^1$ but one faster than the other so that there is a scattering when
they meet.
The difference with the usual case is that the center of mass is
tachyonic and therefore we cannot perform a boost in a reference frame
where it is at rest.
This configuration can be written with $\beta_2, \beta_3, \beta_4>0$
\begin{align}
(k_{1\,\mu}) =& \begin{pmatrix} 0 \\ |m_T| \\ 0 \\ \vec 0 \end{pmatrix},
&
(k_{2\,\mu}) =& \begin{pmatrix} |m_T| \sinh \beta_2 \\ |m_T| \cosh \beta_2
\\ \vec 0 \end{pmatrix},
\nonumber\\
(k_{3\,\mu}) =& \begin{pmatrix} -|m_T| \sinh \beta_3
\\ \vec 0 \end{pmatrix},
&
(k_{4\,\mu}) =& \begin{pmatrix} -|m_T| \sinh \beta_4
\\ \vec 0 \end{pmatrix},
\end{align}
but this not in a canonical form.
%
The canonical configuration can be written as
\begin{align}
(k_{1\,\mu}) =& \begin{pmatrix} E \\ k \\ 0 \\ \vec 0 \end{pmatrix},
&
(k_{2\,\mu}) =& \begin{pmatrix} -E \\ k \\ 0 \\ \vec 0 \end{pmatrix},
&
(k_{3\,\mu})=& \begin{pmatrix} -E_3 \\ -k \\ \vec 0 \end{pmatrix},
&
(k_{4\,\mu}) =& \begin{pmatrix} +E_3 \\ - k \\ \vec 0 \end{pmatrix},
\nonumber\\
&&
E^2-k^2&=-\frac{1}{\ap},
&
&
E_3^2-\frac{1}{\cos^2 \theta}\,k^2&=-\frac{1}{\ap}.
&
\end{align}
Notice that for this configuration $k_2$ has negative energy despite
the fact it is an incoming tachyon and we use the parametrization for
the second component of the momentum as $k \tan \theta$ since it may
take all possible values.

The Mandelstam variables then read
\begin{align}
s=& -4 k^2<0,
\nonumber\\
t=&
2
=
-2\,\sqrt{k^2 + m_T^2}\,
\left(
\sqrt{\frac{1}{\cos^2 \theta} k^2 + m_T^2} - \sqrt{k^2 + m_T^2}
\right)
\nonumber\\
u=&
2
=
2\,\sqrt{k^2 + m_T^2}\,
\left(
\sqrt{\frac{1}{\cos^2 \theta} k^2 + m_T^2} - \sqrt{k^2 + m_T^2}
\right)
.
\end{align}
In figure \ref{fig:some_partial_no_pole}  we show the plot of the Veneziano
amplitude for some values of $s>0$ and one $s<0$ for comparison.

\begin{figure}[!h]
  \centering
    \begin{subfigure}{0.49\textwidth}
    \centering    
   \resizebox{1.0\textwidth}{!}{
\input{Full_Veneziano_no_pole_s_9.63_to_s_69.63_tex.tex}
   }
    \end{subfigure}
    \begin{subfigure}{0.49\textwidth}
    \centering    
   \resizebox{1.0\textwidth}{!}{
\input{Full_Veneziano_with_pole_s_9.50_to_s_69.50_tex.tex}
   }
    \end{subfigure}
\caption{
Partial color ordered Veneziano amplitudes for different values of
$s$.
In particular also for a tachyonic center of mass.
}
\label{fig:some_partial_no_pole}
\end{figure}
%
%



\subsection{ Amplitude "S4zTTT"}

For this amplitude the Mandelstam variables ($\alpha'=2$) satisfy

\begin{align}
s+t+u= 0,
%
%
%
%
\end{align}
where ``$s \ge \sqrt{3}\,i+1$'' (i.e. all $s$ are possible and give $|cos
\theta|<=1$) 
in the color ordered amplitude $ "S4zTTT" $.
%


\subsubsection{ Color ordered amplitude $ "S4zTTT" $} 
 
%
%
The color ordered correlator is given by
\tiny
\begin{dgroup}
\begin{dmath*}
\mbox{C}_{S4zTTT}(\omega ,\mys, \myu)
=
\end{dmath*}
  \begin{dmath*}
  ( 16 )
   ( \left(\omega-1\right)^{2\,\myt+2\,\mys-2} )
   ( \omega^{-2\,\mys-2} )
  \end{dmath*}
\begin{dmath*}
\Bigg(
      + 
 \left(62\,\myt^2-49\,\myt-127\right)\,\omega^4 
\end{dmath*}
\begin{dmath*}
 -\left(294\,\myt^2+340\,\mys\,\myt+29\,\myt+254\,\mys-254\right)\,\omega^3 
\end{dmath*}
\begin{dmath*}
      + 
 3\,\left(98\,\myt^2+268\,\mys\,\myt+127\,\myt+98\,\mys^2+127\,\mys-127\right)\,\omega^2 
\end{dmath*}
\begin{dmath*}
 -\left(340\,\mys\,\myt+254\,\myt+294\,\mys^2+29\,\mys-254\right)\,\omega 
\end{dmath*}
\begin{dmath*}
      + 
 62\,\mys^2 
 -49\,\mys 
 -127 
\Bigg)
\end{dmath*}

%
%
%
\begin{dmath}
\,
\end{dmath}
\end{dgroup}
\normalsize
%
%

%
%

The color ordered amplitude can be written as a product of the Veneziano amplitude
times a polynomial $\mbox{Num}(\mys ,\myu)$
in the Mandelstam variables $\mys$ and $\myu$
divided by another polynomial $\mbox{Den}(\mys ,\myu)$ which comes
from the transformation of the $\Gamma$ function in the denominator
which arises from the moduli space integration to the Veneziano "canonical" one.
Explicitly we get
\begin{equation}
A_{S4zTTT}
=
A_{\mbox{Veneziano} 
}(\mys, \myu)
\frac{\mbox{Num}(\mys ,\myu)}{\mbox{Den}(\mys, \myu)}
,
\end{equation}
where we have
\tiny
\begin{dgroup}
\begin{dmath*}
\mbox{Num}(\mys ,\myu)
=
2784\,\mys\,\left(-\myu\,\left(12\,\myu^2-5\right)-4\,\mys\,\myu^2\,\left(4\,\myu+3\right)-4\,\mys^2\,\myu\,\left(4\,\myu+3\right)\right)
\end{dmath*}
\begin{dmath}
,%
\end{dmath}
\end{dgroup}
\normalsize
and
\tiny
\begin{dgroup}
\begin{dmath}
\mbox{Den}(\mys ,\myu)
=
-\mys\,\left(12\,\myu^2+8\,\myu-1\right)-4\,\mys^2\,\left(3\,\myu+1\right)-\left(\myu+1\right)\,\left(2\,\myu-1\right)\,\left(2\,\myu+1\right)-4\,\mys^3
.
\end{dmath}
\end{dgroup}
\normalsize

The previous quantities in channel $\mys$ can be written
using $\mys$ and $\cos(\theta)$
(``$\mys \ge \sqrt{3}\,i+1$'') with
\begin{align}
\myt =& {{\sqrt{\mys+2}\,\sqrt{\mys^2-2\,\mys+4}\,\cos \theta-\mys^{{{3}\over{2}}}}\over{2\,\sqrt{\mys}}}
\nonumber \\
\myu =& -{{\sqrt{\mys+2}\,\sqrt{\mys^2-2\,\mys+4}\,\cos \theta+\mys^{{{3}\over{2}}}}\over{2\,\sqrt{\mys}}}
\end{align}
as

\tiny
\begin{dgroup}
\begin{dmath*}
\mbox{Num}(s,\cos(\theta))
=
\end{dmath*}
\begin{dmath*}
\Bigg(
      + 
 2784\,\mys\,\left(\mys+2\right)^{{{3}\over{2}}}\,\left(4\,\mys+3\right)\,\left(\mys^2-2\,\mys+4\right)^{{{3}\over{2}}}\,\cos \left(3\,\theta\right) 
\end{dmath*}
\begin{dmath*}
      + 
 5568\,\mys^{{{5}\over{2}}}\,\left(\mys+2\right)\,\left(4\,\mys+3\right)\,\left(\mys^2-2\,\mys+4\right)\,\cos \left(2\,\theta\right) 
\end{dmath*}
\begin{dmath*}
 -2784\,\mys\,\sqrt{\mys+2}\,\sqrt{\mys^2-2\,\mys+4}\,\left(4\,\mys^4-45\,\mys^3-76\,\mys-72\right)\,\cos \theta 
\end{dmath*}
\begin{dmath*}
 -5568\,\mys^{{{5}\over{2}}}\,\left(4\,\mys^4-21\,\mys^3-22\,\mys-24\right)\,\, 
\Bigg)
\end{dmath*}

%
%
%
\begin{dmath}
  \,
  \label{eq:Num}
\end{dmath}
\end{dgroup}
\normalsize

and

\tiny
\begin{dgroup}
\begin{dmath*}
\mbox{Den}(s,\cos(\theta))
=
\end{dmath*}
\begin{dmath*}
\Bigg(
      + 
 \left(\mys+2\right)^{{{3}\over{2}}}\,\left(\mys^2-2\,\mys+4\right)^{{{3}\over{2}}}\,\cos \left(3\,\theta\right) 
\end{dmath*}
\begin{dmath*}
 -2\,\sqrt{\mys}\,\left(\mys+2\right)\,\left(3\,\mys+2\right)\,\left(\mys^2-2\,\mys+4\right)\,\cos \left(2\,\theta\right) 
\end{dmath*}
\begin{dmath*}
      + 
 \sqrt{\mys+2}\,\sqrt{\mys^2-2\,\mys+4}\,\left(15\,\mys^3+16\,\mys^2-4\,\mys+24\right)\,\cos \theta 
\end{dmath*}
\begin{dmath*}
 -2\,\sqrt{\mys}\,\left(\mys+2\right)\,\left(5\,\mys^3-4\,\mys^2+6\,\mys+8\right)\,\, 
\Bigg)
\end{dmath*}

%
%
%
\begin{dmath}
\,
\end{dmath}
\end{dgroup}
\normalsize

We can plot the amplitude for some values.
In particular we can look to the amplitude near or far the resonances
(fig. \ref{fig:S4TTT_special_s})
and for quite different $\mys$ values (fig. \ref{fig:S4TTT_generic_s}).

In particular we can also consider the contribution to the amplitude of $Num(\mys,
\cos \theta)$ ,  the one without the
Veneziano amplitude which shows new poles wrt Veneziano and the
complete amplitude (fig. \ref{fig:S4TTT_generic_s}).

 \begin{figure}[!h]
   \centering
    \begin{subfigure}{0.49\textwidth}
    \centering    
    \resizebox{1.0\textwidth}{!}{\input{S4zTTT.Num.s_min_s_6.01_14.1_22.3_steps_1000_tex.tex}}
    \end{subfigure}
    \hfill
    \begin{subfigure}{0.49\textwidth}
    \centering    
    \resizebox{1.0\textwidth}{!}{\input{S4zTTT.log.Num.s_min_s_6.01_14.1_22.3_steps_1000_tex.tex}}
    \end{subfigure}
    \begin{subfigure}{0.49\textwidth} 
    \centering    
   \resizebox{1.0\textwidth}{!}{\input{S4zTTT.withOUT.s_min_s_6.01_14.1_22.3_steps_1000_tex.tex}}
    \end{subfigure}
    \hfill
    \begin{subfigure}{0.49\textwidth}
    \centering    
   \resizebox{1.0\textwidth}{!}{\input{S4zTTT.log.withOUT.s_min_s_6.01_14.1_22.3_steps_1000_tex.tex}}
    \end{subfigure}
    \begin{subfigure}{0.49\textwidth} 
    \centering    
   \resizebox{1.0\textwidth}{!}{\input{S4zTTT.with.s_min_s_6.01_14.1_22.3_steps_1000_tex.tex}}
    \end{subfigure}
    \hfill
    \begin{subfigure}{0.49\textwidth}
    \centering    
   \resizebox{1.0\textwidth}{!}{\input{S4zTTT.log.with.s_min_s_6.01_14.1_22.3_steps_1000_tex.tex}}
    \end{subfigure}
    \caption{
      Different perspectives on the $S4 T T T$ amplitude.
      We use generic $s$ values, i.e. not close to resonances.
      On the left the Numerator in eq. \eqref{eq:Num},
      then the Numerator and the Denominator with the new poles
      and finally the complete color ordered amplitude with the
      Veneziano contribution.
      On the right there are exactly the same amplitudes but we plot
      their $\log$.
}
\label{fig:S4TTT_generic_s}
\end{figure}

  \begin{figure}[!h]
   \centering
    \begin{subfigure}{0.49\textwidth}
    \centering    
    \resizebox{1.0\textwidth}{!}{\input{S4zTTT.Num.s_10.01_10.25_10.51_steps_1000_tex.tex}}
    \end{subfigure}
    \hfill
    \begin{subfigure}{0.49\textwidth}
    \centering    
    \resizebox{1.0\textwidth}{!}{\input{S4zTTT.log.Num.s_10.01_10.25_10.51_steps_1000_tex.tex}}
    \end{subfigure}
    \begin{subfigure}{0.49\textwidth} 
    \centering    
   \resizebox{1.0\textwidth}{!}{\input{S4zTTT.withOUT.s_10.01_10.25_10.51_steps_1000_tex.tex}}
    \end{subfigure}
    \hfill
    \begin{subfigure}{0.49\textwidth}
    \centering    
   \resizebox{1.0\textwidth}{!}{\input{S4zTTT.log.withOUT.s_10.01_10.25_10.51_steps_1000_tex.tex}}
    \end{subfigure}
    \begin{subfigure}{0.49\textwidth} 
    \centering    
   \resizebox{1.0\textwidth}{!}{\input{S4zTTT.with.s_10.01_10.25_10.51_steps_1000_tex.tex}}
    \end{subfigure}
    \hfill
    \begin{subfigure}{0.49\textwidth}
    \centering    
   \resizebox{1.0\textwidth}{!}{\input{S4zTTT.log.with.s_10.01_10.25_10.51_steps_1000_tex.tex}}
    \end{subfigure}
    \caption{
      Different perspectives on the $S4 T T T$ amplitude.
      We use  two resonant $s$ values and one not resonant to show the difference
      in magnitude of the amplitudes.
      On the left the Numerator in eq. \eqref{eq:Num},
      then the Numerator divided by the Denominator with the new poles
      and finally the complete color ordered amplitude with the
      Veneziano contribution.
      On the right there are exactly the same amplitudes but we plot
      their $\log$.
}
\label{fig:S4TTT_special_s}
\end{figure}

From the perspective of the erratic behavior we can consider the
contribution of $\cos( 0 \theta)$ which corresponds to a spin $0$
exchange with respect to higher spins $\cos( k \theta)$ (fig. \ref{fig:S4TTT_ratios}).
This shows clearly the spin $0$ dominance.

  \begin{figure}[!h]
   \centering
    \centering    
    \resizebox{1.0\textwidth}{!}{\input{S4zTTT.log.ratios.s_0.0_64.1109_tex.tex}}
    \caption{
      Log of the ratio of the coefficient of $\cos( k \theta)$  to the
      terms independent of $\theta$ of the Numerator.
      This shows the dominance of the spin $0$ particles exchange and
      explain why the amplitude is not erratic.
}
\label{fig:S4TTT_ratios}
\end{figure}


\section{ Amplitude "S6zS4zTT" and non cyclically associated amplitudes}

For this amplitude the Mandelstam variables ($\alpha'=2$) satisfy

\begin{align}
s+t+u= 3,
%
%
%
%
\end{align}
where f.x. $s \ge \sqrt{3}\,\sqrt{5}+4$
for the color ordered amplitude "S6zS4zTT" in order to hace
$\cos \theta\in \R$.

For all amplitudes given in the following
the Mandelstam variables are the  ones associated with the corresponding order,
e.g. $s_{1 3 2 4} = t_{1 2 3 4}$,
$t_{1 3 2 4} = s_{1 2 3 4}$ and
$u_{1 3 2 4} = u_{1 2 3 4}$.

We notice that despite all the non cyclically equivalent amplitudes
can be derived from the same correlator (as shown in appendix \ref{app:othercorrsandamps})
they are quite different as the
following ``preview'' shows (fig. \ref{fig:comparison_6400_6040_6004})
 \begin{figure}[h]
   \centering
    \begin{subfigure}{0.30\textwidth}
    \centering    
    \resizebox{1.0\textwidth}{!}{\input{S6zS4zTT.log.Num.s_min_s_14.1_22.3_steps_1000_tex.tex}}
    \end{subfigure}
    \hfill
    \begin{subfigure}{0.30\textwidth}
    \centering    
    \resizebox{1.0\textwidth}{!}{\input{S6zTTS4z.log.Num.s_min_s_14.1_22.3_steps_1000_tex.tex}}
    \end{subfigure}
     \hfill
    \begin{subfigure}{0.30\textwidth}
    \centering    
    \resizebox{1.0\textwidth}{!}{\input{S6zTS4zT.log.Num.s_min_s_14.1_22.3_steps_1000_tex.tex}}
    \end{subfigure}
   \centering
    \begin{subfigure}{0.30\textwidth}
    \centering    
    \resizebox{1.0\textwidth}{!}{\input{S6zS4zTT.log.with.s_min_s_14.1_22.3_steps_1000_tex.tex}}
    \end{subfigure}
    \hfill
    \begin{subfigure}{0.30\textwidth}
    \centering    
    \resizebox{1.0\textwidth}{!}{\input{S6zTTS4z.log.with.s_min_s_14.1_22.3_steps_1000_tex.tex}}
    \end{subfigure}
     \hfill
    \begin{subfigure}{0.30\textwidth}
    \centering    
    \resizebox{1.0\textwidth}{!}{\input{S6zTS4zT.log.with.s_min_s_14.1_22.3_steps_1000_tex.tex}}
    \end{subfigure}
        \caption{
     ``S6zS4zTT'' vs ``S6zTTS4z'' vs ``S6zTS4z1T'' amplitudes with the
          same $s$ values.
          They differ both in the magnitude and in the shape,
          especially in the forward direction.
}
\label{fig:comparison_6400_6040_6004}
\end{figure}

\subsubsection{ Color ordered amplitude $ "S6zS4zTT" $}

%
%
The color ordered correlator is given by
\tiny
\begin{dgroup}
\begin{dmath*}
\mbox{C}_{S6zS4zTT}(\omega ,\mys, \myu)
=
\end{dmath*}
  \begin{dmath*}
  ( 32 )
  ( \left(\omega-1\right)^{2\,\myt+2\,\mys-8} )
  ( \omega^{-2\,\mys-2} )
  \end{dmath*}
\begin{dmath*}
\Bigg(
      + 
 4\,\left(62\,\myt^2-49\,\myt-127\right)\,\left(6750\,\myt^3-18941\,\myt^2-12736\,\myt+24156\right)\,\omega^{10} 
\end{dmath*}
\begin{dmath*}
 -4\,\left(4495500\,\myt^5+4806000\,\mys\,\myt^4-12455664\,\myt^4-464408\,\mys\,\myt^3-4202101\,\myt^3-13384592\,\mys\,\myt^2+36573413\,\myt^2-867656\,\mys\,\myt-6451882\,\myt+6135624\,\mys-24542496\right)\,\omega^9 
\end{dmath*}
\begin{dmath*}
      + 
 2\,\left(37827000\,\myt^5+82829200\,\mys\,\myt^4-59692188\,\myt^4+39575200\,\mys^2\,\myt^3+38987880\,\mys\,\myt^3+63707592\,\myt^3+67366152\,\mys^2\,\myt^2+50572718\,\mys\,\myt^2-252651488\,\myt^2-650332\,\mys^2\,\myt+245724297\,\mys\,\myt+1324626282\,\myt-24295464\,\mys^2-86728386\,\mys-775766080\right)\,\omega^8 
\end{dmath*}
\begin{dmath*}
 -4\,\left(39069000\,\myt^5+136284400\,\mys\,\myt^4+14511420\,\myt^4+135208400\,\mys^2\,\myt^3+204974974\,\mys\,\myt^3+311869450\,\myt^3+37372000\,\mys^3\,\myt^2+271008030\,\mys^2\,\myt^2+843313363\,\mys\,\myt^2-1219044974\,\myt^2+85599584\,\mys^3\,\myt+415788871\,\mys^2\,\myt+273659065\,\mys\,\myt+3580843954\,\myt+45529812\,\mys^3+516301698\,\mys^2+1982527301\,\mys+131522958\right)\,\omega^7 
\end{dmath*}
\begin{dmath*}
      + 
 \left(158760000\,\myt^5+848970800\,\mys\,\myt^4+684755640\,\myt^4+1387549600\,\mys^2\,\myt^3+3105190760\,\mys\,\myt^3+4382350740\,\myt^3+823085600\,\mys^3\,\myt^2+4205527992\,\mys^2\,\myt^2+16296984556\,\mys\,\myt^2-9579145848\,\myt^2+149938800\,\mys^4\,\myt+2008775416\,\mys^3\,\myt+12150834536\,\mys^2\,\myt-1839590420\,\mys\,\myt+28727168688\,\myt+276387120\,\mys^4+2430744768\,\mys^3+14936745120\,\mys^2+60696732276\,\mys-3941902839\right)\,\omega^6 
\end{dmath*}
\begin{dmath*}
 -2\,\left(31752000\,\myt^5+304288200\,\mys\,\myt^4+518098680\,\myt^4+808102000\,\mys^2\,\myt^3+2754617004\,\mys\,\myt^3+4097950524\,\myt^3+808102000\,\mys^3\,\myt^2+4691134080\,\mys^2\,\myt^2+18936360524\,\mys\,\myt^2-565186686\,\myt^2+304288200\,\mys^4\,\myt+2840282604\,\mys^3\,\myt+18023769500\,\mys^2\,\myt-2265383630\,\mys\,\myt+18609760854\,\myt+31752000\,\mys^5+535108680\,\mys^4+4975231332\,\mys^3+25625737410\,\mys^2+118935134307\,\mys-37837102186\right)\,\omega^5 
\end{dmath*}
\begin{dmath*}
      + 
 2\,\left(74969400\,\mys\,\myt^4+252079560\,\myt^4+411542800\,\mys^2\,\myt^3+2232590708\,\mys\,\myt^3+3995888328\,\myt^3+693774800\,\mys^3\,\myt^2+5625685296\,\mys^2\,\myt^2+24061108094\,\mys\,\myt^2+9152451488\,\myt^2+424485400\,\mys^4\,\myt+4604299180\,\mys^3\,\myt+28887830706\,\mys^2\,\myt+3747033376\,\mys\,\myt+27877520700\,\myt+79380000\,\mys^5+1122610320\,\mys^4+9844823880\,\mys^3+52582359074\,\mys^2+284333546133\,\mys-109829288351\right)\,\omega^4 
\end{dmath*}
\begin{dmath*}
 -4\,\left(37372000\,\mys^2\,\myt^3+311178284\,\mys\,\myt^3+786909384\,\myt^3+135208400\,\mys^3\,\myt^2+1570733130\,\mys^2\,\myt^2+7924823242\,\mys\,\myt^2+5840267232\,\myt^2+136284400\,\mys^4\,\myt+1933034374\,\mys^3\,\myt+13272158120\,\mys^2\,\myt+7372532474\,\mys\,\myt+19342913868\,\myt+39069000\,\mys^5+643557420\,\mys^4+5707252724\,\mys^3+35060652780\,\mys^2+206640054681\,\mys-74558252825\right)\,\omega^3 
\end{dmath*}
\begin{dmath*}
      + 
 \left(79150400\,\mys^3\,\myt^2+1272139704\,\mys^2\,\myt^2+8144761444\,\mys\,\myt^2+9381359460\,\myt^2+165658400\,\mys^4\,\myt+3065342160\,\mys^3\,\myt+25310129596\,\mys^2\,\myt+30198759822\,\mys\,\myt+61649042436\,\myt+75654000\,\mys^5+1543707624\,\mys^4+15441194580\,\mys^3+120991128398\,\mys^2+713755809318\,\mys-218156684699\right)\,\omega^2 
\end{dmath*}
\begin{dmath*}
 -2\,\left(9612000\,\mys^4\,\myt+225628184\,\mys^3\,\myt+2331309696\,\mys^2\,\myt+5027964966\,\mys\,\myt+9977013424\,\myt+8991000\,\mys^5+226512672\,\mys^4+2715406102\,\mys^3+29849242224\,\mys^2+169776841413\,\mys-39908053696\right)\,\omega 
\end{dmath*}
\begin{dmath*}
      + 
 2\,\left(837000\,\mys^5+25623316\,\mys^4+377638706\,\mys^3+6166444857\,\mys^2+34466056430\,\mys-4988506712\right) 
\Bigg)
\end{dmath*}

%
%
%
\begin{dmath}
\label{eq:S6S4corr}
\,
\end{dmath}
\end{dgroup}
\normalsize
%
%

%
%

The color ordered amplitude can be written as a product of the Veneziano amplitude
times a polynomial $\mbox{Num}(\mys ,\myu)$
in the Mandelstam variables $\mys$ and $\myu$
(defined according to the ordering of states in the color ordered amplitude,
i.e
$\mys = \mys_{\mbox{S6zS4zTT}}$,
$\myt = \myt_{\mbox{S6zS4zTT}}$,
and
$\myu= \myu_{\mbox{S6zS4zTT}}$
)
divided by another polynomial $\mbox{Den}(\mys ,\myu)$ which comes
from the transformation of the $\Gamma$ function in the denominator
which arises from the moduli space integration to the Veneziano "canonical" one.
Explicitly we get
\begin{equation}
A_{S6zS4zTT}
=
A_{\mbox{Veneziano} 
}(\mys, \myu)
\frac{\mbox{Num}(\mys ,\myu)}{\mbox{Den}(\mys, \myu)}
,
\end{equation}
where we have
\tiny
\begin{dgroup}
\begin{dmath*}
\mbox{Num}(\mys ,\myu)
=
370383360\,\mys\,\myu\,\left(2\,\myu-1\right)\,\left(\mys\,\left(26136\,\myu^6-32184\,\myu^5-571358\,\myu^4+1807188\,\myu^3-1514533\,\myu^2-236934\,\myu+520614\right)+3\,\left(22560\,\myu^6-203040\,\myu^5+664580\,\myu^4-941880\,\myu^3+433019\,\myu^2+198363\,\myu-175072\right)+\mys^2\,\left(11800\,\myu^6-27792\,\myu^5+71726\,\myu^4-462832\,\myu^3+718484\,\myu^2-30817\,\myu-258330\right)+2\,\mys^3\,\left(1128\,\myu^6+7548\,\myu^5-34522\,\myu^4+53498\,\myu^3-75623\,\myu^2+15643\,\myu+35352\right)+2\,\mys^4\,\left(160\,\myu^6+1944\,\myu^5+3008\,\myu^4-17612\,\myu^3+9351\,\myu^2-1480\,\myu-4884\right)+4\,\mys^5\,\left(240\,\myu^5+376\,\myu^4+582\,\myu^3-1403\,\myu^2-1262\,\myu-360\right)+8\,\mys^6\,\left(4\,\myu+3\right)\,\left(5\,\myu+4\right)\,\left(6\,\myu^2-7\,\myu+6\right)+16\,\mys^7\,\myu\,\left(4\,\myu+3\right)\,\left(5\,\myu+4\right)\right)
\end{dmath*}
\begin{dmath}
,%
\end{dmath}
\end{dgroup}
\normalsize
and
\tiny
\begin{dgroup}
\begin{dmath}
\mbox{Den}(\mys ,\myu)
=
-2\,\mys\,\left(64\,\myu^7-392\,\myu^6+672\,\myu^5-70\,\myu^4-602\,\myu^3+273\,\myu^2+61\,\myu-21\right)-\mys^2\,\left(448\,\myu^6-2352\,\myu^5+3360\,\myu^4-280\,\myu^3-1806\,\myu^2+546\,\myu+61\right)-14\,\mys^3\,\left(64\,\myu^5-280\,\myu^4+320\,\myu^3-20\,\myu^2-86\,\myu+13\right)-7\,\mys^4\,\left(160\,\myu^4-560\,\myu^3+480\,\myu^2-20\,\myu-43\right)-28\,\mys^5\,\left(32\,\myu^3-84\,\myu^2+48\,\myu-1\right)-112\,\mys^6\,\left(4\,\myu^2-7\,\myu+2\right)-16\,\mys^7\,\left(8\,\myu-7\right)-\left(\myu-2\right)\,\left(\myu-1\right)\,\myu\,\left(\myu+1\right)\,\left(2\,\myu-7\right)\,\left(2\,\myu-3\right)\,\left(2\,\myu-1\right)\,\left(2\,\myu+1\right)-16\,\mys^8
.
\end{dmath}
\end{dgroup}
\normalsize

The previous quantities in channel $\mys$ can be written
using $\mys$ and $\cos(\theta)$
($\mys \ge \sqrt{3}\,\sqrt{5}+4$) with
\begin{align}
\myt =& {{\sqrt{\mys+2}\,\sqrt{\mys^2-8\,\mys+1}\,\cos \theta+\left(3-\mys\right)\,\sqrt{\mys}}\over{2\,\sqrt{\mys}}}
\nonumber \\
\myu =& -{{\sqrt{\mys+2}\,\sqrt{\mys^2-8\,\mys+1}\,\cos \theta+\left(\mys-3\right)\,\sqrt{\mys}}\over{2\,\sqrt{\mys}}}
\end{align}
as

\tiny
\begin{dgroup}
\begin{dmath*}
\mbox{Num}(s,\cos(\theta))
=
\end{dmath*}
\begin{dmath*}
\Bigg(
      + 
 370383360\,\mys^{{{3}\over{2}}}\,\left(\mys+2\right)^4\,\left(\mys^2-8\,\mys+1\right)^4\,\left(40\,\mys^4+282\,\mys^3+1475\,\mys^2+3267\,\mys+8460\right)\,\cos \left(8\,\theta\right) 
\end{dmath*}
\begin{dmath*}
      + 
 740766720\,\mys^2\,\left(\mys+2\right)^{{{7}\over{2}}}\,\left(2\,\mys-5\right)\,\left(\mys^2-8\,\mys+1\right)^{{{7}\over{2}}}\,\left(40\,\mys^4+282\,\mys^3+1475\,\mys^2+3267\,\mys+8460\right)\,\cos \left(7\,\theta\right) 
\end{dmath*}
\begin{dmath*}
      + 
 1481533440\,\mys^{{{3}\over{2}}}\,\left(\mys+2\right)^3\,\left(\mys^2-8\,\mys+1\right)^3\,\left(288\,\mys^6-466\,\mys^5-10729\,\mys^4-60563\,\mys^3-181652\,\mys^2-335202\,\mys+33840\right)\,\cos \left(6\,\theta\right) 
\end{dmath*}
\begin{dmath*}
 -740766720\,\mys^2\,\left(\mys+2\right)^{{{5}\over{2}}}\,\left(2\,\mys-5\right)\,\left(\mys^2-8\,\mys+1\right)^{{{5}\over{2}}}\,\left(200\,\mys^7-782\,\mys^6+2267\,\mys^5+23519\,\mys^4+213449\,\mys^3+539001\,\mys^2+1423482\,\mys-118440\right)\,\cos \left(5\,\theta\right) 
\end{dmath*}
\begin{dmath*}
 -1481533440\,\mys^{{{3}\over{2}}}\,\left(\mys+2\right)^2\,\left(\mys^2-8\,\mys+1\right)^2\,\left(200\,\mys^{10}-1246\,\mys^9+571\,\mys^8+19481\,\mys^7+133776\,\mys^6-959922\,\mys^5-3993067\,\mys^4-19423165\,\mys^3-16379432\,\mys^2+4595364\,\mys-236880\right)\,\cos \left(4\,\theta\right) 
\end{dmath*}
\begin{dmath*}
      + 
 740766720\,\mys^2\,\left(\mys+2\right)^{{{3}\over{2}}}\,\left(2\,\mys-5\right)\,\left(\mys^2-8\,\mys+1\right)^{{{3}\over{2}}}\,\left(360\,\mys^{10}-4758\,\mys^9+28035\,\mys^8-47493\,\mys^7-16274\,\mys^6+1234432\,\mys^5+11713695\,\mys^4+45519275\,\mys^3+79259324\,\mys^2-16194372\,\mys+710640\right)\,\cos \left(3\,\theta\right) 
\end{dmath*}
\begin{dmath*}
      + 
 1481533440\,\mys^{{{3}\over{2}}}\,\left(\mys+2\right)\,\left(\mys^2-8\,\mys+1\right)\,\left(640\,\mys^{13}-10624\,\mys^{12}+72098\,\mys^{11}-218895\,\mys^{10}+257427\,\mys^9+523574\,\mys^8+8822062\,\mys^7-25764079\,\mys^6-269346621\,\mys^5-902214476\,\mys^4-308769010\,\mys^3+206328488\,\mys^2-26621496\,\mys+947520\right)\,\cos \left(2\,\theta\right) 
\end{dmath*}
\begin{dmath*}
 -740766720\,\mys^2\,\sqrt{\mys+2}\,\left(2\,\mys-5\right)\,\sqrt{\mys^2-8\,\mys+1}\,\left(200\,\mys^{13}-4174\,\mys^{12}+37251\,\mys^{11}-164353\,\mys^{10}+387379\,\mys^9-351115\,\mys^8-6472129\,\mys^7+40771945\,\mys^6+512884529\,\mys^5+1937229237\,\mys^4+1947384626\,\mys^3-727863724\,\mys^2+75620040\,\mys-2368800\right)\,\cos \theta 
\end{dmath*}
\begin{dmath*}
 -370383360\,\mys^{{{3}\over{2}}}\,\left(1800\,\mys^{16}-42798\,\mys^{15}+416423\,\mys^{14}-2047153\,\mys^{13}+5051664\,\mys^{12}-4601000\,\mys^{11}-14772014\,\mys^{10}+149390746\,\mys^9+680172884\,\mys^8-2031102698\,\mys^7-22731336689\,\mys^6-39707907441\,\mys^5-2552403316\,\mys^4+8784970408\,\mys^3-2035774928\,\mys^2+170528880\,\mys-4737600\right)\,\, 
\Bigg)
\end{dmath*}

%
%
%
\begin{dmath}
  \,
  \label{eq:Num6400}
\end{dmath}
\end{dgroup}
\normalsize

and

\tiny
\begin{dgroup}
\begin{dmath*}
\mbox{Den}(s,\cos(\theta))
=
\end{dmath*}
\begin{dmath*}
\Bigg(
 -\sqrt{\mys}\,\left(\mys+2\right)^4\,\left(\mys^2-8\,\mys+1\right)^4\,\cos \left(8\,\theta\right) 
\end{dmath*}
\begin{dmath*}
      + 
 4\,\mys\,\left(\mys+2\right)^{{{7}\over{2}}}\,\left(4\,\mys+5\right)\,\left(\mys^2-8\,\mys+1\right)^{{{7}\over{2}}}\,\cos \left(7\,\theta\right) 
\end{dmath*}
\begin{dmath*}
 -8\,\sqrt{\mys}\,\left(\mys+2\right)^3\,\left(\mys^2-8\,\mys+1\right)^3\,\left(15\,\mys^3+29\,\mys^2-8\,\mys+2\right)\,\cos \left(6\,\theta\right) 
\end{dmath*}
\begin{dmath*}
      + 
 28\,\mys\,\left(\mys+2\right)^{{{5}\over{2}}}\,\left(\mys^2-8\,\mys+1\right)^{{{5}\over{2}}}\,\left(20\,\mys^4+41\,\mys^3-66\,\mys^2-107\,\mys+10\right)\,\cos \left(5\,\theta\right) 
\end{dmath*}
\begin{dmath*}
 -28\,\sqrt{\mys}\,\left(\mys+2\right)^2\,\left(\mys^2-8\,\mys+1\right)^2\,\left(65\,\mys^6+104\,\mys^5-582\,\mys^4-1140\,\mys^3-151\,\mys^2-36\,\mys+4\right)\,\cos \left(4\,\theta\right) 
\end{dmath*}
\begin{dmath*}
      + 
 28\,\mys\,\left(\mys+2\right)^{{{3}\over{2}}}\,\left(\mys^2-8\,\mys+1\right)^{{{3}\over{2}}}\,\left(156\,\mys^7+91\,\mys^6-2668\,\mys^5-4562\,\mys^4+2836\,\mys^3+5135\,\mys^2-1252\,\mys+60\right)\,\cos \left(3\,\theta\right) 
\end{dmath*}
\begin{dmath*}
 -8\,\sqrt{\mys}\,\left(\mys+2\right)\,\left(\mys^2-8\,\mys+1\right)\,\left(1001\,\mys^9-1001\,\mys^8-26334\,\mys^7-24920\,\mys^6+117705\,\mys^5+187635\,\mys^4+28552\,\mys^3-3430\,\mys^2-840\,\mys+56\right)\,\cos \left(2\,\theta\right) 
\end{dmath*}
\begin{dmath*}
      + 
 4\,\mys\,\sqrt{\mys+2}\,\sqrt{\mys^2-8\,\mys+1}\,\left(2860\,\mys^{10}-9009\,\mys^9-97482\,\mys^8+45913\,\mys^7+960106\,\mys^6+1099161\,\mys^5-696326\,\mys^4-824157\,\mys^3+356370\,\mys^2-41580\,\mys+1400\right)\,\cos \theta 
\end{dmath*}
\begin{dmath*}
 -\sqrt{\mys}\,\left(\mys+2\right)\,\left(6435\,\mys^{11}-50622\,\mys^{10}-136224\,\mys^9+994512\,\mys^8+1998946\,\mys^7-3693252\,\mys^6-7694572\,\mys^5-577240\,\mys^4+411347\,\mys^3+6034\,\mys^2-6300\,\mys+280\right)\,\, 
\Bigg)
\end{dmath*}

%
%
%
\begin{dmath}
\,
\end{dmath}
\end{dgroup}
\normalsize

We can plot the amplitude for some values.
In particular we can look to the amplitude near or far the resonances
(fig. \ref{fig:S6zS4zTT_special_s})
and for quite different $\mys$ values (fig. \ref{fig:S6zS4zTT_generic_s}).
In particular we can also consider the contribution to the amplitude of $Num(\mys,
\cos \theta)$,  the one without the
Veneziano amplitude which shows new poles wrt Veneziano and the
complete amplitude (fig. \ref{fig:S6zS4zTT_generic_s}).

 \begin{figure}[!h]
   \centering
    \begin{subfigure}{0.49\textwidth}
    \centering    
    \resizebox{1.0\textwidth}{!}{\input{S6zS4zTT.Num.s_min_s_14.1_22.3_steps_1000_tex.tex}}
    \end{subfigure}
    \hfill
    \begin{subfigure}{0.49\textwidth}
    \centering    
    \resizebox{1.0\textwidth}{!}{\input{S6zS4zTT.log.Num.s_min_s_14.1_22.3_steps_1000_tex.tex}}
    \end{subfigure}
    \begin{subfigure}{0.49\textwidth} 
    \centering    
   \resizebox{1.0\textwidth}{!}{\input{S6zS4zTT.withOUT.s_min_s_14.1_22.3_steps_1000_tex.tex}}
    \end{subfigure}
    \hfill
    \begin{subfigure}{0.49\textwidth}
    \centering    
   \resizebox{1.0\textwidth}{!}{\input{S6zS4zTT.log.withOUT.s_min_s_14.1_22.3_steps_1000_tex.tex}}
    \end{subfigure}
    \begin{subfigure}{0.49\textwidth} 
    \centering    
   \resizebox{1.0\textwidth}{!}{\input{S6zS4zTT.with.s_min_s_14.1_22.3_steps_1000_tex.tex}}
    \end{subfigure}
    \hfill
    \begin{subfigure}{0.49\textwidth}
    \centering    
   \resizebox{1.0\textwidth}{!}{\input{S6zS4zTT.log.with.s_min_s_14.1_22.3_steps_1000_tex.tex}}
    \end{subfigure}
    \caption{
      Different perspectives on the ``S6zS4zTT'' amplitude.
      We use generic $s$ values, i.e. not close to resonances.
      On the left the Numerator in eq. \eqref{eq:Num6400},
      then the Numerator divided by the Denominator with the new poles
      and finally the complete color ordered amplitude with the
      Veneziano contribution.
      On the right there are exactly the same amplitudes but we plot
      their $\log$.
}
\label{fig:S6zS4zTT_generic_s}
\end{figure}

  \begin{figure}[!h]
   \centering
    \begin{subfigure}{0.49\textwidth}
    \centering    
    \resizebox{1.0\textwidth}{!}{\input{S6zS4zTT.Num.s_10.01_10.25_10.51_steps_1000_tex.tex}}
    \end{subfigure}
    \hfill
    \begin{subfigure}{0.49\textwidth}
    \centering    
    \resizebox{1.0\textwidth}{!}{\input{S6zS4zTT.log.Num.s_10.01_10.25_10.51_steps_1000_tex.tex}}
    \end{subfigure}
    \begin{subfigure}{0.49\textwidth} 
    \centering    
   \resizebox{1.0\textwidth}{!}{\input{S6zS4zTT.withOUT.s_10.01_10.25_10.51_steps_1000_tex.tex}}
    \end{subfigure}
    \hfill
    \begin{subfigure}{0.49\textwidth}
    \centering    
   \resizebox{1.0\textwidth}{!}{\input{S6zS4zTT.log.withOUT.s_10.01_10.25_10.51_steps_1000_tex.tex}}
    \end{subfigure}
    \begin{subfigure}{0.49\textwidth} 
    \centering    
   \resizebox{1.0\textwidth}{!}{\input{S6zS4zTT.with.s_10.01_10.25_10.51_steps_1000_tex.tex}}
    \end{subfigure}
    \hfill
    \begin{subfigure}{0.49\textwidth}
    \centering    
   \resizebox{1.0\textwidth}{!}{\input{S6zS4zTT.log.with.s_10.01_10.25_10.51_steps_1000_tex.tex}}
    \end{subfigure}
    \caption{
      Different perspectives on the ``S6zS4zTT'' amplitude.
      We use  two resonant $s$ values and one not resonant to show the difference
      in magnitude of the amplitudes.
      On the left the Numerator in eq. \eqref{eq:Num6400},
      then the Numerator divided by the Denominator with the new poles
      and finally the complete color ordered amplitude with the
      Veneziano contribution.
      On the right there are exactly the same amplitudes but we plot
      their $\log$.
}
\label{fig:S6zS4zTT_special_s}
\end{figure}

From the perspective of the erratic behavior we can consider the
contribution of $\cos( 0 \theta)$ which corresponds to a spin $0$
exchange with respect to higher spins $\cos( k \theta)$ (fig. \ref{fig:S6zS4zTT_ratios}).
This shows clearly the spin $0$ dominance.

  \begin{figure}[!h]
   \centering
    \centering    
    \resizebox{1.0\textwidth}{!}{\input{S6zS4zTT.log.ratios.s_7.872983346207417_78.72983346207417_steps_1000_tex.tex}}
    \caption{
      Log of the ratio of the coefficient of $\cos( k \theta)$  to the
      terms independent of $\theta$ of the Numerator.
      This shows the dominance of the spin $0$ particles exchange and
      explain why the amplitude is not erratic.
}
\label{fig:S6zS4zTT_ratios}
\end{figure}

\subsubsection{ Color ordered amplitude "S6zTTS4z" }

%
%
The color ordered correlator can be derived from \eqref{eq:S6S4corr}
as described in appendix \ref{app:othercorrsandamps} and is given by
\tiny
\begin{dgroup}
\begin{dmath*}
\mbox{C}_{S6zTTS4z}(\omega ,\mys, \myu)
=
\end{dmath*}
  \begin{dmath*}
  ( 32 )
  ( \left(\omega-1\right)^{2\,\myt+2\,\mys-8} )
  ( \omega^{-2\,\mys-2} )
  \end{dmath*}
\begin{dmath*}
\Bigg(
      + 
 4\,\left(62\,\myt^2-49\,\myt-127\right)\,\left(6750\,\myt^3-18941\,\myt^2-12736\,\myt+24156\right)\,\omega^{10} 
\end{dmath*}
\begin{dmath*}
 -4\,\left(4495500\,\myt^5+4806000\,\mys\,\myt^4-17477664\,\myt^4-464408\,\mys\,\myt^3-14976997\,\myt^3-13384592\,\mys\,\myt^2+47985137\,\myt^2-867656\,\mys\,\myt+19528754\,\myt+6135624\,\mys-24542496\right)\,\omega^9 
\end{dmath*}
\begin{dmath*}
      + 
 2\,\left(37827000\,\myt^5+82829200\,\mys\,\myt^4-199417188\,\myt^4+39575200\,\mys^2\,\myt^3-150066120\,\mys\,\myt^3-123468306\,\myt^3+67366152\,\mys^2\,\myt^2-696992950\,\mys\,\myt^2+695524861\,\myt^2-650332\,\mys^2\,\myt-306031041\,\mys\,\myt+2600920407\,\myt-24295464\,\mys^2+342942402\,\mys-1420272262\right)\,\omega^8 
\end{dmath*}
\begin{dmath*}
 -4\,\left(39069000\,\myt^5+136284400\,\mys\,\myt^4-277979580\,\myt^4+135208400\,\mys^2\,\myt^3-525233726\,\mys\,\myt^3+153126164\,\myt^3+37372000\,\mys^3\,\myt^2-81092670\,\mys^2\,\myt^2-1063501763\,\mys\,\myt^2+992403340\,\myt^2+85599584\,\mys^3\,\myt-1052197019\,\mys^2\,\myt+2034552499\,\mys\,\myt+5607238501\,\myt+45529812\,\mys^3-1023251862\,\mys^2+7460529557\,\mys-17120821699\right)\,\omega^7 
\end{dmath*}
\begin{dmath*}
      + 
 \left(158760000\,\myt^5+848970800\,\mys\,\myt^4-1341459360\,\myt^4+1387549600\,\mys^2\,\myt^3-4981348840\,\mys\,\myt^3+1784780040\,\myt^3+823085600\,\mys^3\,\myt^2-4195601208\,\mys^2\,\myt^2+413715552\,\mys\,\myt^2+10785503248\,\myt^2+149938800\,\mys^4\,\myt-305704184\,\mys^3\,\myt-8340760084\,\mys^2\,\myt+35794637448\,\mys\,\myt-29300852348\,\myt+276387120\,\mys^4-4472555472\,\mys^3+24447846360\,\mys^2-54045497944\,\mys-4920611951\right)\,\omega^6 
\end{dmath*}
\begin{dmath*}
 -2\,\left(31752000\,\myt^5+304288200\,\mys\,\myt^4-182227320\,\myt^4+808102000\,\mys^2\,\myt^3-2066108796\,\mys\,\myt^3-1075703028\,\myt^3+808102000\,\mys^3\,\myt^2-3965568120\,\mys^2\,\myt^2+1878346694\,\mys\,\myt^2+7087843182\,\myt^2+304288200\,\mys^4\,\myt-2066108796\,\mys^3\,\myt+1878346694\,\mys^2\,\myt+4698151444\,\mys\,\myt+19851964209\,\myt+31752000\,\mys^5-182227320\,\mys^4-1075703028\,\mys^3+7087843182\,\mys^2+19851964209\,\mys-105574761892\right)\,\omega^5 
\end{dmath*}
\begin{dmath*}
      + 
 \left(149938800\,\mys\,\myt^4+276387120\,\myt^4+823085600\,\mys^2\,\myt^3-305704184\,\mys\,\myt^3-4472555472\,\myt^3+1387549600\,\mys^3\,\myt^2-4195601208\,\mys^2\,\myt^2-8340760084\,\mys\,\myt^2+24447846360\,\myt^2+848970800\,\mys^4\,\myt-4981348840\,\mys^3\,\myt+413715552\,\mys^2\,\myt+35794637448\,\mys\,\myt-54045497944\,\myt+158760000\,\mys^5-1341459360\,\mys^4+1784780040\,\mys^3+10785503248\,\mys^2-29300852348\,\mys-4920611951\right)\,\omega^4 
\end{dmath*}
\begin{dmath*}
 -4\,\left(37372000\,\mys^2\,\myt^3+85599584\,\mys\,\myt^3+45529812\,\myt^3+135208400\,\mys^3\,\myt^2-81092670\,\mys^2\,\myt^2-1052197019\,\mys\,\myt^2-1023251862\,\myt^2+136284400\,\mys^4\,\myt-525233726\,\mys^3\,\myt-1063501763\,\mys^2\,\myt+2034552499\,\mys\,\myt+7460529557\,\myt+39069000\,\mys^5-277979580\,\mys^4+153126164\,\mys^3+992403340\,\mys^2+5607238501\,\mys-17120821699\right)\,\omega^3 
\end{dmath*}
\begin{dmath*}
      + 
 2\,\left(39575200\,\mys^3\,\myt^2+67366152\,\mys^2\,\myt^2-650332\,\mys\,\myt^2-24295464\,\myt^2+82829200\,\mys^4\,\myt-150066120\,\mys^3\,\myt-696992950\,\mys^2\,\myt-306031041\,\mys\,\myt+342942402\,\myt+37827000\,\mys^5-199417188\,\mys^4-123468306\,\mys^3+695524861\,\mys^2+2600920407\,\mys-1420272262\right)\,\omega^2 
\end{dmath*}
\begin{dmath*}
 -4\,\left(4806000\,\mys^4\,\myt-464408\,\mys^3\,\myt-13384592\,\mys^2\,\myt-867656\,\mys\,\myt+6135624\,\myt+4495500\,\mys^5-17477664\,\mys^4-14976997\,\mys^3+47985137\,\mys^2+19528754\,\mys-24542496\right)\,\omega 
\end{dmath*}
\begin{dmath*}
      + 
 4\,\left(62\,\mys^2-49\,\mys-127\right)\,\left(6750\,\mys^3-18941\,\mys^2-12736\,\mys+24156\right) 
\Bigg)
\end{dmath*}

%
%
%
\begin{dmath}
\,
\end{dmath}
\end{dgroup}
\normalsize
%
%

%
%

The color ordered amplitude can be written as a product of the Veneziano amplitude
times a polynomial $\mbox{Num}(\mys ,\myu)$
in the Mandelstam variables $\mys$ and $\myu$
(defined according to the ordering of states in the color ordered amplitude,
i.e
$\mys = \mys_{\mbox{S6zTTS4z}}= \myt_{\mbox{S6zS4zTT}}$,
$\myt = \myt_{\mbox{S6zTTS4z}}= \myu_{\mbox{S6zS4zTT}}$,
and
$\myu= \myu_{\mbox{S6zTTS4z}}= \mys_{\mbox{S6zS4zTT}}$
)
divided by another polynomial $\mbox{Den}(\mys ,\myu)$ which comes
from the transformation of the $\Gamma$ function in the denominator
which arises from the moduli space integration to the Veneziano "canonical" one.
Explicitly we get
\begin{equation}
A_{S6zTTS4z}
=
A_{\mbox{Veneziano} 
}(\mys, \myu)
\frac{\mbox{Num}(\mys ,\myu)}{\mbox{Den}(\mys, \myu)}
,
\end{equation}
where we have
\tiny
\begin{dgroup}
\begin{dmath*}
\mbox{Num}(\mys ,\myu)
=
370383360\,\mys\,\left(2\,\mys-1\right)\,\myu\,\left(\mys^2\,\left(496\,\myu^7-200\,\myu^6-5612\,\myu^5+18702\,\myu^4-151246\,\myu^3+718484\,\myu^2-1514533\,\myu+1299057\right)+2\,\mys^4\,\left(480\,\myu^6+752\,\myu^5+3008\,\myu^4-34522\,\myu^3+35863\,\myu^2-285679\,\myu+996870\right)+\mys\,\left(\myu-3\right)\,\left(192\,\myu^6+1392\,\myu^5-872\,\myu^4-5576\,\myu^3+14558\,\myu^2+12857\,\myu-198363\right)+6\,\left(96\,\myu^6-240\,\myu^5-1628\,\myu^4+11784\,\myu^3-43055\,\myu^2+86769\,\myu-87536\right)+4\,\mys^3\,\left(\myu-3\right)\,\left(80\,\myu^6+332\,\myu^5+1578\,\myu^4-4072\,\myu^3+14533\,\myu^2-72109\,\myu+235470\right)+24\,\mys^5\,\left(\myu-3\right)\,\left(40\,\myu^4+282\,\myu^3+1475\,\myu^2+3267\,\myu+8460\right)+8\,\mys^6\,\left(40\,\myu^4+282\,\myu^3+1475\,\myu^2+3267\,\myu+8460\right)\right)
\end{dmath*}
\begin{dmath}
,%
\end{dmath}
\end{dgroup}
\normalsize
and
\tiny
\begin{dgroup}
\begin{dmath}
\mbox{Den}(\mys ,\myu)
=
-2\,\mys\,\left(64\,\myu^7-392\,\myu^6+672\,\myu^5-70\,\myu^4-602\,\myu^3+273\,\myu^2+61\,\myu-21\right)-\mys^2\,\left(448\,\myu^6-2352\,\myu^5+3360\,\myu^4-280\,\myu^3-1806\,\myu^2+546\,\myu+61\right)-14\,\mys^3\,\left(64\,\myu^5-280\,\myu^4+320\,\myu^3-20\,\myu^2-86\,\myu+13\right)-7\,\mys^4\,\left(160\,\myu^4-560\,\myu^3+480\,\myu^2-20\,\myu-43\right)-28\,\mys^5\,\left(32\,\myu^3-84\,\myu^2+48\,\myu-1\right)-112\,\mys^6\,\left(4\,\myu^2-7\,\myu+2\right)-16\,\mys^7\,\left(8\,\myu-7\right)-\left(\myu-2\right)\,\left(\myu-1\right)\,\myu\,\left(\myu+1\right)\,\left(2\,\myu-7\right)\,\left(2\,\myu-3\right)\,\left(2\,\myu-1\right)\,\left(2\,\myu+1\right)-16\,\mys^8
.
\end{dmath}
\end{dgroup}
\normalsize

The previous quantities in channel $\mys$ can be written
using $\mys$ and $\cos(\theta)$
(``$\mys \ge \sqrt{5}\,i+2$'') with
\begin{align}
\myt =& {{\sqrt{\mys^2-4\,\mys+9}\,\sqrt{\mys^2-2\,\mys+4}\,\cos \theta-\mys^2+3\,\mys+6}\over{2\,\mys}}
\nonumber \\
\myu =& -{{\sqrt{\mys^2-4\,\mys+9}\,\sqrt{\mys^2-2\,\mys+4}\,\cos \theta+\mys^2-3\,\mys+6}\over{2\,\mys}}
\end{align}
as

\tiny
\begin{dgroup}
\begin{dmath*}
\mbox{Num}(s,\cos(\theta))
=
\end{dmath*}
\begin{dmath*}
\Bigg(
      + 
 370383360\,\mys^2\,\left(2\,\mys-1\right)\,\left(4\,\mys+3\right)\,\left(5\,\mys+4\right)\,\left(\mys^2-4\,\mys+9\right)^4\,\left(\mys^2-2\,\mys+4\right)^4\,\cos \left(8\,\theta\right) 
\end{dmath*}
\begin{dmath*}
      + 
 740766720\,\mys^2\,\left(2\,\mys-1\right)\,\left(4\,\mys+3\right)\,\left(5\,\mys+4\right)\,\left(\mys^2-4\,\mys+9\right)^{{{7}\over{2}}}\,\left(\mys^2-2\,\mys+4\right)^{{{7}\over{2}}}\,\left(2\,\mys^2-17\,\mys+42\right)\,\cos \left(7\,\theta\right) 
\end{dmath*}
\begin{dmath*}
 -1481533440\,\mys^2\,\left(2\,\mys-1\right)\,\left(\mys^2-4\,\mys+9\right)^3\,\left(\mys^2-2\,\mys+4\right)^3\,\left(96\,\mys^5-4063\,\mys^4+8724\,\mys^3+2770\,\mys^2-17436\,\mys-9936\right)\,\cos \left(6\,\theta\right) 
\end{dmath*}
\begin{dmath*}
 -740766720\,\mys^2\,\left(2\,\mys-1\right)\,\left(\mys^2-4\,\mys+9\right)^{{{5}\over{2}}}\,\left(\mys^2-2\,\mys+4\right)^{{{5}\over{2}}}\,\left(200\,\mys^8-3582\,\mys^7+15265\,\mys^6+63782\,\mys^5-459131\,\mys^4+421776\,\mys^3+537960\,\mys^2-643176\,\mys-489888\right)\,\cos \left(5\,\theta\right) 
\end{dmath*}
\begin{dmath*}
 -1481533440\,\mys^2\,\left(2\,\mys-1\right)\,\left(\mys^2-4\,\mys+9\right)^2\,\left(\mys^2-2\,\mys+4\right)^2\,\left(100\,\mys^{10}-2613\,\mys^9+37082\,\mys^8-249586\,\mys^7+174910\,\mys^6+2455355\,\mys^5-6448468\,\mys^4+1821428\,\mys^3+7320768\,\mys^2-3629232\,\mys-4245696\right)\,\cos \left(4\,\theta\right) 
\end{dmath*}
\begin{dmath*}
      + 
 740766720\,\mys^2\,\left(2\,\mys-1\right)\,\left(\mys^2-4\,\mys+9\right)^{{{3}\over{2}}}\,\left(\mys^2-2\,\mys+4\right)^{{{3}\over{2}}}\,\left(360\,\mys^{12}-9798\,\mys^{11}+88257\,\mys^{10}-51150\,\mys^9-4999022\,\mys^8+20833790\,\mys^7+12540213\,\mys^6-162574018\,\mys^5+213518620\,\mys^4+43433736\,\mys^3-232763760\,\mys^2+42407712\,\mys+110388096\right)\,\cos \left(3\,\theta\right) 
\end{dmath*}
\begin{dmath*}
      + 
 1481533440\,\mys^2\,\left(2\,\mys-1\right)\,\left(\mys^2-4\,\mys+9\right)\,\left(\mys^2-2\,\mys+4\right)\,\left(320\,\mys^{14}-10672\,\mys^{13}+167153\,\mys^{12}-1435784\,\mys^{11}+7117760\,\mys^{10}-11435740\,\mys^9-49934763\,\mys^8+131491432\,\mys^7+402384062\,\mys^6-1497530916\,\mys^5+1068685192\,\mys^4+824120976\,\mys^3-1245712608\,\mys^2-52519104\,\mys+560431872\right)\,\cos \left(2\,\theta\right) 
\end{dmath*}
\begin{dmath*}
 -740766720\,\mys^2\,\left(2\,\mys-1\right)\,\sqrt{\mys^2-4\,\mys+9}\,\sqrt{\mys^2-2\,\mys+4}\,\left(200\,\mys^{16}-6974\,\mys^{15}+92937\,\mys^{14}-431362\,\mys^{13}-3550281\,\mys^{12}+42940834\,\mys^{11}-144828025\,\mys^{10}+6604026\,\mys^9+820917445\,\mys^8+2011552284\,\mys^7-18787282012\,\mys^6+34186203112\,\mys^5-10899861648\,\mys^4-23583486816\,\mys^3+18797339520\,\mys^2+5227618176\,\mys-9127033344\right)\,\cos \theta 
\end{dmath*}
\begin{dmath*}
 -370383360\,\mys^2\,\left(2\,\mys-1\right)\,\left(900\,\mys^{18}-35949\,\mys^{17}+675100\,\mys^{16}-7407564\,\mys^{15}+53962368\,\mys^{14}-249238174\,\mys^{13}+660829776\,\mys^{12}-985184788\,\mys^{11}+1291208540\,\mys^{10}-440845389\,\mys^9-19767241708\,\mys^8+103923865624\,\mys^7-243411968160\,\mys^6+246122062352\,\mys^5+16593158208\,\mys^4-211794265728\,\mys^3+98459089920\,\mys^2+63427339008\,\mys-60526642176\right)\,\, 
\Bigg)
\end{dmath*}

%
%
%
\begin{dmath}
  \,
  \label{eq:Num6004}
\end{dmath}
\end{dgroup}
\normalsize

and

\tiny
\begin{dgroup}
\begin{dmath*}
\mbox{Den}(s,\cos(\theta))
=
\end{dmath*}
\begin{dmath*}
\Bigg(
 -\left(\mys^2-4\,\mys+9\right)^4\,\left(\mys^2-2\,\mys+4\right)^4\,\cos \left(8\,\theta\right) 
\end{dmath*}
\begin{dmath*}
      + 
 4\,\left(\mys^2-4\,\mys+9\right)^{{{7}\over{2}}}\,\left(\mys^2-2\,\mys+4\right)^{{{7}\over{2}}}\,\left(4\,\mys^2+5\,\mys-24\right)\,\cos \left(7\,\theta\right) 
\end{dmath*}
\begin{dmath*}
 -8\,\left(\mys^2-4\,\mys+9\right)^3\,\left(\mys^2-2\,\mys+4\right)^3\,\left(15\,\mys^4+29\,\mys^3-140\,\mys^2-244\,\mys+540\right)\,\cos \left(6\,\theta\right) 
\end{dmath*}
\begin{dmath*}
      + 
 28\,\left(\mys^2-4\,\mys+9\right)^{{{5}\over{2}}}\,\left(\mys^2-2\,\mys+4\right)^{{{5}\over{2}}}\,\left(20\,\mys^6+41\,\mys^5-234\,\mys^4-647\,\mys^3+1054\,\mys^2+3156\,\mys-4320\right)\,\cos \left(5\,\theta\right) 
\end{dmath*}
\begin{dmath*}
 -28\,\left(\mys^2-4\,\mys+9\right)^2\,\left(\mys^2-2\,\mys+4\right)^2\,\left(65\,\mys^8+104\,\mys^7-894\,\mys^6-2580\,\mys^5+3881\,\mys^4+20940\,\mys^3-7124\,\mys^2-87984\,\mys+84240\right)\,\cos \left(4\,\theta\right) 
\end{dmath*}
\begin{dmath*}
      + 
 28\,\left(\mys^2-4\,\mys+9\right)^{{{3}\over{2}}}\,\left(\mys^2-2\,\mys+4\right)^{{{3}\over{2}}}\,\left(156\,\mys^{10}+91\,\mys^9-2356\,\mys^8-5306\,\mys^7+11260\,\mys^6+58895\,\mys^5+14060\,\mys^4-334236\,\mys^3-312624\,\mys^2+1690416\,\mys-1213056\right)\,\cos \left(3\,\theta\right) 
\end{dmath*}
\begin{dmath*}
 -8\,\left(\mys^2-4\,\mys+9\right)\,\left(\mys^2-2\,\mys+4\right)\,\left(1001\,\mys^{12}-1001\,\mys^{11}-14322\,\mys^{10}-15218\,\mys^9+83685\,\mys^8+287343\,\mys^7+50980\,\mys^6-1745128\,\mys^5-3913980\,\mys^4+5349008\,\mys^3+38486448\,\mys^2-83026944\,\mys+46702656\right)\,\cos \left(2\,\theta\right) 
\end{dmath*}
\begin{dmath*}
      + 
 4\,\sqrt{\mys^2-4\,\mys+9}\,\sqrt{\mys^2-2\,\mys+4}\,\left(2860\,\mys^{14}-9009\,\mys^{13}-25410\,\mys^{12}+26509\,\mys^{11}+182686\,\mys^{10}+272181\,\mys^9-330422\,\mys^8-2365137\,\mys^7-4404618\,\mys^6-2646504\,\mys^5+34049624\,\mys^4+318982944\,\mys^3-1327566240\,\mys^2+1759133376\,\mys-800616960\right)\,\cos \theta 
\end{dmath*}
\begin{dmath*}
 -\left(6435\,\mys^{16}-37752\,\mys^{15}+30228\,\mys^{14}+112224\,\mys^{13}-39998\,\mys^{12}-215152\,\mys^{11}-340756\,\mys^{10}+11424\,\mys^9+1063107\,\mys^8+9628136\,\mys^7-37517896\,\mys^6-390403328\,\mys^5+4189269392\,\mys^4-16012007424\,\mys^3+30330039168\,\mys^2-28662087168\,\mys+10808328960\right)\,\, 
\Bigg)
\end{dmath*}

%
%
%
\begin{dmath}
\,
\end{dmath}
\end{dgroup}
\normalsize

We can plot the amplitude for some values.
In particular we can look to the amplitude near or far the resonances
(fig. \ref{fig:S6zTTS4z_special_s})
and for quite different $\mys$ values (fig. \ref{fig:S6zTTS4z_generic_s}).
We can also consider the contribution to the amplitude of $Num(\mys,
\cos \theta)$,  the one without the
Veneziano amplitude which shows new poles wrt Veneziano and the
complete amplitude (fig. \ref{fig:S6zTTS4z_generic_s}).

 \begin{figure}[!h]
   \centering
    \begin{subfigure}{0.49\textwidth}
    \centering    
    \resizebox{1.0\textwidth}{!}{\input{S6zTTS4z.Num.s_min_s_14.1_22.3_steps_1000_tex.tex}}
    \end{subfigure}
    \hfill
    \begin{subfigure}{0.49\textwidth}
    \centering    
    \resizebox{1.0\textwidth}{!}{\input{S6zTTS4z.log.Num.s_min_s_14.1_22.3_steps_1000_tex.tex}}
    \end{subfigure}
    \begin{subfigure}{0.49\textwidth} 
    \centering    
   \resizebox{1.0\textwidth}{!}{\input{S6zTTS4z.withOUT.s_min_s_14.1_22.3_steps_1000_tex.tex}}
    \end{subfigure}
    \hfill
    \begin{subfigure}{0.49\textwidth}
    \centering    
   \resizebox{1.0\textwidth}{!}{\input{S6zTTS4z.log.withOUT.s_min_s_14.1_22.3_steps_1000_tex.tex}}
    \end{subfigure}
    \begin{subfigure}{0.49\textwidth} 
    \centering    
   \resizebox{1.0\textwidth}{!}{\input{S6zTTS4z.with.s_min_s_14.1_22.3_steps_1000_tex.tex}}
    \end{subfigure}
    \hfill
    \begin{subfigure}{0.49\textwidth}
    \centering    
   \resizebox{1.0\textwidth}{!}{\input{S6zTTS4z.log.with.s_min_s_14.1_22.3_steps_1000_tex.tex}}
    \end{subfigure}
    \caption{
      Different perspectives on the ``S6zTTS4z'' amplitude.
      We use generic $s$ values, i.e. not close to resonances.
      On the left the Numerator in eq. \eqref{eq:Num6004},
      then the Numerator divided by the Denominator with the new poles
      and finally the complete color ordered amplitude with the
      Veneziano contribution.
      On the right there are exactly the same amplitudes but we plot
      their $\log$.
}
\label{fig:S6zTTS4z_generic_s}
\end{figure}

  \begin{figure}[!h]
   \centering
    \begin{subfigure}{0.49\textwidth}
    \centering    
    \resizebox{1.0\textwidth}{!}{\input{S6zTTS4z.Num.s_10.01_10.25_10.51_steps_1000_tex.tex}}
    \end{subfigure}
    \hfill
    \begin{subfigure}{0.49\textwidth}
    \centering    
    \resizebox{1.0\textwidth}{!}{\input{S6zTTS4z.log.Num.s_10.01_10.25_10.51_steps_1000_tex.tex}}
    \end{subfigure}
    \begin{subfigure}{0.49\textwidth} 
    \centering    
   \resizebox{1.0\textwidth}{!}{\input{S6zTTS4z.withOUT.s_10.01_10.25_10.51_steps_1000_tex.tex}}
    \end{subfigure}
    \hfill
    \begin{subfigure}{0.49\textwidth}
    \centering    
   \resizebox{1.0\textwidth}{!}{\input{S6zTTS4z.log.withOUT.s_10.01_10.25_10.51_steps_1000_tex.tex}}
    \end{subfigure}
    \begin{subfigure}{0.49\textwidth} 
    \centering    
   \resizebox{1.0\textwidth}{!}{\input{S6zTTS4z.with.s_10.01_10.25_10.51_steps_1000_tex.tex}}
    \end{subfigure}
    \hfill
    \begin{subfigure}{0.49\textwidth}
    \centering    
   \resizebox{1.0\textwidth}{!}{\input{S6zTTS4z.log.with.s_10.01_10.25_10.51_steps_1000_tex.tex}}
    \end{subfigure}
    \caption{
      Different perspectives on the ``S6zTTS4z'' amplitude.
      We use  two resonant $s$ values and one not resonant to show the difference
      in magnitude of the amplitudes.
      On the left the Numerator in eq. \eqref{eq:Num6004},
      then the Numerator divided by the Denominator with the new poles
      and finally the complete color ordered amplitude with the
      Veneziano contribution.
      On the right there are exactly the same amplitudes but we plot
      their $\log$.
}
\label{fig:S6zTTS4z_special_s}
\end{figure}

From the perspective of the erratic behavior we can consider the
contribution of $\cos( 0 \theta)$ which corresponds to a spin $0$
exchange with respect to higher spins $\cos( k \theta)$
(fig. \ref{fig:S6zTTS4z_ratios}).
This shows clearly the spin $0$ dominance.

  \begin{figure}[!h]
   \centering
    \centering    
    \resizebox{1.0\textwidth}{!}{\input{S6zTTS4z.log.ratios.s_0.0_64.1109_steps_1000_tex.tex}}
    \caption{
      Log of the ratio of the coefficient of $\cos( k \theta)$  to the
      terms independent of $\theta$ of the Numerator.
      This shows the dominance of the spin $0$ particles exchange and
      explain why the amplitude is not erratic.
}
\label{fig:S6zTTS4z_ratios}
\end{figure}

\subsubsection{ Color ordered amplitude "S6zTS4zT"}

%
%
The color ordered correlator can also be derived from \eqref{eq:S6S4corr}
as described in appendix \ref{app:othercorrsandamps} and is given by
\tiny
\begin{dgroup}
\begin{dmath*}
\mbox{C}_{S6zTS4zT}(\omega ,\mys, \myu)
=
\end{dmath*}
  \begin{dmath*}
  ( 32 )
  ( \left(1-\omega\right)^{2\,\myt + 2\,\mys-8} )
  ( \omega^{-2\,\mys-2} )
  \end{dmath*}
\begin{dmath*}
\Bigg(
      + 
 2\,\left(837000\,\myt^5+25623316\,\myt^4+377638706\,\myt^3+6166444857\,\myt^2+34466056430\,\myt-4988506712\right)\,\omega^{10} 
\end{dmath*}
\begin{dmath*}
 -2\,\left(8991000\,\myt^5+9612000\,\mys\,\myt^4+226512672\,\myt^4+225628184\,\mys\,\myt^3+2715406102\,\myt^3+2331309696\,\mys\,\myt^2+29849242224\,\myt^2+5027964966\,\mys\,\myt+169776841413\,\myt+9977013424\,\mys-39908053696\right)\,\omega^9 
\end{dmath*}
\begin{dmath*}
      + 
 \left(75654000\,\myt^5+165658400\,\mys\,\myt^4+1543707624\,\myt^4+79150400\,\mys^2\,\myt^3+3065342160\,\mys\,\myt^3+15441194580\,\myt^3+1272139704\,\mys^2\,\myt^2+25310129596\,\mys\,\myt^2+120991128398\,\myt^2+8144761444\,\mys^2\,\myt+30198759822\,\mys\,\myt+713755809318\,\myt+9381359460\,\mys^2+61649042436\,\mys-218156684699\right)\,\omega^8 
\end{dmath*}
\begin{dmath*}
 -4\,\left(39069000\,\myt^5+136284400\,\mys\,\myt^4+643557420\,\myt^4+135208400\,\mys^2\,\myt^3+1933034374\,\mys\,\myt^3+5707252724\,\myt^3+37372000\,\mys^3\,\myt^2+1570733130\,\mys^2\,\myt^2+13272158120\,\mys\,\myt^2+35060652780\,\myt^2+311178284\,\mys^3\,\myt+7924823242\,\mys^2\,\myt+7372532474\,\mys\,\myt+206640054681\,\myt+786909384\,\mys^3+5840267232\,\mys^2+19342913868\,\mys-74558252825\right)\,\omega^7 
\end{dmath*}
\begin{dmath*}
      + 
 2\,\left(79380000\,\myt^5+424485400\,\mys\,\myt^4+1122610320\,\myt^4+693774800\,\mys^2\,\myt^3+4604299180\,\mys\,\myt^3+9844823880\,\myt^3+411542800\,\mys^3\,\myt^2+5625685296\,\mys^2\,\myt^2+28887830706\,\mys\,\myt^2+52582359074\,\myt^2+74969400\,\mys^4\,\myt+2232590708\,\mys^3\,\myt+24061108094\,\mys^2\,\myt+3747033376\,\mys\,\myt+284333546133\,\myt+252079560\,\mys^4+3995888328\,\mys^3+9152451488\,\mys^2+27877520700\,\mys-109829288351\right)\,\omega^6 
\end{dmath*}
\begin{dmath*}
 -2\,\left(31752000\,\myt^5+304288200\,\mys\,\myt^4+535108680\,\myt^4+808102000\,\mys^2\,\myt^3+2840282604\,\mys\,\myt^3+4975231332\,\myt^3+808102000\,\mys^3\,\myt^2+4691134080\,\mys^2\,\myt^2+18023769500\,\mys\,\myt^2+25625737410\,\myt^2+304288200\,\mys^4\,\myt+2754617004\,\mys^3\,\myt+18936360524\,\mys^2\,\myt-2265383630\,\mys\,\myt+118935134307\,\myt+31752000\,\mys^5+518098680\,\mys^4+4097950524\,\mys^3-565186686\,\mys^2+18609760854\,\mys-37837102186\right)\,\omega^5 
\end{dmath*}
\begin{dmath*}
      + 
 \left(149938800\,\mys\,\myt^4+276387120\,\myt^4+823085600\,\mys^2\,\myt^3+2008775416\,\mys\,\myt^3+2430744768\,\myt^3+1387549600\,\mys^3\,\myt^2+4205527992\,\mys^2\,\myt^2+12150834536\,\mys\,\myt^2+14936745120\,\myt^2+848970800\,\mys^4\,\myt+3105190760\,\mys^3\,\myt+16296984556\,\mys^2\,\myt-1839590420\,\mys\,\myt+60696732276\,\myt+158760000\,\mys^5+684755640\,\mys^4+4382350740\,\mys^3-9579145848\,\mys^2+28727168688\,\mys-3941902839\right)\,\omega^4 
\end{dmath*}
\begin{dmath*}
 -4\,\left(37372000\,\mys^2\,\myt^3+85599584\,\mys\,\myt^3+45529812\,\myt^3+135208400\,\mys^3\,\myt^2+271008030\,\mys^2\,\myt^2+415788871\,\mys\,\myt^2+516301698\,\myt^2+136284400\,\mys^4\,\myt+204974974\,\mys^3\,\myt+843313363\,\mys^2\,\myt+273659065\,\mys\,\myt+1982527301\,\myt+39069000\,\mys^5+14511420\,\mys^4+311869450\,\mys^3-1219044974\,\mys^2+3580843954\,\mys+131522958\right)\,\omega^3 
\end{dmath*}
\begin{dmath*}
      + 
 2\,\left(39575200\,\mys^3\,\myt^2+67366152\,\mys^2\,\myt^2-650332\,\mys\,\myt^2-24295464\,\myt^2+82829200\,\mys^4\,\myt+38987880\,\mys^3\,\myt+50572718\,\mys^2\,\myt+245724297\,\mys\,\myt-86728386\,\myt+37827000\,\mys^5-59692188\,\mys^4+63707592\,\mys^3-252651488\,\mys^2+1324626282\,\mys-775766080\right)\,\omega^2 
\end{dmath*}
\begin{dmath*}
 -4\,\left(4806000\,\mys^4\,\myt-464408\,\mys^3\,\myt-13384592\,\mys^2\,\myt-867656\,\mys\,\myt+6135624\,\myt+4495500\,\mys^5-12455664\,\mys^4-4202101\,\mys^3+36573413\,\mys^2-6451882\,\mys-24542496\right)\,\omega 
\end{dmath*}
\begin{dmath*}
      + 
 4\,\left(62\,\mys^2-49\,\mys-127\right)\,\left(6750\,\mys^3-18941\,\mys^2-12736\,\mys+24156\right) 
\Bigg)
\end{dmath*}

%
%
%
\begin{dmath}
\,
\end{dmath}
\end{dgroup}
\normalsize
%
%

%
%

The color ordered amplitude can be written as a product of the Veneziano amplitude
times a polynomial $\mbox{Num}(\mys ,\myu)$
in the Mandelstam variables $\mys$ and $\myu$
(defined according to the ordering of states in the color ordered amplitude,
i.e
$\mys = \mys_{\mbox{S6zTS4zT}}= \myu_{\mbox{S6zS4zTT}}$,
$\myt = \myt_{\mbox{S6zTS4zT}}= \mys_{\mbox{S6zS4zTT}}$,
and
$\myu= \myu_{\mbox{S6zTS4zT}}= \myt_{\mbox{S6zS4zTT}}$
)
divided by another polynomial $\mbox{Den}(\mys ,\myu)$ which comes
from the transformation of the $\Gamma$ function in the denominator
which arises from the moduli space integration to the Veneziano "canonical" one.
Explicitly we get
\begin{equation}
A_{S6zTS4zT}
=
A_{\mbox{Veneziano} 
}(\mys, \myu)
\frac{\mbox{Num}(\mys ,\myu)}{\mbox{Den}(\mys, \myu)}
,
\end{equation}
where we have
\tiny
\begin{dgroup}
\begin{dmath*}
\mbox{Num}(\mys ,\myu)
=
370383360\,\mys\,\left(2\,\mys-1\right)\,\left(-\mys^2\,\myu\,\left(2\,\myu-1\right)\,\left(496\,\myu^7-8872\,\myu^6+55444\,\myu^5-155242\,\myu^4+103858\,\myu^3+160840\,\myu^2-294389\,\myu-161574\right)-\mys\,\myu\,\left(2\,\myu-1\right)\,\left(192\,\myu^7-4848\,\myu^6+42472\,\myu^5-168280\,\myu^4+340838\,\myu^3-294389\,\myu^2-94878\,\myu+81306\right)-2\,\mys^3\,\myu\,\left(2\,\myu-1\right)\,\left(160\,\myu^7-1808\,\myu^6+6084\,\myu^5-918\,\myu^4-28730\,\myu^3+51929\,\myu^2+170419\,\myu+67464\right)-2\,\mys^4\,\myu\,\left(2\,\myu-1\right)\,\left(640\,\myu^6-6664\,\myu^5+21232\,\myu^4-918\,\myu^3-77621\,\myu^2-84140\,\myu-23196\right)+6\,\myu\,\left(2\,\myu-1\right)\,\left(96\,\myu^6-1488\,\myu^5+7732\,\myu^4-22488\,\myu^3+26929\,\myu^2-13551\,\myu-16760\right)-4\,\mys^5\,\myu\,\left(2\,\myu-1\right)\,\left(480\,\myu^5-3332\,\myu^4+3042\,\myu^3+13861\,\myu^2+10618\,\myu+2232\right)-8\,\mys^6\,\myu\,\left(2\,\myu-1\right)\,\left(4\,\myu+3\right)\,\left(5\,\myu+4\right)\,\left(8\,\myu^2-35\,\myu-6\right)-16\,\mys^7\,\myu^2\,\left(2\,\myu-1\right)\,\left(4\,\myu+3\right)\,\left(5\,\myu+4\right)\right)
\end{dmath*}
\begin{dmath}
,%
\end{dmath}
\end{dgroup}
\normalsize
and
\tiny
\begin{dgroup}
\begin{dmath*}
\mbox{Den}(\mys ,\myu)
=
\left(\myu+\mys-2\right)\,\left(\myu+\mys-1\right)\,\left(\myu+\mys\right)\,\left(\myu+\mys+1\right)\,\left(2\,\myu+2\,\mys-7\right)\,\left(2\,\myu+2\,\mys-5\right)\,\left(2\,\myu+2\,\mys-3\right)\,\left(2\,\myu+2\,\mys-1\right)\,\left(2\,\myu+2\,\mys+1\right)
.
\end{dmath*}
\begin{dmath}
\, ~
\end{dmath}
\end{dgroup}
\normalsize

The previous quantities in channel $\mys$ can be written
using $\mys$ and $\cos(\theta)$
(``$\mys \ge \sqrt{5}\,i+2$'') with
\begin{align}
\myt =& {{\sqrt{\mys^2-4\,\mys+9}\,\sqrt{\mys^2-2\,\mys+4}\,\cos \theta-\mys^2+3\,\mys-6}\over{2\,\mys}}
\nonumber \\
\myu =& -{{\sqrt{\mys^2-4\,\mys+9}\,\sqrt{\mys^2-2\,\mys+4}\,\cos \theta+\mys^2-3\,\mys-6}\over{2\,\mys}}
\end{align}
as

\tiny
\begin{dgroup}
\begin{dmath*}
\mbox{Num}(s,\cos(\theta))
=
\end{dmath*}
\begin{dmath*}
\Bigg(
      + 
 370383360\,\mys^2\,\left(2\,\mys-1\right)\,\left(4\,\mys+3\right)\,\left(5\,\mys+4\right)\,\left(\mys^2-4\,\mys+9\right)^{{{9}\over{2}}}\,\left(\mys^2-2\,\mys+4\right)^{{{9}\over{2}}}\,\cos \left(9\,\theta\right) 
\end{dmath*}
\begin{dmath*}
      + 
 740766720\,\mys^2\,\left(2\,\mys-1\right)\,\left(4\,\mys+3\right)\,\left(5\,\mys+4\right)\,\left(\mys^2-4\,\mys+9\right)^4\,\left(\mys^2-2\,\mys+4\right)^4\,\left(\mys^2+9\,\mys-48\right)\,\cos \left(8\,\theta\right) 
\end{dmath*}
\begin{dmath*}
 -370383360\,\mys^2\,\left(2\,\mys-1\right)\,\left(\mys^2-4\,\mys+9\right)^{{{7}\over{2}}}\,\left(\mys^2-2\,\mys+4\right)^{{{7}\over{2}}}\,\left(140\,\mys^6-2879\,\mys^5-4382\,\mys^4+41397\,\mys^3-27454\,\mys^2-112932\,\mys-52272\right)\,\cos \left(7\,\theta\right) 
\end{dmath*}
\begin{dmath*}
 -5926133760\,\mys^2\,\left(2\,\mys-1\right)\,\left(\mys^2-4\,\mys+9\right)^3\,\left(\mys^2-2\,\mys+4\right)^3\,\left(20\,\mys^8-191\,\mys^7-4396\,\mys^6+14945\,\mys^5+35231\,\mys^4-106839\,\mys^3-16062\,\mys^2+175248\,\mys+93312\right)\,\cos \left(6\,\theta\right) 
\end{dmath*}
\begin{dmath*}
      + 
 1481533440\,\mys^2\,\left(2\,\mys-1\right)\,\left(\mys^2-4\,\mys+9\right)^{{{5}\over{2}}}\,\left(\mys^2-2\,\mys+4\right)^{{{5}\over{2}}}\,\left(100\,\mys^{10}-3865\,\mys^9+62490\,\mys^8-32336\,\mys^7-1070846\,\mys^6+1331449\,\mys^5+6461380\,\mys^4-9591364\,\mys^3-6488352\,\mys^2+11697264\,\mys+7542720\right)\,\cos \left(5\,\theta\right) 
\end{dmath*}
\begin{dmath*}
      + 
 2963066880\,\mys^2\,\left(2\,\mys-1\right)\,\left(\mys^2-4\,\mys+9\right)^2\,\left(\mys^2-2\,\mys+4\right)^2\,\left(140\,\mys^{12}-3587\,\mys^{11}+21445\,\mys^{10}+373390\,\mys^9-1585626\,\mys^8-3776715\,\mys^7+18816861\,\mys^6+4443376\,\mys^5-85404324\,\mys^4+60667440\,\mys^3+84535344\,\mys^2-67967424\,\mys-57480192\right)\,\cos \left(4\,\theta\right) 
\end{dmath*}
\begin{dmath*}
 -1481533440\,\mys^2\,\left(2\,\mys-1\right)\,\left(\mys^2-4\,\mys+9\right)^{{{3}\over{2}}}\,\left(\mys^2-2\,\mys+4\right)^{{{3}\over{2}}}\,\left(140\,\mys^{14}-7379\,\mys^{13}+208896\,\mys^{12}-2926143\,\mys^{11}+2826570\,\mys^{10}+67653223\,\mys^9-156604108\,\mys^8-291086101\,\mys^7+736210838\,\mys^6+979131800\,\mys^5-2910582264\,\mys^4+634058208\,\mys^3+2670239520\,\mys^2-1071330624\,\mys-1375605504\right)\,\cos \left(3\,\theta\right) 
\end{dmath*}
\begin{dmath*}
 -5926133760\,\mys^2\,\left(2\,\mys-1\right)\,\left(\mys^2-4\,\mys+9\right)\,\left(\mys^2-2\,\mys+4\right)\,\left(140\,\mys^{16}-5273\,\mys^{15}+94616\,\mys^{14}-874151\,\mys^{13}-4477463\,\mys^{12}+69570660\,\mys^{11}-179182312\,\mys^{10}-149370699\,\mys^9+584589601\,\mys^8+1186792543\,\mys^7-1510544210\,\mys^6-5679378804\,\mys^5+8214523656\,\mys^4+1489937616\,\mys^3-7367192352\,\mys^2+1152403200\,\mys+3318921216\right)\,\cos \left(2\,\theta\right) 
\end{dmath*}
\begin{dmath*}
      + 
 740766720\,\mys^2\,\left(2\,\mys-1\right)\,\sqrt{\mys^2-4\,\mys+9}\,\sqrt{\mys^2-2\,\mys+4}\,\left(140\,\mys^{18}-8783\,\mys^{17}+311936\,\mys^{16}-6687688\,\mys^{15}+178631116\,\mys^{14}-1607755974\,\mys^{13}+5631814924\,\mys^{12}-5542966536\,\mys^{11}-5960692680\,\mys^{10}-3974142307\,\mys^9+16021216772\,\mys^8+74454584200\,\mys^7+65147556416\,\mys^6-604179978832\,\mys^5+475250997312\,\mys^4+321438941568\,\mys^3-477345623040\,\mys^2-22016966400\,\mys+210402137088\right)\,\cos \theta 
\end{dmath*}
\begin{dmath*}
      + 
 740766720\,\mys^2\,\left(2\,\mys-1\right)\,\left(700\,\mys^{20}-31975\,\mys^{19}+751345\,\mys^{18}-11332856\,\mys^{17}+169516372\,\mys^{16}-1509292346\,\mys^{15}+7172393654\,\mys^{14}-18054371620\,\mys^{13}+21910846000\,\mys^{12}-9143306471\,\mys^{11}+678084465\,\mys^{10}+997624164\,\mys^9+11217359752\,\mys^8+152572344976\,\mys^7-1329093069264\,\mys^6+2594227336512\,\mys^5-881104291200\,\mys^4-1789992214272\,\mys^3+1427419977984\,\mys^2+394864284672\,\mys-691733053440\right)\,\, 
\Bigg)
\end{dmath*}

%
%
%
\begin{dmath}
  \,
  \label{eq:Num6040}
\end{dmath}
\end{dgroup}
\normalsize

and

\tiny
\begin{dgroup}
\begin{dmath*}
\mbox{Den}(s,\cos(\theta))
=
\end{dmath*}
\begin{dmath*}
\Bigg(
 -\left(\mys^2-4\,\mys+9\right)^{{{9}\over{2}}}\,\left(\mys^2-2\,\mys+4\right)^{{{9}\over{2}}}\,\cos \left(9\,\theta\right) 
\end{dmath*}
\begin{dmath*}
      + 
 2\,\left(\mys^2-4\,\mys+9\right)^4\,\left(\mys^2-2\,\mys+4\right)^4\,\left(9\,\mys^2+8\,\mys+54\right)\,\cos \left(8\,\theta\right) 
\end{dmath*}
\begin{dmath*}
 -\left(\mys^2-4\,\mys+9\right)^{{{7}\over{2}}}\,\left(\mys^2-2\,\mys+4\right)^{{{7}\over{2}}}\,\left(153\,\mys^4+202\,\mys^3+1893\,\mys^2+1230\,\mys+5508\right)\,\cos \left(7\,\theta\right) 
\end{dmath*}
\begin{dmath*}
      + 
 16\,\left(\mys^2-4\,\mys+9\right)^3\,\left(\mys^2-2\,\mys+4\right)^3\,\left(51\,\mys^6+66\,\mys^5+930\,\mys^4+798\,\mys^3+5596\,\mys^2+2484\,\mys+11016\right)\,\cos \left(6\,\theta\right) 
\end{dmath*}
\begin{dmath*}
 -4\,\left(\mys^2-4\,\mys+9\right)^{{{5}\over{2}}}\,\left(\mys^2-2\,\mys+4\right)^{{{5}\over{2}}}\,\left(765\,\mys^8+620\,\mys^7+17880\,\mys^6+10980\,\mys^5+159587\,\mys^4+69360\,\mys^3+647796\,\mys^2+153360\,\mys+991440\right)\,\cos \left(5\,\theta\right) 
\end{dmath*}
\begin{dmath*}
      + 
 56\,\left(\mys^2-4\,\mys+9\right)^2\,\left(\mys^2-2\,\mys+4\right)^2\,\left(153\,\mys^{10}-20\,\mys^9+4320\,\mys^8-480\,\mys^7+50165\,\mys^6-3444\,\mys^5+302226\,\mys^4-3928\,\mys^3+940536\,\mys^2+12960\,\mys+1189728\right)\,\cos \left(4\,\theta\right) 
\end{dmath*}
\begin{dmath*}
 -4\,\left(\mys^2-4\,\mys+9\right)^{{{3}\over{2}}}\,\left(\mys^2-2\,\mys+4\right)^{{{3}\over{2}}}\,\left(4641\,\mys^{12}-7098\,\mys^{11}+157521\,\mys^{10}-201810\,\mys^9+2271955\,\mys^8-2328438\,\mys^7+17932115\,\mys^6-13403334\,\mys^5+82078416\,\mys^4-38396568\,\mys^3+205051392\,\mys^2-44579808\,\mys+216530496\right)\,\cos \left(3\,\theta\right) 
\end{dmath*}
\begin{dmath*}
      + 
 16\,\left(\mys^2-4\,\mys+9\right)\,\left(\mys^2-2\,\mys+4\right)\,\left(1989\,\mys^{14}-6734\,\mys^{13}+84084\,\mys^{12}-228690\,\mys^{11}+1508549\,\mys^{10}-3293262\,\mys^9+14995698\,\mys^8-25663366\,\mys^7+89745540\,\mys^6-113669388\,\mys^5+323851192\,\mys^4-272072304\,\mys^3+649827360\,\mys^2-275970240\,\mys+556792704\right)\,\cos \left(2\,\theta\right) 
\end{dmath*}
\begin{dmath*}
 -2\,\sqrt{\mys^2-4\,\mys+9}\,\sqrt{\mys^2-2\,\mys+4}\,\left(21879\,\mys^{16}-124696\,\mys^{15}+1218360\,\mys^{14}-5065368\,\mys^{13}+27609974\,\mys^{12}-88673256\,\mys^{11}+342666264\,\mys^{10}-868221384\,\mys^9+2582002923\,\mys^8-5133954624\,\mys^7+12167333720\,\mys^6-18329858400\,\mys^5+34961617488\,\mys^4-36609826176\,\mys^3+55567264896\,\mys^2-31544308224\,\mys+36748318464\right)\,\cos \theta 
\end{dmath*}
\begin{dmath*}
      + 
 2\,\left(12155\,\mys^{18}-102960\,\mys^{17}+939510\,\mys^{16}-5084640\,\mys^{15}+27061674\,\mys^{14}-107857680\,\mys^{13}+415846552\,\mys^{12}-1292458320\,\mys^{11}+3877200963\,\mys^{10}-9604198656\,\mys^9+23026481498\,\mys^8-45370517136\,\mys^7+87254867040\,\mys^6-133052528000\,\mys^5+201751517664\,\mys^4-221013619200\,\mys^3+252354465792\,\mys^2-158522158080\,\mys+122494394880\right)\,\, 
\Bigg)
\end{dmath*}

%
%
%
\begin{dmath}
\,
\end{dmath}
\end{dgroup}
\normalsize

We can plot the amplitude for some values.
In particular we can look to the amplitude near or far the resonances
(fig. \ref{fig:S6zTS4zT_special_s})
and for quite different $\mys$ values (fig. \ref{fig:S6zTS4zT_generic_s}).
We can also consider the contribution to the amplitude of $Num(\mys,
\cos \theta)$,  the one without the
Veneziano amplitude which shows new poles wrt Veneziano and the
complete amplitude (fig. \ref{fig:S6zTS4zT_generic_s}).

 \begin{figure}[!h]
   \centering
    \begin{subfigure}{0.49\textwidth}
    \centering    
    \resizebox{1.0\textwidth}{!}{\input{S6zTS4zT.Num.s_min_s_14.1_22.3_steps_1000_tex.tex}}
    \end{subfigure}
    \hfill
    \begin{subfigure}{0.49\textwidth}
    \centering    
    \resizebox{1.0\textwidth}{!}{\input{S6zTS4zT.log.Num.s_min_s_14.1_22.3_steps_1000_tex.tex}}
    \end{subfigure}
    \begin{subfigure}{0.49\textwidth} 
    \centering    
   \resizebox{1.0\textwidth}{!}{\input{S6zTS4zT.withOUT.s_min_s_14.1_22.3_steps_1000_tex.tex}}
    \end{subfigure}
    \hfill
    \begin{subfigure}{0.49\textwidth}
    \centering    
   \resizebox{1.0\textwidth}{!}{\input{S6zTS4zT.log.withOUT.s_min_s_14.1_22.3_steps_1000_tex.tex}}
    \end{subfigure}
    \begin{subfigure}{0.49\textwidth} 
    \centering    
   \resizebox{1.0\textwidth}{!}{\input{S6zTS4zT.with.s_min_s_14.1_22.3_steps_1000_tex.tex}}
    \end{subfigure}
    \hfill
    \begin{subfigure}{0.49\textwidth}
    \centering    
   \resizebox{1.0\textwidth}{!}{\input{S6zTS4zT.log.with.s_min_s_14.1_22.3_steps_1000_tex.tex}}
    \end{subfigure}
    \caption{
      Different perspectives on the ``S6zTS4zT'' amplitude.
      We use generic $s$ values, i.e. not close to resonances.
      On the left the Numerator in eq. \eqref{eq:Num6040},
      then the Numerator divided by the Denominator with the new poles
      and finally the complete color ordered amplitude with the
      Veneziano contribution.
      On the right there are exactly the same amplitudes but we plot
      their $\log$.
}
\label{fig:S6zTS4zT_generic_s}
\end{figure}

  \begin{figure}[!h]
   \centering
    \begin{subfigure}{0.49\textwidth}
    \centering    
    \resizebox{1.0\textwidth}{!}{\input{S6zTS4zT.Num.s_10.01_10.25_10.51_steps_1000_tex.tex}}
    \end{subfigure}
    \hfill
    \begin{subfigure}{0.49\textwidth}
    \centering    
    \resizebox{1.0\textwidth}{!}{\input{S6zTS4zT.log.Num.s_10.01_10.25_10.51_steps_1000_tex.tex}}
    \end{subfigure}
    \begin{subfigure}{0.49\textwidth} 
    \centering    
   \resizebox{1.0\textwidth}{!}{\input{S6zTS4zT.withOUT.s_10.01_10.25_10.51_steps_1000_tex.tex}}
    \end{subfigure}
    \hfill
    \begin{subfigure}{0.49\textwidth}
    \centering    
   \resizebox{1.0\textwidth}{!}{\input{S6zTS4zT.log.withOUT.s_10.01_10.25_10.51_steps_1000_tex.tex}}
    \end{subfigure}
    \begin{subfigure}{0.49\textwidth} 
    \centering    
   \resizebox{1.0\textwidth}{!}{\input{S6zTS4zT.with.s_10.01_10.25_10.51_steps_1000_tex.tex}}
    \end{subfigure}
    \hfill
    \begin{subfigure}{0.49\textwidth}
    \centering    
   \resizebox{1.0\textwidth}{!}{\input{S6zTS4zT.log.with.s_10.01_10.25_10.51_steps_1000_tex.tex}}
    \end{subfigure}
    \caption{
      Different perspectives on the ``S6zTS4zT'' amplitude.
      We use  two resonant $s$ values and one not resonant to show the difference
      in magnitude of the amplitudes.
      On the left the Numerator in eq. \eqref{eq:Num6040},
      then the Numerator divided by the Denominator with the new poles
      and finally the complete color ordered amplitude with the
      Veneziano contribution.
      On the right there are exactly the same amplitudes but we plot
      their $\log$.
}
\label{fig:S6zTS4zT_special_s}
\end{figure}

From the perspective of the erratic behavior we can consider the
contribution of $\cos( 0 \theta)$ which corresponds to a spin $0$
exchange with respect to higher spins $\cos( k \theta)$ (fig. \ref{fig:S6zTS4zT_ratios}).
This shows clearly the spin $0$ dominance.

  \begin{figure}[!h]
   \centering
    \centering    
    \resizebox{1.0\textwidth}{!}{\input{S6zTS4zT.log.ratios.s_0.0_64.1109_steps_1000_tex.tex}}
    \caption{
      Log of the ratio of the coefficient of $\cos( k \theta)$  to the
      terms independent of $\theta$ of the Numerator.
      This shows the dominance of the spin $0$ particles exchange and
      explain why the amplitude is not erratic.
}
\label{fig:S6zTS4zT_ratios}
\end{figure}

\subsection{ ``S4zS4zS4zS4z``} 

This is the first amplitude with four identical massive scalars
(excluding the trivial case of Kaluza-Klein scalars derived from
massless fields upon compactification)
which can be computed in string.

For this amplitude the Mandelstam variables ($\alpha'=2$) satisfy

\begin{align}
s+t+u= 6,
%
%
%
%
\end{align}
where $s \ge 6$
in the color ordered amplitude $ "S4zS4zS4zS4z" $.

\subsubsection{ Color ordered amplitude $ "S4zS4zS4zS4z" $}

%
%
The color ordered correlator is given by
\tiny
\begin{dgroup}
\begin{dmath*}
\mbox{C}_{S4zS4zS4zS4z}(\omega ,\mys, \myu)
=
\end{dmath*}
  \begin{dmath*}
  ( 16384 )
  ( \left(\omega-1\right)^{2\,\myt+2\,\mys-14} )
  ( \omega^{-2\,\mys-2} )
  \end{dmath*}
\begin{dmath*}
\Bigg(
      + 
 \left(7688\,\myt^4+151822\,\myt^3+1118925\,\myt^2+15872400\,\myt+81745200\right)^2\,\omega^{16} 
\end{dmath*}
\begin{dmath*}
 -4\,\left(280273728\,\myt^8+324126080\,\mys\,\myt^7+9491528912\,\myt^7+11590154016\,\mys\,\myt^6+140264024552\,\myt^6+189050931264\,\mys\,\myt^5+1747174138614\,\myt^5+2293418324412\,\mys\,\myt^4+17581168968579\,\myt^4+19608982377930\,\mys\,\myt^3+123870734344035\,\myt^3+124029403560450\,\mys\,\myt^2+675027532492650\,\myt^2+354765176028000\,\mys\,\myt+4125674521836000\,\myt+13364555446080000\right)\,\omega^{15} 
\end{dmath*}
\begin{dmath*}
      + 
 2\,\left(4547667264\,\myt^8+10754835456\,\mys\,\myt^7+129481915248\,\myt^7+5892944256\,\mys^2\,\myt^6+326732566416\,\mys\,\myt^6+1629393905404\,\myt^6+188237921712\,\mys^2\,\myt^5+4517179677196\,\mys\,\myt^5+16729836019308\,\myt^5+2810843238508\,\mys^2\,\myt^4+43761383528358\,\mys\,\myt^4+139215330718605\,\myt^4+26225168093226\,\mys^2\,\myt^3+293996376351603\,\mys\,\myt^3+817091474997159\,\myt^3+157212649242009\,\mys^2\,\myt^2+1392136855093440\,\mys\,\myt^2+3722308837299396\,\myt^2+474971697138744\,\mys^2\,\myt+3248228376967392\,\mys\,\myt+21727356869374056\,\myt+776580919379856\,\mys^2-4659485516279136\,\mys+98216214471642600\right)\,\omega^{14} 
\end{dmath*}
\begin{dmath*}
 -2\,\left(20578682880\,\myt^8+75406967296\,\mys\,\myt^7+486214388640\,\myt^7+84325650432\,\mys^2\,\myt^6+1913517239440\,\mys\,\myt^6+5277350373336\,\myt^6+28716693760\,\mys^3\,\myt^5+2252017397712\,\mys^2\,\myt^5+22153863803932\,\mys\,\myt^5+43348643000856\,\myt^5+795201459712\,\mys^3\,\myt^4+27526783188100\,\mys^2\,\myt^4+166671966993672\,\mys\,\myt^4+312655686973200\,\myt^4+10005129103968\,\mys^3\,\myt^3+192933533634120\,\mys^2\,\myt^3+885528082667331\,\mys\,\myt^3+1405730488585788\,\myt^3+66269012611500\,\mys^3\,\myt^2+819798283353987\,\mys^2\,\myt^2+3232406684709234\,\mys\,\myt^2+5316388060298958\,\myt^2+210487234775616\,\mys^3\,\myt+1726071203880360\,\mys^2\,\myt+5499394787212128\,\mys\,\myt+30434380687895652\,\myt+197102957926464\,\mys^3+3662139814320816\,\mys^2-28151014652094600\,\mys+217001468531149200\right)\,\omega^{13} 
\end{dmath*}
\begin{dmath*}
      + 
 \left(113485627584\,\myt^8+582034876416\,\mys\,\myt^7+2247101511840\,\myt^7+1008026329600\,\mys^2\,\myt^6+12288404043520\,\mys\,\myt^6+22715821938744\,\myt^6+698459430912\,\mys^3\,\myt^5+22185653605824\,\mys^2\,\myt^5+122708622571240\,\mys\,\myt^5+145080402756696\,\myt^5+162571206016\,\mys^4\,\myt^4+15744268287808\,\mys^3\,\myt^4+222001155562468\,\mys^2\,\myt^4+723226711788282\,\mys\,\myt^4+874905626397942\,\myt^4+3678751891872\,\mys^4\,\myt^3+155181028574088\,\mys^3\,\myt^3+1166486695147350\,\mys^2\,\myt^3+3016397975532057\,\mys\,\myt^3+1728863854262370\,\myt^3+34629686462484\,\mys^4\,\myt^2+715102860849228\,\mys^3\,\myt^2+3752086844598147\,\mys^2\,\myt^2+5070123795453174\,\mys\,\myt^2+6927705866700144\,\myt^2+134580967224768\,\mys^4\,\myt+1511455268386944\,\mys^3\,\myt+4033292908028490\,\mys^2\,\myt+35613918426060\,\mys\,\myt+51886363844576520\,\myt+195046411500576\,\mys^4+221781515037120\,\mys^3+19131181430594184\,\mys^2-152973348659646840\,\mys+732016439979458400\right)\,\omega^{12} 
\end{dmath*}
\begin{dmath*}
 -2\,\left(97582786560\,\myt^8+674311663872\,\mys\,\myt^7+1737086634432\,\myt^7+1638075064320\,\mys^2\,\myt^6+12237253213824\,\mys\,\myt^6+19061052312960\,\myt^6+1755835476480\,\mys^3\,\myt^5+29934019043328\,\mys^2\,\myt^5+117242790991748\,\mys\,\myt^5+85673922007680\,\myt^5+828145390848\,\mys^4\,\myt^4+31891558395360\,\mys^3\,\myt^4+261301725325088\,\mys^2\,\myt^4+550349432171772\,\mys\,\myt^4+316503938645334\,\myt^4+136172709120\,\mys^5\,\myt^3+14658743491776\,\mys^4\,\myt^3+252815453744648\,\mys^3\,\myt^3+1094706653446044\,\mys^2\,\myt^3+1276723329324903\,\mys\,\myt^3-1293584332611258\,\myt^3+2266031381184\,\mys^5\,\myt^2+102998146264812\,\mys^4\,\myt^2+855392699971188\,\mys^3\,\myt^2+2400361101519075\,\mys^2\,\myt^2-5344149583749492\,\mys\,\myt^2-2868002032829994\,\myt^2+13239823038048\,\mys^5\,\myt+260580382267104\,\mys^4\,\myt+1330683877860318\,\mys^3\,\myt-3720095937045576\,\mys^2\,\myt-16699207325055912\,\mys\,\myt+32636642213459016\,\myt+22173930465408\,\mys^5+252530277520608\,\mys^4-1063475976285792\,\mys^3+14257987772110344\,\mys^2-109416573061423848\,\mys+599381653965280200\right)\,\omega^{11} 
\end{dmath*}
\begin{dmath*}
      + 
 2\,\left(102258628416\,\myt^8+957705348096\,\mys\,\myt^7+1977566675616\,\myt^7+3154959590016\,\mys^2\,\myt^6+16724577739824\,\mys\,\myt^6+26383294941696\,\myt^6+4751543983104\,\mys^3\,\myt^5+50884265333136\,\mys^2\,\myt^5+177346677792132\,\mys\,\myt^5+70249513118904\,\myt^5+3459917038976\,\mys^4\,\myt^4+71293320596256\,\mys^3\,\myt^4+443879882057310\,\mys^2\,\myt^4+626223922262298\,\mys\,\myt^4+38488178472030\,\myt^4+1146475287552\,\mys^5\,\myt^3+47905810275632\,\mys^4\,\myt^3+505903872652532\,\mys^3\,\myt^3+1530611921487078\,\mys^2\,\myt^3-278417569473279\,\mys\,\myt^3-4140592029725724\,\myt^3+132508462464\,\mys^6\,\myt^2+14207905052496\,\mys^5\,\myt^2+268782575545710\,\mys^4\,\myt^2+1385741933954532\,\mys^3\,\myt^2+815562429430221\,\mys^2\,\myt^2-18273840192136755\,\mys\,\myt^2-14404363150648932\,\myt^2+1376628260112\,\mys^6\,\myt+59289271037424\,\mys^5\,\myt+487998737269200\,\mys^4\,\myt+1173979715509125\,\mys^3\,\myt-16779585632035194\,\mys^2\,\myt-45818297668088244\,\mys\,\myt+95916775013472024\,\myt+3749851003392\,\mys^6+54459299498688\,\mys^5+439364682486432\,\mys^4-4497518151127728\,\mys^3+13847429201356368\,\mys^2-65202285769735752\,\mys+985049868822402600\right)\,\omega^{10} 
\end{dmath*}
\begin{dmath*}
 -4\,\left(29884728384\,\myt^8+405283882752\,\mys\,\myt^7+877913884512\,\myt^7+1834731601920\,\mys^2\,\myt^6+8609725427040\,\mys\,\myt^6+13984320542976\,\myt^6+3768794110848\,\mys^3\,\myt^5+30628686935424\,\mys^2\,\myt^5+106013392249068\,\mys\,\myt^5+27761406793704\,\myt^5+3857270446080\,\mys^4\,\myt^4+51622470081504\,\mys^3\,\myt^4+297440845382244\,\mys^2\,\myt^4+290762808891642\,\mys\,\myt^4-19690129919844\,\myt^4+1964282987264\,\mys^5\,\myt^3+44069677993040\,\mys^4\,\myt^3+379755009496304\,\mys^3\,\myt^3+819026571826656\,\mys^2\,\myt^3-786233024280963\,\mys\,\myt^3-3565690655734188\,\myt^3+452233152000\,\mys^6\,\myt^2+18447399793512\,\mys^5\,\myt^2+230683527135000\,\mys^4\,\myt^2+836403501080916\,\mys^3\,\myt^2-968412644597538\,\mys^2\,\myt^2-15445809502484931\,\mys\,\myt^2-10912426362004284\,\myt^2+34560570240\,\mys^7\,\myt+3276986825448\,\mys^6\,\myt+62705620266120\,\mys^5\,\myt+325006183255680\,\mys^4\,\myt-184324583683584\,\mys^3\,\myt-14712436975534110\,\mys^2\,\myt-40147788205492812\,\mys\,\myt+93043728594970920\,\myt+160339471488\,\mys^7+6007498607232\,\mys^6+39722489672400\,\mys^5+165951535310280\,\mys^4-3628674817283592\,\mys^3-899276116511358\,\mys^2+39088993391824770\,\mys+696892925520976500\right)\,\omega^9 
\end{dmath*}
\begin{dmath*}
      + 
 3\,\left(9961576128\,\myt^8+247209725952\,\mys\,\myt^7+722959001856\,\myt^7+1666522357248\,\mys^2\,\myt^6+8904756351744\,\mys\,\myt^6+14373953350560\,\myt^6+4760856692736\,\mys^3\,\myt^5+37451137145472\,\mys^2\,\myt^5+125692532883360\,\mys\,\myt^5+48754345717728\,\myt^5+6670023979392\,\mys^4\,\myt^4+73152438520992\,\mys^3\,\myt^4+403847175956956\,\mys^2\,\myt^4+410371267709790\,\mys\,\myt^4+58974533142684\,\myt^4+4760856692736\,\mys^5\,\myt^3+73152438520992\,\mys^4\,\myt^3+586434140737264\,\mys^3\,\myt^3+1076149738303314\,\mys^2\,\myt^3-1008519402804297\,\mys\,\myt^3-5637101446017648\,\myt^3+1666522357248\,\mys^6\,\myt^2+37451137145472\,\mys^5\,\myt^2+403847175956956\,\mys^4\,\myt^2+1076149738303314\,\mys^3\,\myt^2-1948131773705520\,\mys^2\,\myt^2-24444984118435146\,\mys\,\myt^2-10860462354840120\,\myt^2+247209725952\,\mys^7\,\myt+8904756351744\,\mys^6\,\myt+125692532883360\,\mys^5\,\myt+410371267709790\,\mys^4\,\myt-1008519402804297\,\mys^3\,\myt-24444984118435146\,\mys^2\,\myt-60574161032208072\,\mys\,\myt+123916570728472512\,\myt+9961576128\,\mys^8+722959001856\,\mys^7+14373953350560\,\mys^6+48754345717728\,\mys^5+58974533142684\,\mys^4-5637101446017648\,\mys^3-10860462354840120\,\mys^2+123916570728472512\,\mys+1068246660489534000\right)\,\omega^8 
\end{dmath*}
\begin{dmath*}
 -4\,\left(34560570240\,\mys\,\myt^7+160339471488\,\myt^7+452233152000\,\mys^2\,\myt^6+3276986825448\,\mys\,\myt^6+6007498607232\,\myt^6+1964282987264\,\mys^3\,\myt^5+18447399793512\,\mys^2\,\myt^5+62705620266120\,\mys\,\myt^5+39722489672400\,\myt^5+3857270446080\,\mys^4\,\myt^4+44069677993040\,\mys^3\,\myt^4+230683527135000\,\mys^2\,\myt^4+325006183255680\,\mys\,\myt^4+165951535310280\,\myt^4+3768794110848\,\mys^5\,\myt^3+51622470081504\,\mys^4\,\myt^3+379755009496304\,\mys^3\,\myt^3+836403501080916\,\mys^2\,\myt^3-184324583683584\,\mys\,\myt^3-3628674817283592\,\myt^3+1834731601920\,\mys^6\,\myt^2+30628686935424\,\mys^5\,\myt^2+297440845382244\,\mys^4\,\myt^2+819026571826656\,\mys^3\,\myt^2-968412644597538\,\mys^2\,\myt^2-14712436975534110\,\mys\,\myt^2-899276116511358\,\myt^2+405283882752\,\mys^7\,\myt+8609725427040\,\mys^6\,\myt+106013392249068\,\mys^5\,\myt+290762808891642\,\mys^4\,\myt-786233024280963\,\mys^3\,\myt-15445809502484931\,\mys^2\,\myt-40147788205492812\,\mys\,\myt+39088993391824770\,\myt+29884728384\,\mys^8+877913884512\,\mys^7+13984320542976\,\mys^6+27761406793704\,\mys^5-19690129919844\,\mys^4-3565690655734188\,\mys^3-10912426362004284\,\mys^2+93043728594970920\,\mys+696892925520976500\right)\,\omega^7 
\end{dmath*}
\begin{dmath*}
      + 
 2\,\left(132508462464\,\mys^2\,\myt^6+1376628260112\,\mys\,\myt^6+3749851003392\,\myt^6+1146475287552\,\mys^3\,\myt^5+14207905052496\,\mys^2\,\myt^5+59289271037424\,\mys\,\myt^5+54459299498688\,\myt^5+3459917038976\,\mys^4\,\myt^4+47905810275632\,\mys^3\,\myt^4+268782575545710\,\mys^2\,\myt^4+487998737269200\,\mys\,\myt^4+439364682486432\,\myt^4+4751543983104\,\mys^5\,\myt^3+71293320596256\,\mys^4\,\myt^3+505903872652532\,\mys^3\,\myt^3+1385741933954532\,\mys^2\,\myt^3+1173979715509125\,\mys\,\myt^3-4497518151127728\,\myt^3+3154959590016\,\mys^6\,\myt^2+50884265333136\,\mys^5\,\myt^2+443879882057310\,\mys^4\,\myt^2+1530611921487078\,\mys^3\,\myt^2+815562429430221\,\mys^2\,\myt^2-16779585632035194\,\mys\,\myt^2+13847429201356368\,\myt^2+957705348096\,\mys^7\,\myt+16724577739824\,\mys^6\,\myt+177346677792132\,\mys^5\,\myt+626223922262298\,\mys^4\,\myt-278417569473279\,\mys^3\,\myt-18273840192136755\,\mys^2\,\myt-45818297668088244\,\mys\,\myt-65202285769735752\,\myt+102258628416\,\mys^8+1977566675616\,\mys^7+26383294941696\,\mys^6+70249513118904\,\mys^5+38488178472030\,\mys^4-4140592029725724\,\mys^3-14404363150648932\,\mys^2+95916775013472024\,\mys+985049868822402600\right)\,\omega^6 
\end{dmath*}
\begin{dmath*}
 -2\,\left(136172709120\,\mys^3\,\myt^5+2266031381184\,\mys^2\,\myt^5+13239823038048\,\mys\,\myt^5+22173930465408\,\myt^5+828145390848\,\mys^4\,\myt^4+14658743491776\,\mys^3\,\myt^4+102998146264812\,\mys^2\,\myt^4+260580382267104\,\mys\,\myt^4+252530277520608\,\myt^4+1755835476480\,\mys^5\,\myt^3+31891558395360\,\mys^4\,\myt^3+252815453744648\,\mys^3\,\myt^3+855392699971188\,\mys^2\,\myt^3+1330683877860318\,\mys\,\myt^3-1063475976285792\,\myt^3+1638075064320\,\mys^6\,\myt^2+29934019043328\,\mys^5\,\myt^2+261301725325088\,\mys^4\,\myt^2+1094706653446044\,\mys^3\,\myt^2+2400361101519075\,\mys^2\,\myt^2-3720095937045576\,\mys\,\myt^2+14257987772110344\,\myt^2+674311663872\,\mys^7\,\myt+12237253213824\,\mys^6\,\myt+117242790991748\,\mys^5\,\myt+550349432171772\,\mys^4\,\myt+1276723329324903\,\mys^3\,\myt-5344149583749492\,\mys^2\,\myt-16699207325055912\,\mys\,\myt-109416573061423848\,\myt+97582786560\,\mys^8+1737086634432\,\mys^7+19061052312960\,\mys^6+85673922007680\,\mys^5+316503938645334\,\mys^4-1293584332611258\,\mys^3-2868002032829994\,\mys^2+32636642213459016\,\mys+599381653965280200\right)\,\omega^5 
\end{dmath*}
\begin{dmath*}
      + 
 \left(162571206016\,\mys^4\,\myt^4+3678751891872\,\mys^3\,\myt^4+34629686462484\,\mys^2\,\myt^4+134580967224768\,\mys\,\myt^4+195046411500576\,\myt^4+698459430912\,\mys^5\,\myt^3+15744268287808\,\mys^4\,\myt^3+155181028574088\,\mys^3\,\myt^3+715102860849228\,\mys^2\,\myt^3+1511455268386944\,\mys\,\myt^3+221781515037120\,\myt^3+1008026329600\,\mys^6\,\myt^2+22185653605824\,\mys^5\,\myt^2+222001155562468\,\mys^4\,\myt^2+1166486695147350\,\mys^3\,\myt^2+3752086844598147\,\mys^2\,\myt^2+4033292908028490\,\mys\,\myt^2+19131181430594184\,\myt^2+582034876416\,\mys^7\,\myt+12288404043520\,\mys^6\,\myt+122708622571240\,\mys^5\,\myt+723226711788282\,\mys^4\,\myt+3016397975532057\,\mys^3\,\myt+5070123795453174\,\mys^2\,\myt+35613918426060\,\mys\,\myt-152973348659646840\,\myt+113485627584\,\mys^8+2247101511840\,\mys^7+22715821938744\,\mys^6+145080402756696\,\mys^5+874905626397942\,\mys^4+1728863854262370\,\mys^3+6927705866700144\,\mys^2+51886363844576520\,\mys+732016439979458400\right)\,\omega^4 
\end{dmath*}
\begin{dmath*}
 -2\,\left(28716693760\,\mys^5\,\myt^3+795201459712\,\mys^4\,\myt^3+10005129103968\,\mys^3\,\myt^3+66269012611500\,\mys^2\,\myt^3+210487234775616\,\mys\,\myt^3+197102957926464\,\myt^3+84325650432\,\mys^6\,\myt^2+2252017397712\,\mys^5\,\myt^2+27526783188100\,\mys^4\,\myt^2+192933533634120\,\mys^3\,\myt^2+819798283353987\,\mys^2\,\myt^2+1726071203880360\,\mys\,\myt^2+3662139814320816\,\myt^2+75406967296\,\mys^7\,\myt+1913517239440\,\mys^6\,\myt+22153863803932\,\mys^5\,\myt+166671966993672\,\mys^4\,\myt+885528082667331\,\mys^3\,\myt+3232406684709234\,\mys^2\,\myt+5499394787212128\,\mys\,\myt-28151014652094600\,\myt+20578682880\,\mys^8+486214388640\,\mys^7+5277350373336\,\mys^6+43348643000856\,\mys^5+312655686973200\,\mys^4+1405730488585788\,\mys^3+5316388060298958\,\mys^2+30434380687895652\,\mys+217001468531149200\right)\,\omega^3 
\end{dmath*}
\begin{dmath*}
      + 
 2\,\left(5892944256\,\mys^6\,\myt^2+188237921712\,\mys^5\,\myt^2+2810843238508\,\mys^4\,\myt^2+26225168093226\,\mys^3\,\myt^2+157212649242009\,\mys^2\,\myt^2+474971697138744\,\mys\,\myt^2+776580919379856\,\myt^2+10754835456\,\mys^7\,\myt+326732566416\,\mys^6\,\myt+4517179677196\,\mys^5\,\myt+43761383528358\,\mys^4\,\myt+293996376351603\,\mys^3\,\myt+1392136855093440\,\mys^2\,\myt+3248228376967392\,\mys\,\myt-4659485516279136\,\myt+4547667264\,\mys^8+129481915248\,\mys^7+1629393905404\,\mys^6+16729836019308\,\mys^5+139215330718605\,\mys^4+817091474997159\,\mys^3+3722308837299396\,\mys^2+21727356869374056\,\mys+98216214471642600\right)\,\omega^2 
\end{dmath*}
\begin{dmath*}
 -4\,\left(324126080\,\mys^7\,\myt+11590154016\,\mys^6\,\myt+189050931264\,\mys^5\,\myt+2293418324412\,\mys^4\,\myt+19608982377930\,\mys^3\,\myt+124029403560450\,\mys^2\,\myt+354765176028000\,\mys\,\myt+280273728\,\mys^8+9491528912\,\mys^7+140264024552\,\mys^6+1747174138614\,\mys^5+17581168968579\,\mys^4+123870734344035\,\mys^3+675027532492650\,\mys^2+4125674521836000\,\mys+13364555446080000\right)\,\omega 
\end{dmath*}
\begin{dmath*}
      + 
 \left(7688\,\mys^4+151822\,\mys^3+1118925\,\mys^2+15872400\,\mys+81745200\right)^2 
\Bigg)
\end{dmath*}

%
%
%
\begin{dmath}
\,
\end{dmath}
\end{dgroup}
\normalsize
%
%

%
%

The color ordered amplitude can be written as a product of the Veneziano amplitude
times a polynomial $\mbox{Num}(\mys ,\myu)$
in the Mandelstam variables $\mys$ and $\myu$
divided by another polynomial $\mbox{Den}(\mys ,\myu)$ which comes
from the transformation of the $\Gamma$ function in the denominator
which arises from the moduli space integration to the Veneziano "canonical" one.
Explicitly we get
\begin{equation}
A_{S4zS4zS4zS4z}
=
A_{\mbox{Veneziano} 
}(\mys, \myu)
\frac{\mbox{Num}(\mys ,\myu)}{\mbox{Den}(\mys, \myu)}
,
\end{equation}
where we have
\tiny
\begin{dgroup}
\begin{dmath*}
\mbox{Num}(\mys ,\myu)
=
60072668430336\,
\Bigg(65536\,\mys^8\,\myu^{16}
+720896\,\mys^7\,\myu^{16}+4808704\,\mys^6\,\myu^{16}+18870272\,\mys^5\,\myu^{16}+59822336\,\mys^4\,\myu^{16}+132541440\,\mys^3\,\myu^{16}+280662016\,\mys^2\,\myu^{16}+280627200\,\mys\,\myu^{16}+466560000\,\myu^{16}+524288\,\mys^9\,\myu^{15}+2621440\,\mys^8\,\myu^{15}+3866624\,\mys^7\,\myu^{15}-79855616\,\mys^6\,\myu^{15}-427194368\,\mys^5\,\myu^{15}-1811140608\,\mys^4\,\myu^{15}-4116692992\,\mys^3\,\myu^{15}-11226759168\,\mys^2\,\myu^{15}-9737625600\,\mys\,\myu^{15}-22394880000\,\myu^{15}+1835008\,\mys^{10}\,\myu^{14}-1114112\,\mys^9\,\myu^{14}-36225024\,\mys^8\,\myu^{14}-343339008\,\mys^7\,\myu^{14}+224235520\,\mys^6\,\myu^{14}+2746905600\,\mys^5\,\myu^{14}+24349196544\,\mys^4\,\myu^{14}+49200138752\,\mys^3\,\myu^{14}+208961029120\,\mys^2\,\myu^{14}+131670144000\,\mys\,\myu^{14}+487555200000\,\myu^{14}+3670016\,\mys^{11}\,\myu^{13}-20643840\,\mys^{10}\,\myu^{13}-54247424\,\mys^9\,\myu^{13}-323964928\,\mys^8\,\myu^{13}+4260135936\,\mys^7\,\myu^{13}+2103628800\,\mys^6\,\myu^{13}+15909392128\,\mys^5\,\myu^{13}-205283943424\,\mys^4\,\myu^{13}-197373016064\,\mys^3\,\myu^{13}-2434804915200\,\mys^2\,\myu^{13}-685480320000\,\mys\,\myu^{13}-6368544000000\,\myu^{13}+4587520\,\mys^{12}\,\myu^{12}-44498944\,\mys^{11}\,\myu^{12}+50683904\,\mys^{10}\,\myu^{12}-246611968\,\mys^9\,\myu^{12}+8053071872\,\mys^8\,\myu^{12}-26526434304\,\mys^7\,\myu^{12}+15361376768\,\mys^6\,\myu^{12}-404336535808\,\mys^5\,\myu^{12}+1404645235040\,\mys^4\,\myu^{12}-1669415137024\,\mys^3\,\myu^{12}+20086738845184\,\mys^2\,\myu^{12}-2892655895040\,\mys\,\myu^{12}+55623924720000\,\myu^{12}+3670016\,\mys^{13}\,\myu^{11}-44498944\,\mys^{12}\,\myu^{11}+139296768\,\mys^{11}\,\myu^{11}-597110784\,\mys^{10}\,\myu^{11}+10292295680\,\mys^9\,\myu^{11}-58374865920\,\mys^8\,\myu^{11}+134001096960\,\mys^7\,\myu^{11}-543310280192\,\mys^6\,\myu^{11}+3446407628864\,\mys^5\,\myu^{11}-9334533828480\,\mys^4\,\myu^{11}+29418655534080\,\mys^3\,\myu^{11}-124444569074688\,\mys^2\,\myu^{11}+71370302480640\,\mys\,\myu^{11}-342563981760000\,\myu^{11}+1835008\,\mys^{14}\,\myu^{10}-20643840\,\mys^{13}\,\myu^{10}+50683904\,\mys^{12}\,\myu^{10}-597110784\,\mys^{11}\,\myu^{10}+11576215040\,\mys^{10}\,\myu^{10}-74134760448\,\mys^9\,\myu^{10}+234781563136\,\mys^8\,\myu^{10}-820243819520\,\mys^7\,\myu^{10}+4527104101728\,\mys^6\,\myu^{10}-18197800128128\,\mys^5\,\myu^{10}+57184376915600\,\mys^4\,\myu^{10}-214023641868384\,\mys^3\,\myu^{10}+590662322763008\,\mys^2\,\myu^{10}-551157482445120\,\mys\,\myu^{10}+1527531291000000\,\myu^{10}+524288\,\mys^{15}\,\myu^9-1114112\,\mys^{14}\,\myu^9-54247424\,\mys^{13}\,\myu^9-246611968\,\mys^{12}\,\myu^9+10292295680\,\mys^{11}\,\myu^9-74134760448\,\mys^{10}\,\myu^9+257958004736\,\mys^9\,\myu^9-834111267584\,\mys^8\,\myu^9+4243162974528\,\mys^7\,\myu^9-18798385860480\,\mys^6\,\myu^9+70200626314320\,\mys^5\,\myu^9-281440460528064\,\mys^4\,\myu^9+976202659907008\,\mys^3\,\myu^9-2134830793346880\,\mys^2\,\myu^9+2533775736686400\,\mys\,\myu^9-4979462202000000\,\myu^9+65536\,\mys^{16}\,\myu^8+2621440\,\mys^{15}\,\myu^8-36225024\,\mys^{14}\,\myu^8-323964928\,\mys^{13}\,\myu^8+8053071872\,\mys^{12}\,\myu^8-58374865920\,\mys^{11}\,\myu^8+234781563136\,\mys^{10}\,\myu^8-834111267584\,\mys^9\,\myu^8+3483949111680\,\mys^8\,\myu^8-12473517746048\,\mys^7\,\myu^8+45000029315376\,\mys^6\,\myu^8-231649926099552\,\mys^5\,\myu^8+1038783352971777\,\mys^4\,\myu^8-3035983327098716\,\mys^3\,\myu^8+5742033498692240\,\mys^2\,\myu^8-7699957610342760\,\mys\,\myu^8+11810922136942500\,\myu^8+720896\,\mys^{16}\,\myu^7+3866624\,\mys^{15}\,\myu^7-343339008\,\mys^{14}\,\myu^7+4260135936\,\mys^{13}\,\myu^7-26526434304\,\mys^{12}\,\myu^7+134001096960\,\mys^{11}\,\myu^7-820243819520\,\mys^{10}\,\myu^7+4243162974528\,\mys^9\,\myu^7-12473517746048\,\mys^8\,\myu^7+22072809135840\,\mys^7\,\myu^7-108864407202048\,\mys^6\,\myu^7+744090434511876\,\mys^5\,\myu^7-2811697635400712\,\mys^4\,\myu^7+6522130466079584\,\mys^3\,\myu^7-11066778745879584\,\mys^2\,\myu^7+15761353043238480\,\mys\,\myu^7-19949667077340000\,\myu^7+4808704\,\mys^{16}\,\myu^6-79855616\,\mys^{15}\,\myu^6+224235520\,\mys^{14}\,\myu^6+2103628800\,\mys^{13}\,\myu^6+15361376768\,\mys^{12}\,\myu^6-543310280192\,\mys^{11}\,\myu^6+4527104101728\,\mys^{10}\,\myu^6-18798385860480\,\mys^9\,\myu^6+45000029315376\,\mys^8\,\myu^6-108864407202048\,\mys^7\,\myu^6+528676719665158\,\mys^6\,\myu^6-2174218594855052\,\mys^5\,\myu^6+5463553226113312\,\mys^4\,\myu^6-9379280110724060\,\mys^3\,\myu^6+14340315332220812\,\mys^2\,\myu^6-21112178776750080\,\mys\,\myu^6+22820232892275000\,\myu^6+18870272\,\mys^{16}\,\myu^5-427194368\,\mys^{15}\,\myu^5+2746905600\,\mys^{14}\,\myu^5+15909392128\,\mys^{13}\,\myu^5-404336535808\,\mys^{12}\,\myu^5+3446407628864\,\mys^{11}\,\myu^5-18197800128128\,\mys^{10}\,\myu^5+70200626314320\,\mys^9\,\myu^5-231649926099552\,\mys^8\,\myu^5+744090434511876\,\mys^7\,\myu^5-2174218594855052\,\mys^6\,\myu^5+4814916435653216\,\mys^5\,\myu^5-7243079016085044\,\mys^4\,\myu^5+8117569234024836\,\mys^3\,\myu^5-10848079504152360\,\mys^2\,\myu^5+16672013496088200\,\mys\,\myu^5-15586762668750000\,\myu^5
\end{dmath*}
\begin{dmath*}
+59822336\,\mys^{16}\,\myu^4-1811140608\,\mys^{15}\,\myu^4+24349196544\,\mys^{14}\,\myu^4-205283943424\,\mys^{13}\,\myu^4+1404645235040\,\mys^{12}\,\myu^4-9334533828480\,\mys^{11}\,\myu^4+57184376915600\,\mys^{10}\,\myu^4-281440460528064\,\mys^9\,\myu^4+1038783352971777\,\mys^8\,\myu^4-2811697635400712\,\mys^7\,\myu^4+5463553226113312\,\mys^6\,\myu^4-7243079016085044\,\mys^5\,\myu^4+5866226444534828\,\mys^4\,\myu^4-2606948769780120\,\mys^3\,\myu^4+2387066457922740\,\mys^2\,\myu^4-4886260629267600\,\mys\,\myu^4+3520138901152500\,\myu^4
\end{dmath*}
\begin{dmath*}
+132541440\,\mys^{16}\,\myu^3-4116692992\,\mys^{15}\,\myu^3+49200138752\,\mys^{14}\,\myu^3-197373016064\,\mys^{13}\,\myu^3-1669415137024\,\mys^{12}\,\myu^3+29418655534080\,\mys^{11}\,\myu^3-214023641868384\,\mys^{10}\,\myu^3+976202659907008\,\mys^9\,\myu^3-3035983327098716\,\mys^8\,\myu^3+6522130466079584\,\mys^7\,\myu^3-9379280110724060\,\mys^6\,\myu^3+8117569234024836\,\mys^5\,\myu^3-2606948769780120\,\mys^4\,\myu^3-2199135231976320\,\mys^3\,\myu^3+3244877942385600\,\mys^2\,\myu^3-3059313063360120\,\mys\,\myu^3+3094803557730000\,\myu^3
\end{dmath*}
\begin{dmath*}
+280662016\,\mys^{16}\,\myu^2-11226759168\,\mys^{15}\,\myu^2+208961029120\,\mys^{14}\,\myu^2-2434804915200\,\mys^{13}\,\myu^2+20086738845184\,\mys^{12}\,\myu^2-124444569074688\,\mys^{11}\,\myu^2+590662322763008\,\mys^{10}\,\myu^2-2134830793346880\,\mys^9\,\myu^2+5742033498692240\,\mys^8\,\myu^2-11066778745879584\,\mys^7\,\myu^2+14340315332220812\,\mys^6\,\myu^2-10848079504152360\,\mys^5\,\myu^2+2387066457922740\,\mys^4\,\myu^2+3244877942385600\,\mys^3\,\myu^2-3717924264970740\,\mys^2\,\myu^2+3166534651865400\,\mys\,\myu^2-2496866297850000\,\myu^2+280627200\,\mys^{16}\,\myu-9737625600\,\mys^{15}\,\myu+131670144000\,\mys^{14}\,\myu-685480320000\,\mys^{13}\,\myu-2892655895040\,\mys^{12}\,\myu+71370302480640\,\mys^{11}\,\myu-551157482445120\,\mys^{10}\,\myu+2533775736686400\,\mys^9\,\myu-7699957610342760\,\mys^8\,\myu+15761353043238480\,\mys^7\,\myu-21112178776750080\,\mys^6\,\myu+16672013496088200\,\mys^5\,\myu-4886260629267600\,\mys^4\,\myu-3059313063360120\,\mys^3\,\myu+3166534651865400\,\mys^2\,\myu-1256227019874000\,\mys\,\myu+531972441000000\,\myu+466560000\,\mys^{16}-22394880000\,\mys^{15}+487555200000\,\mys^{14}-6368544000000\,\mys^{13}+55623924720000\,\mys^{12}-342563981760000\,\mys^{11}+1527531291000000\,\mys^{10}-4979462202000000\,\mys^9+11810922136942500\,\mys^8-19949667077340000\,\mys^7+22820232892275000\,\mys^6-15586762668750000\,\mys^5+3520138901152500\,\mys^4+3094803557730000\,\mys^3-2496866297850000\,\mys^2+531972441000000\,\mys\Bigg)
\end{dmath*}
\begin{dmath}
  ,%
  \label{eq:NumS4S4S4S4}
\end{dmath}
\end{dgroup}
\normalsize
and
\tiny
\begin{dgroup}
\begin{dmath*}
\mbox{Den}(\mys ,\myu)
=
\left(\myu+\mys-6\right)\,\left(\myu+\mys-5\right)\,\left(\myu+\mys-4\right)\,\left(\myu+\mys-3\right)\,\left(\myu+\mys-2\right)\,\left(\myu+\mys-1\right)\,\left(\myu+\mys\right)\,\left(\myu+\mys+1\right)\,\left(2\,\myu+2\,\mys-13\right)\,\left(2\,\myu+2\,\mys-11\right)\,\left(2\,\myu+2\,\mys-9\right)\,\left(2\,\myu+2\,\mys-7\right)\,\left(2\,\myu+2\,\mys-5\right)\,\left(2\,\myu+2\,\mys-3\right)\,\left(2\,\myu+2\,\mys-1\right)\,\left(2\,\myu+2\,\mys+1\right)
.
\end{dmath*}
\begin{dmath}
\, ~
\end{dmath}
\end{dgroup}
\normalsize

The previous quantities in channel $\mys$ can be written
using $\mys$ and $\cos(\theta)$
($\mys \ge 6$) with
\begin{align}
\myt =& {{\left(\mys-6\right)\,\cos \theta-\mys+6}\over{2}}
\nonumber \\
\myu =& -{{\left(\mys-6\right)\,\cos \theta+\mys-6}\over{2}}
\end{align}

We can plot the amplitude for some values.
In particular we can look to the amplitude near or far the resonances
(fig. \ref{fig:S4zS4zS4zS4z_special_s}).
and for quite different $\mys$ values (fig. \ref{fig:S4zS4zS4zS4z_generic_s}).
We can also consider the contribution to the amplitude of $Num(\mys,
\cos \theta)$,  the one without the
Veneziano amplitude which shows new poles wrt Veneziano and the
complete amplitude (fig. \ref{fig:S4zS4zS4zS4z_generic_s}).

 \begin{figure}[!h]
   \centering
    \begin{subfigure}{0.49\textwidth}
    \centering    
    \resizebox{1.0\textwidth}{!}{\input{S4zS4zS4zS4z.Num.s_6.21_14.1_22.3_steps_1000_tex.tex}}
    \end{subfigure}
    \hfill
    \begin{subfigure}{0.49\textwidth}
    \centering    
    \resizebox{1.0\textwidth}{!}{\input{S4zS4zS4zS4z.log.Num.s_6.21_14.1_22.3_steps_1000_tex.tex}}
    \end{subfigure}
    \begin{subfigure}{0.49\textwidth} 
    \centering    
   \resizebox{1.0\textwidth}{!}{\input{S4zS4zS4zS4z.withOUT.s_6.21_14.1_22.3_steps_1000_tex.tex}}
    \end{subfigure}
    \hfill
    \begin{subfigure}{0.49\textwidth}
    \centering    
   \resizebox{1.0\textwidth}{!}{\input{S4zS4zS4zS4z.log.withOUT.s_6.21_14.1_22.3_steps_1000_tex.tex}}
    \end{subfigure}
    \begin{subfigure}{0.49\textwidth} 
    \centering    
   \resizebox{1.0\textwidth}{!}{\input{S4zS4zS4zS4z.with.s_6.21_14.1_22.3_steps_1000_tex.tex}}
    \end{subfigure}
    \hfill
    \begin{subfigure}{0.49\textwidth}
    \centering    
   \resizebox{1.0\textwidth}{!}{\input{S4zS4zS4zS4z.log.with.s_6.21_14.1_22.3_steps_1000_tex.tex}}
    \end{subfigure}
    \caption{
      Different perspectives on the $S4 S4 S4 S4$ amplitude.
      We use generic $s$ values, i.e. not close to resonances.
      On the left the Numerator in eq. \eqref{eq:NumS4S4S4S4},
      then the Numerator divided by the Denominator with the new poles
      and finally the complete color ordered amplitude with the
      Veneziano contribution.
      On the right there are exactly the same amplitudes but we plot
      their $\log$.
}
\label{fig:S4zS4zS4zS4z_generic_s}
\end{figure}

  \begin{figure}[!h]
   \centering
    \begin{subfigure}{0.49\textwidth}
    \centering    
    \resizebox{1.0\textwidth}{!}{\input{S4zS4zS4zS4z.Num.s_10.01_10.25_10.51_steps_1000_tex.tex}}
    \end{subfigure}
    \hfill
    \begin{subfigure}{0.49\textwidth}
    \centering    
    \resizebox{1.0\textwidth}{!}{\input{S4zS4zS4zS4z.log.Num.s_10.01_10.25_10.51_steps_1000_tex.tex}}
    \end{subfigure}
    \begin{subfigure}{0.49\textwidth} 
    \centering    
   \resizebox{1.0\textwidth}{!}{\input{S4zS4zS4zS4z.withOUT.s_10.01_10.25_10.51_steps_1000_tex.tex}}
    \end{subfigure}
    \hfill
    \begin{subfigure}{0.49\textwidth}
    \centering    
   \resizebox{1.0\textwidth}{!}{\input{S4zS4zS4zS4z.log.withOUT.s_10.01_10.25_10.51_steps_1000_tex.tex}}
    \end{subfigure}
    \begin{subfigure}{0.49\textwidth} 
    \centering    
   \resizebox{1.0\textwidth}{!}{\input{S4zS4zS4zS4z.with.s_10.01_10.25_10.51_steps_1000_tex.tex}}
    \end{subfigure}
    \hfill
    \begin{subfigure}{0.49\textwidth}
    \centering    
   \resizebox{1.0\textwidth}{!}{\input{S4zS4zS4zS4z.log.with.s_10.01_10.25_10.51_steps_1000_tex.tex}}
    \end{subfigure}
    \caption{
      Different perspectives on the $S4 T T T$ amplitude.
      We use  two resonant $s$ values and one not resonant to show the difference
      in magnitude of the amplitudes.
      On the left the Numerator in eq. \eqref{eq:NumS4S4S4S4},
      then the Numerator divided by the Denominator with the new poles
      and finally the complete color ordered amplitude with the
      Veneziano contribution.
      On the right there are exactly the same amplitudes but we plot
      their $\log$.
}
\label{fig:S4zS4zS4zS4z_special_s}
\end{figure}

From the perspective of the erratic behavior we can consider the
contribution of $\cos( 0 \theta)$ which corresponds to a spin $0$
exchange with respect to higher spins $\cos( k \theta)$ (fig. \ref{fig:S4zS4zS4zS4z_ratios}).
This shows clearly the spin $0$ dominance.

  \begin{figure}[!h]
   \centering
    \centering    
    \resizebox{1.0\textwidth}{!}{\input{S4zS4zS4zS4z.log.ratios.s_6_60_tex.tex}}
    \caption{
      Log of the ratio of the coefficient of $\cos( k \theta)$  to the
      terms independent of $\theta$ of the Numerator.
      This shows the dominance of the spin $0$ particles exchange and
      explain why the amplitude is not erratic.
}
\label{fig:S4zS4zS4zS4z_ratios}
\end{figure}

\section*{Acknowledgments}
We would like to thank Carlo Angelantonj, Dripto Biswas, Chrysoula
Markou and Raffaele Marotta for
discussions.
This research is partially supported by the MUR PRIN contract
2020KR4KN2 “String Theory as a bridge between Gauge Theories and
Quantum Gravity” and by the INFN project ST\&FI “String Theory \&
Fundamental Interactions”.

\appendix

\section{Amplitudes for non cyclic orderings  }
\label{app:othercorrsandamps}

In this appendix we would like to discuss how to derive the color
ordered non cyclic correlators and amplitudes when
it is known
one $SL(2,\R)$ gauge fixed color ordered correlator
like
\begin{align}
C_{1234}(x_r, s,t,u, \alpha_r)
&=
\langle 0 | V^{(matt)}(x_1; k_1, \alpha_1)\, V^{(matt)}(x_2; k_2, \alpha_2)\,
\nonumber\\
&\phantom{\langle 0 | } \times
V^{(matt)}(x_3; k_3, \alpha_3)\, V^{(matt)}(x_4; k_4, \alpha_4)\, | 0\rangle
\nonumber\\
&\times ~x_{12} x_{14} x_{24}
~d x_3
\nonumber\\
=&
f_{1234}(\omega, s,t,u,\alpha_r)~ d \omega
,
\end{align}
where 
$V^{(matt)}(x_r; \alpha_r)$ is the matter vertex operator for the state
$\alpha_r$ ($r=1,2,3,4$) with momentum $k_r$.
and $x_1 > x_2> x_3> x_4$.
We have also defined the anharmonic ratio
\begin{equation}
\omega=
\omega_{1234}=
\frac{x_{12} x_{34}}{x_{13} x_{24}}
.
\end{equation}

The color ordered amplitude is then obtained as
\begin{align}
A_{1234}( s,t,u, \alpha_r)
&=
\int_0^1 d \omega~f_{1234}(\omega, s,t,u,\alpha_r)
.
\end{align}

We would now compute the correlators and amplitudes associated with
different color orderings.

In the case where all $\alpha_r$ are equal, as in Veneziano amplitude,
the task is immediate since it amounts to use the maps among the
different Mandelstam variables in the different orderings, i.e
\begin{align}
s_{1342}=&t, &t_{1342}=&u, &u_{1342}=&s, 
\nonumber\\
s_{1423}=&u, &t_{1423}=&s, &u_{1423}=&t,
\nonumber\\
s_{1243}=&s, &t_{1243}=&u, &u_{1243}=&t,
\nonumber\\
s_{1432}=&u, &t_{1432}=&t, &u_{1432}=&t,
\nonumber\\
s_{1324}=&t, &t_{1324}=&s, &u_{1324}=&u
.
\end{align}
For example
\begin{equation}
A_{1342}(s_{1342}, t_{1342}, t_{1342}, \alpha_r=\alpha)
=
A_{1234}(s=u_{1342}, t=u_{1342}, u=t_{1342}, \alpha_r=\alpha)
.
\end{equation}

In all the other cases where some $\alpha_r$ differ there is no obvious
solution which may be implemented on the amplitude.

There is however a solution which requires the knowledge of the correlator.

Consider in facts the correlator
\begin{align}
C_{1423}&(x_r, s_{1423},t_{1423},u_{1423}, \alpha_r)
\nonumber\\
=&
\langle 0 | V^{(matt)}(x_1; k_1, \alpha_1)\, V^{(matt)}(x_4;
k_4, \alpha_4)\,
V^{(matt)}(x_2; k_2, \alpha_2)\, V^{(matt)}(x_3; k_3, \alpha_3)\, | 0\rangle
\nonumber\\
&\times ~x_{14} x_{13} x_{43}
~d x_2
\nonumber\\
=&
f_{1423}(\omega_{1423}, s_{1423},t_{1423},u_{1423}, \alpha_r)~ d \omega_{1423}
,
\end{align}
where and $x_1 > x_3> x_4> x_2$ and we have gauge fixed $SL(2,\R)$ in
the natural usual way which however differs from the original ordering.
We have also defined in a natural way
\begin{equation}
\omega_{1423}=
\frac{x_{14} x_{23}}{x_{12} x_{43}}
=\frac{1}{1-\omega}
,
\end{equation}
so that the amplitude reads
\begin{align}
A_{1423}( s_{1423},t_{1423},u_{1423}, \alpha_r)
&=
\int_0^1 d \omega_{1423}~f_{1423}(\omega_{1423}, s_{1423},t_{1423},u_{1423},\alpha_r)
.
\end{align}

Now the vertices can be written as
\begin{equation}
V^{(matt)}(x_r; k_r, \alpha_r)
=
~e^{i 2 k_r \cdot L(x_r)}
,
\end{equation}
where $\mbox{Pol}_{\alpha_r}$ is a polynomial in its arguments.

This means that the original amplitude can be written as 
\begin{align}
C_{1234}(x_r, s,t,u, \alpha_r)
&=
\delta^D(\sum_r k_r)\,
\sum c_{n,m} \prod_a \partial_{x_r}^{m_a}  \partial_{x_s}^{n_a} \ln(x_{r s})
\nonumber\\
&\times
x_{12}^{2\ap k_1\cdot k_2}\,x_{13}^{2\ap k_1\cdot k_3}\,x_{14}^{2\ap k_1\cdot k_4}\,
x_{23}^{2\ap k_2\cdot k_3}\,x_{24}^{2\ap k_2\cdot k_4}\,
x_{34}^{2\ap k_3\cdot k_4}
\nonumber\\
&\times ~x_{12} x_{14} x_{24}
~d x_3
,
\end{align}
for certain computable $c_{n,m}$s.
In an analogous way
\begin{align}
C_{1423}(x_r, s_{1423},t_{1423},u_{1423}, \alpha_r)
&=
\delta^D(\sum_r k_r)\,
\sum c_{n,m} \prod_a \partial_{x_r}^{m_a}  \partial_{x_s}^{n_a} \ln(x_{r s})
\nonumber\\
&\times
x_{14}^{2\ap k_1\cdot k_4}\,x_{12}^{2\ap k_1\cdot k_2}\,x_{13}^{2\ap k_1\cdot k_3}\,
x_{42}^{2\ap k_4\cdot k_2}\,x_{43}^{2\ap k_4\cdot k_3}\,
x_{23}^{2\ap k_2\cdot k_3}
\nonumber\\
&\times ~x_{14} x_{13} x_{43}
~d x_2
,
\end{align}
where the expression has exactly the same $c_{n,m}$s as before since
only one derivative is sufficient to make $\ln(x_{r s})$
and $\ln(x_{s r})$ indistinguishable.

Therefore we can compute the ratio
\begin{align}
&
\frac{
f_{1423}(\omega_{1423}, s_{1423},t_{1423},u_{1423}, \alpha_r)~ d \omega_{1423}
}{
f_{1234}(\omega=\frac{1}{1-\omega_{1423}},
s=t_{1423},t=u_{1423},u=s_{1423}, \alpha_r)~
\frac{1}{(1-\omega_{1423})^2}~ d \omega_{1423}
}
\nonumber\\
=&
\frac{
x_{42}^{2\ap k_4\cdot k_2}\,x_{43}^{2\ap k_4\cdot k_3}\,
~x_{14} x_{13} x_{43} ~d x_2
}{
x_{24}^{2\ap k_2\cdot k_4}\,x_{34}^{2\ap k_3\cdot k_4}\,
~x_{12} x_{14} x_{24} ~d x_3
}
\nonumber\\
=&
e^{\pm 2i \pi \ap k_4\cdot (k_2+k_3)}
(-1)
,
\end{align}
where
\begin{equation}
\frac{
x_{13} x_{43} ~d x_2
}{
~x_{12}x_{24} ~d x_3
}
=-1
,
\end{equation}
when we take $x_1, x_4$ and $\omega$ fixed.

We get therefore
\begin{align}
f_{1423}&
(\omega_{1423}, s_{1423},t_{1423},u_{1423}, \alpha_r)~ d \omega_{1423}
\nonumber\\
=&
-e^{\pm 2i \pi \ap k_4\cdot (k_2+k_3)}
f_{1234}(\omega=\frac{1}{1-\omega_{1423}},
s=t_{1423},t=u_{1423},u=s_{1423}, \alpha_r)~
\frac{1}{(1-\omega_{1423})^2}~ d \omega_{1423}
\nonumber\\
=&
f_{1234}(\omega=\frac{1}{1-\omega_{1423}},
s=t_{1423},t=u_{1423},u=s_{1423}, \alpha_r)~
\frac{1}{(1-\omega_{1423})^2}~ d \omega_{1423}|_{\mbox{positive bases}}
,
\end{align}
where the last equality is when we use $|\omega|$ and $|1-\omega|$  
and we can forget the phases.
This last prescription can be verified on Veneziano correlator where
the phases are already present.

Finally we can compute the amplitude
\begin{align}
A_{1423}( s_{1423},t_{1423},u_{1423}, \alpha_r)
&=
\int_0^1 d \omega_{1423}~f_{1423}(\omega_{1423}, s_{1423},t_{1423},u_{1423},\alpha_r)|
\nonumber\\
&=
\int_0^1 
\frac{d \omega_{1423}}{(1-\omega_{1423})^2}
f_{1234}(\omega=\frac{1}{1-\omega_{1423}},
s=t_{1423},t=u_{1423},u=s_{1423}, \alpha_r)|
\nonumber\\
&=
\int_1^\infty d \omega~
f_{1234}(\omega,
s=t_{1423},t=u_{1423},u=s_{1423}, \alpha_r)|~
,
\end{align}
or when expressing this color ordered amplitude
using the basic $s,t$ and $u$ of the $A_{1234}$ amplitude
\begin{align}
A_{1423}( u,s,t, \alpha_r)
%
&=
\int_1^\infty d \omega~
f_{1234}(\omega,
s,t,u, \alpha_r)
,
\end{align}
which is the expression needed for computing the full amplitude which
is obtained by summing over all orderings and
expressed using $s,t$ and $u$ only .

Notice that when a non trivial polynomial in $s,u$ and $\omega$ arises
in the correlator it is not obvious to devise any change of $s,u$
which maps one color ordered amplitude into another.

The other amplitudes can be obtained in a similar way.
Explicitly we get
\begin{align}
\int_0^1 d \omega_{1423}\,
f_{1423}&
(\omega_{1423}= \frac{1}{1-\omega}, s_{1423},t_{1423},u_{1423}, \alpha_r)
\nonumber\\
=&
\int_1^{\infty}  d \omega
f_{1234}(
\omega=\frac{1}{1-\omega_{1423}},
s=t_{1423},t=u_{1423},u=s_{1423}, \alpha_r)~
,
\nonumber\\
\int_0^1 d \omega_{1342}\,
f_{1342}&
(\omega_{1342}, s_{1342},t_{1342},u_{1342}, \alpha_r)
\nonumber\\
=&
\int_{-\infty}^0  d \omega
f_{1234}(
\omega=1-\frac{1}{\omega_{1342}},
s=u_{1342},t=s_{1342},u=t_{1342}, \alpha_r)~
,
\end{align}
and the anticyclic ordering
\begin{align}
\int_0^1 d \omega_{1243}\,
f_{1243}&
(\omega_{1243}, s_{1243},t_{1243},u_{1243}, \alpha_r)
\nonumber\\
=&
\int_{-\infty}^0  d \omega
f_{1234}(
\omega=\frac{-\omega_{1243}}{1-\omega_{1243}},
s=s_{1243},t=u_{1243},u=t_{1243}, \alpha_r)~
,
\nonumber\\
\int_0^1 d \omega_{1432}\,
f_{1432}&
(\omega_{1432}, s_{1432},t_{1432}, \alpha_r)
\nonumber\\
=&
\int_0^1  d \omega
f_{1234}(
\omega={1-\omega_{1432}},
s=u_{1432}, t=t_{1432}, u=s_{1432}, \alpha_r)~
\nonumber\\
\int_0^1 d \omega_{1324}\,
f_{1324}&
(\omega_{1324}, s_{1324},t_{1324}, \alpha_r)
\nonumber\\
=&
\int_1^{\infty}  d \omega
f_{1234}(
\omega=\frac{1}{\omega_{1324}},
s=t_{1324},t=s_{1324},u=u_{1324}, \alpha_r)~
.
\end{align}

The full $U(1)$ amplitude is obtained by summing all color ordered
amplitudes expressed using $s$, $t$ and $u$ from the basic amplitude
$A_{1234}$ this means
\begin{align}
A(s,t,u)
=&
A_{1234}(s&, &t&, &u)&
\nonumber\\
&+ A_{1423}(s_{1432}=u,& &t_{1432}=t,& &u_{1432}=t)&
\nonumber\\                                         
&+ A_{1342}(s_{1324}=t,& &t_{1324}=s,& &u_{1324}=u)&
\nonumber\\                                         
&+ A_{1243}(s_{1243}=s,& &t_{1243}=u,& &u_{1243}=t)&
\nonumber\\                                         
&+ A_{1432}(s_{1432}=u,& &t_{1432}=t,& &u_{1432}=t)&
\nonumber\\                                         
&+ A_{1324}(s_{1324}=t,& &t_{1324}=s,& &u_{1324}=u)&
.
\end{align}     

If we are interested in the complete correlator and we want it to be
function of $\omega$ so that $0\le \omega \le 1$ we have to take all
the possible correlators and to make the Mandelstam replacement as in
the case of the amplitude but then we have to simply rename
$\omega_{1432}$ and so on $\omega$,
explicitly
\begin{align}
f(s,t,u, \omega )
=&
f_{1234}(s&, &t,& &u,& &\omega_{1234}=\omega )&
\nonumber\\
&+ f_{1423}(s_{1432}=u,& &t_{1432}=t,& &u_{1432}=t,&&\omega_{1432}=\omega )&
\nonumber\\                                                                 
&+ f_{1342}(s_{1324}=t,& &t_{1324}=s,& &u_{1324}=u,&&\omega_{1342}=\omega )&
\nonumber\\                                                                 
&+ f_{1243}(s_{1243}=s,& &t_{1243}=u,& &u_{1243}=t,&&\omega_{1243}=\omega )&
\nonumber\\                                                                 
&+ f_{1432}(s_{1432}=u,& &t_{1432}=t,& &u_{1432}=t,&&\omega_{1432}=\omega )&
\nonumber\\                                                                 
&+ f_{1324}(s_{1324}=t,& &t_{1324}=s,& &u_{1324}=u,&&\omega_{1324}=\omega )&
.
\end{align}     

\section{Level 10 states}
\label{app:level10states}

%
%
%
The first level $10$ scalar with
the maximum number of terms the highest
occurrences of zero modes
eliminated using the null states has $157$ terms
out of $279$ possible and reads
{\footnotesize
\begin{align*}
|S10z1\rangle
=&
-6935292000\,\left(10 , 0\right) 
+94471732800\,\left(9 , 1\right) 
+1395489600\,\left(1 , 0\right)\,\left(9 , 0\right) 
\nonumber\\ 
&
-456913920600\,\left(8 , 2\right) 
-3532347000\,\left(1 , 0\right)\,\left(8 , 1\right) 
-15851576700\,\left(2 , 0\right)\,\left(8 , 0\right) 
\nonumber\\ 
&
+16188732000\,\left(1 , 1\right)\,\left(8 , 0\right) 
+269482500\,\left(1 , 0\right)^2\,\left(8 , 0\right) 
+1017474393600\,\left(7 , 3\right) 
\nonumber\\ 
&
-33493788000\,\left(1 , 0\right)\,\left(7 , 2\right) 
+68103216000\,\left(2 , 0\right)\,\left(7 , 1\right) 
-52980480000\,\left(1 , 1\right)\,\left(7 , 1\right) 
\nonumber\\ 
&
-1805580000\,\left(1 , 0\right)^2\,\left(7 , 1\right) 
+25637875200\,\left(3 , 0\right)\,\left(7 , 0\right) 
-32628708000\,\left(2 , 1\right)\,\left(7 , 0\right) 
\nonumber\\ 
&
-801360000\,\left(1 , 0\right)\,\left(2 , 0\right)\,\left(7 , 0\right) 
+626220000\,\left(1 , 0\right)\,\left(1 , 1\right)\,\left(7 , 0\right) 
-1256174740800\,\left(6 , 4\right) 
\nonumber\\ 
&
+116534544000\,\left(1 , 0\right)\,\left(6 , 3\right) 
-116617158000\,\left(2 , 0\right)\,\left(6 , 2\right) 
+177150240000\,\left(1 , 1\right)\,\left(6 , 2\right) 
\nonumber\\ 
&
+6561480000\,\left(1 , 0\right)^2\,\left(6 , 2\right) 
-71086512000\,\left(3 , 0\right)\,\left(6 , 1\right) 
-6006960000\,\left(2 , 1\right)\,\left(6 , 1\right) 
\nonumber\\ 
&
+1345650000\,\left(1 , 0\right)\,\left(2 , 0\right)\,\left(6 , 1\right) 
-1587600000\,\left(1 , 0\right)\,\left(1 , 1\right)\,\left(6 , 1\right) 
+11300654400\,\left(4 , 0\right)\,\left(6 , 0\right) 
\nonumber\\ 
&
-255510192000\,\left(3 , 1\right)\,\left(6 , 0\right) 
-5662720000\,\left(1 , 0\right)\,\left(3 , 0\right)\,\left(6 , 0\right) 
+254080638000\,\left(2 , 2\right)\,\left(6 , 0\right) 
\nonumber\\ 
&
+854670000\,\left(1 , 0\right)\,\left(2 , 1\right)\,\left(6 , 0\right) 
+7813260000\,\left(2 , 0\right)^2\,\left(6 , 0\right) 
-5427840000\,\left(1 , 1\right)\,\left(2 , 0\right)\,\left(6 , 0\right) 
\nonumber\\ 
&
+1587600000\,\left(1 , 1\right)^2\,\left(6 , 0\right) 
+607282147584\,\left(5 , 5\right) 
-81345942720\,\left(1 , 0\right)\,\left(5 , 4\right) 
\nonumber\\ 
&
+2634992640\,\left(2 , 0\right)\,\left(5 , 3\right) 
-551036203200\,\left(1 , 1\right)\,\left(5 , 3\right) 
-20082748000\,\left(1 , 0\right)^2\,\left(5 , 3\right) 
\nonumber\\ 
&
+225351473760\,\left(3 , 0\right)\,\left(5 , 2\right) 
+247693312800\,\left(2 , 1\right)\,\left(5 , 2\right) 
+12886098000\,\left(1 , 0\right)\,\left(2 , 0\right)\,\left(5 , 2\right) 
\nonumber\\ 
&
-5267808000\,\left(1 , 0\right)\,\left(1 , 1\right)\,\left(5 , 2\right) 
-104496000\,\left(1 , 0\right)^3\,\left(5 , 2\right) 
-89828383680\,\left(4 , 0\right)\,\left(5 , 1\right) 
\nonumber\\ 
&
+865169323200\,\left(3 , 1\right)\,\left(5 , 1\right) 
+29446648000\,\left(1 , 0\right)\,\left(3 , 0\right)\,\left(5 , 1\right) 
-790017472800\,\left(2 , 2\right)\,\left(5 , 1\right) 
\nonumber\\ 
&
-5845392000\,\left(1 , 0\right)\,\left(2 , 1\right)\,\left(5 , 1\right) 
-36082698000\,\left(2 , 0\right)^2\,\left(5 , 1\right) 
+15876000000\,\left(1 , 1\right)\,\left(2 , 0\right)\,\left(5 , 1\right) 
\nonumber\\ 
&
+104496000\,\left(1 , 0\right)^2\,\left(2 , 0\right)\,\left(5 , 1\right) 
-23408646912\,\left(5 , 0\right)^2 
+368756136000\,\left(4 , 1\right)\,\left(5 , 0\right) 
\nonumber\\ 
&
+6258067200\,\left(1 , 0\right)\,\left(4 , 0\right)\,\left(5 , 0\right) 
-319433436000\,\left(3 , 2\right)\,\left(5 , 0\right) 
+9348102400\,\left(1 , 0\right)\,\left(3 , 1\right)\,\left(5 , 0\right) 
\nonumber\\ 
&
-9549187200\,\left(2 , 0\right)\,\left(3 , 0\right)\,\left(5 , 0\right) 
-5376162400\,\left(1 , 1\right)\,\left(3 , 0\right)\,\left(5 , 0\right) 
-12435219600\,\left(1 , 0\right)\,\left(2 , 2\right)\,\left(5 , 0\right) 
\nonumber\\ 
&
+17850699600\,\left(2 , 0\right)\,\left(2 , 1\right)\,\left(5 , 0\right) 
-4762800000\,\left(1 , 1\right)\,\left(2 , 1\right)\,\left(5 , 0\right) 
+104496000\,\left(1 , 0\right)^2\,\left(2 , 1\right)\,\left(5 , 0\right) 
\nonumber\\ 
&
-104496000\,\left(1 , 0\right)\,\left(1 , 1\right)\,\left(2 , 0\right)\,\left(5 , 0\right) 
+59737860000\,\left(2 , 0\right)\,\left(4 , 4\right) 
+412610955750\,\left(1 , 1\right)\,\left(4 , 4\right) 
\nonumber\\ 
&
+14943553275\,\left(1 , 0\right)^2\,\left(4 , 4\right) 
-167485920000\,\left(3 , 0\right)\,\left(4 , 3\right) 
-215509203000\,\left(2 , 1\right)\,\left(4 , 3\right) 
\nonumber\\ 
&
-12604951100\,\left(1 , 0\right)\,\left(2 , 0\right)\,\left(4 , 3\right) 
+6371568000\,\left(1 , 0\right)\,\left(1 , 1\right)\,\left(4 , 3\right) 
+123480000\,\left(1 , 0\right)^3\,\left(4 , 3\right) 
\nonumber\\ 
&
-286098435000\,\left(4 , 0\right)\,\left(4 , 2\right) 
-271032372000\,\left(3 , 1\right)\,\left(4 , 2\right) 
-43951472400\,\left(1 , 0\right)\,\left(3 , 0\right)\,\left(4 , 2\right) 
\nonumber\\ 
&
+55566000000\,\left(2 , 2\right)\,\left(4 , 2\right) 
+61435500\,\left(1 , 0\right)\,\left(2 , 1\right)\,\left(4 , 2\right) 
+2976750000\,\left(2 , 0\right)^2\,\left(4 , 2\right) 
\nonumber\\ 
&
+54671452500\,\left(1 , 1\right)\,\left(2 , 0\right)\,\left(4 , 2\right) 
+446460000\,\left(1 , 0\right)^2\,\left(2 , 0\right)\,\left(4 , 2\right) 
-28894320000\,\left(1 , 1\right)^2\,\left(4 , 2\right) 
\nonumber\\ 
&
-370440000\,\left(1 , 0\right)^2\,\left(1 , 1\right)\,\left(4 , 2\right) 
-771289485750\,\left(4 , 1\right)^2 
-40007615550\,\left(1 , 0\right)\,\left(4 , 0\right)\,\left(4 , 1\right) 
\end{align*}

\begin{align}
&
+1332626535000\,\left(3 , 2\right)\,\left(4 , 1\right) 
+15484392000\,\left(1 , 0\right)\,\left(3 , 1\right)\,\left(4 , 1\right) 
+73883607500\,\left(2 , 0\right)\,\left(3 , 0\right)\,\left(4 , 1\right) 
\nonumber\\ 
&
-33498360000\,\left(1 , 1\right)\,\left(3 , 0\right)\,\left(4 , 1\right) 
-123480000\,\left(1 , 0\right)^2\,\left(3 , 0\right)\,\left(4 , 1\right) 
+8829124500\,\left(1 , 0\right)\,\left(2 , 2\right)\,\left(4 , 1\right) 
\nonumber\\ 
&
-55227112500\,\left(2 , 0\right)\,\left(2 , 1\right)\,\left(4 , 1\right) 
+28894320000\,\left(1 , 1\right)\,\left(2 , 1\right)\,\left(4 , 1\right) 
-446460000\,\left(1 , 0\right)\,\left(2 , 0\right)^2\,\left(4 , 1\right) 
\nonumber\\ 
&
+370440000\,\left(1 , 0\right)\,\left(1 , 1\right)\,\left(2 , 0\right)\,\left(4 , 1\right) 
-6952837500\,\left(2 , 0\right)\,\left(4 , 0\right)^2 
+11311477275\,\left(1 , 1\right)\,\left(4 , 0\right)^2 
\nonumber\\ 
&
+347766720000\,\left(3 , 3\right)\,\left(4 , 0\right) 
+37648107500\,\left(1 , 0\right)\,\left(3 , 2\right)\,\left(4 , 0\right) 
-2610368400\,\left(2 , 0\right)\,\left(3 , 1\right)\,\left(4 , 0\right) 
\nonumber\\ 
&
+11642400000\,\left(1 , 1\right)\,\left(3 , 1\right)\,\left(4 , 0\right) 
-123480000\,\left(1 , 0\right)^2\,\left(3 , 1\right)\,\left(4 , 0\right) 
+8643600000\,\left(3 , 0\right)^2\,\left(4 , 0\right) 
\nonumber\\ 
&
-19766203100\,\left(2 , 1\right)\,\left(3 , 0\right)\,\left(4 , 0\right) 
+123480000\,\left(1 , 0\right)\,\left(1 , 1\right)\,\left(3 , 0\right)\,\left(4 , 0\right) 
-2976750000\,\left(2 , 0\right)\,\left(2 , 2\right)\,\left(4 , 0\right) 
\nonumber\\ 
&
-23835724500\,\left(1 , 1\right)\,\left(2 , 2\right)\,\left(4 , 0\right) 
+165900000\,\left(1 , 0\right)^2\,\left(2 , 2\right)\,\left(4 , 0\right) 
+15500824500\,\left(2 , 1\right)^2\,\left(4 , 0\right) 
\nonumber\\ 
&
-778260000\,\left(1 , 0\right)\,\left(2 , 0\right)\,\left(2 , 1\right)\,\left(4 , 0\right) 
+370440000\,\left(1 , 0\right)\,\left(1 , 1\right)\,\left(2 , 1\right)\,\left(4 , 0\right) 
+612360000\,\left(1 , 1\right)\,\left(2 , 0\right)^2\,\left(4 , 0\right) 
\nonumber\\ 
&
-370440000\,\left(1 , 1\right)^2\,\left(2 , 0\right)\,\left(4 , 0\right) 
-414892800000\,\left(3 , 1\right)\,\left(3 , 3\right) 
+17012800000\,\left(1 , 0\right)\,\left(3 , 0\right)\,\left(3 , 3\right) 
%
\nonumber\\ 
&
+556646958000\,\left(2 , 2\right)\,\left(3 , 3\right) 
+6659590000\,\left(1 , 0\right)\,\left(2 , 1\right)\,\left(3 , 3\right) 
+26403816600\,\left(2 , 0\right)^2\,\left(3 , 3\right) 
\nonumber\\ 
&
-68300806000\,\left(1 , 1\right)\,\left(2 , 0\right)\,\left(3 , 3\right) 
-658560000\,\left(1 , 0\right)^2\,\left(2 , 0\right)\,\left(3 , 3\right) 
+28153440000\,\left(1 , 1\right)^2\,\left(3 , 3\right) 
\nonumber\\ 
&
+411600000\,\left(1 , 0\right)^2\,\left(1 , 1\right)\,\left(3 , 3\right) 
-575909838000\,\left(3 , 2\right)^2 
-30367750000\,\left(1 , 0\right)\,\left(3 , 1\right)\,\left(3 , 2\right) 
\nonumber\\ 
&
-65582521200\,\left(2 , 0\right)\,\left(3 , 0\right)\,\left(3 , 2\right) 
+41941606000\,\left(1 , 1\right)\,\left(3 , 0\right)\,\left(3 , 2\right) 
-21840000\,\left(1 , 0\right)^2\,\left(3 , 0\right)\,\left(3 , 2\right) 
\nonumber\\ 
&
+5927040000\,\left(1 , 0\right)\,\left(2 , 2\right)\,\left(3 , 2\right) 
+41114304000\,\left(2 , 0\right)\,\left(2 , 1\right)\,\left(3 , 2\right) 
-26671680000\,\left(1 , 1\right)\,\left(2 , 1\right)\,\left(3 , 2\right) 
\nonumber\\ 
&
+47040000\,\left(1 , 0\right)^2\,\left(2 , 1\right)\,\left(3 , 2\right) 
+816480000\,\left(1 , 0\right)\,\left(2 , 0\right)^2\,\left(3 , 2\right) 
-540960000\,\left(1 , 0\right)\,\left(1 , 1\right)\,\left(2 , 0\right)\,\left(3 , 2\right) 
\nonumber\\ 
&
+27552406000\,\left(2 , 0\right)\,\left(3 , 1\right)^2 
-28153440000\,\left(1 , 1\right)\,\left(3 , 1\right)^2 
-34574400000\,\left(3 , 0\right)^2\,\left(3 , 1\right) 
\nonumber\\ 
&
+22514954000\,\left(2 , 1\right)\,\left(3 , 0\right)\,\left(3 , 1\right) 
+1338960000\,\left(1 , 0\right)\,\left(2 , 0\right)\,\left(3 , 0\right)\,\left(3 , 1\right) 
-823200000\,\left(1 , 0\right)\,\left(1 , 1\right)\,\left(3 , 0\right)\,\left(3 , 1\right) 
\nonumber\\ 
&
-17406144000\,\left(2 , 0\right)\,\left(2 , 2\right)\,\left(3 , 1\right) 
+38525760000\,\left(1 , 1\right)\,\left(2 , 2\right)\,\left(3 , 1\right) 
-47040000\,\left(1 , 0\right)^2\,\left(2 , 2\right)\,\left(3 , 1\right) 
\nonumber\\ 
&
-11854080000\,\left(2 , 1\right)^2\,\left(3 , 1\right) 
+47040000\,\left(1 , 0\right)\,\left(2 , 0\right)\,\left(2 , 1\right)\,\left(3 , 1\right) 
-816480000\,\left(2 , 0\right)^3\,\left(3 , 1\right) 
\nonumber\\ 
&
+493920000\,\left(1 , 1\right)\,\left(2 , 0\right)^2\,\left(3 , 1\right) 
+40166544600\,\left(2 , 2\right)\,\left(3 , 0\right)^2 
+21840000\,\left(1 , 0\right)\,\left(2 , 1\right)\,\left(3 , 0\right)^2 
\nonumber\\ 
&
-680400000\,\left(1 , 1\right)\,\left(2 , 0\right)\,\left(3 , 0\right)^2 
+411600000\,\left(1 , 1\right)^2\,\left(3 , 0\right)^2 
-29635200000\,\left(2 , 1\right)\,\left(2 , 2\right)\,\left(3 , 0\right) 
\nonumber\\ 
&
-816480000\,\left(1 , 0\right)\,\left(2 , 0\right)\,\left(2 , 2\right)\,\left(3 , 0\right) 
+540960000\,\left(1 , 0\right)\,\left(1 , 1\right)\,\left(2 , 2\right)\,\left(3 , 0\right) 
-47040000\,\left(1 , 0\right)\,\left(2 , 1\right)^2\,\left(3 , 0\right) 
\nonumber\\ 
&
+816480000\,\left(2 , 0\right)^2\,\left(2 , 1\right)\,\left(3 , 0\right) 
-493920000\,\left(1 , 1\right)\,\left(2 , 0\right)\,\left(2 , 1\right)\,\left(3 , 0\right) 
-8890560000\,\left(1 , 1\right)\,\left(2 , 2\right)^2 
\nonumber\\ 
&
+8890560000\,\left(2 , 1\right)^2\,\left(2 , 2\right) 
,
\end{align}
}%


%
%
%
The second level $10$ scalar with
the maximum number of terms the highest
occurrences of zero modes
eliminated using the null states has again $157$ terms
out of $279$ possible and reads
{\footnotesize
\begin{align*}
|S10z2\rangle
=&
+65112321600\,\left(9 , 1\right) 
\nonumber\\ 
&
+705211200\,\left(1 , 0\right)\,\left(9 , 0\right) 
-316414042200\,\left(8 , 2\right) 
-2600955000\,\left(1 , 0\right)\,\left(8 , 1\right) 
\nonumber\\ 
&
-12279627900\,\left(2 , 0\right)\,\left(8 , 0\right) 
+9414972000\,\left(1 , 1\right)\,\left(8 , 0\right) 
-55093500\,\left(1 , 0\right)^2\,\left(8 , 0\right) 
\nonumber\\ 
&
+690708211200\,\left(7 , 3\right) 
-25510428000\,\left(1 , 0\right)\,\left(7 , 2\right) 
+56394288000\,\left(2 , 0\right)\,\left(7 , 1\right) 
\nonumber\\ 
&
-22498560000\,\left(1 , 1\right)\,\left(7 , 1\right) 
-112140000\,\left(1 , 0\right)^2\,\left(7 , 1\right) 
+18773798400\,\left(3 , 0\right)\,\left(7 , 0\right) 
\nonumber\\ 
&
-20774628000\,\left(2 , 1\right)\,\left(7 , 0\right) 
-349776000\,\left(1 , 0\right)\,\left(2 , 0\right)\,\left(7 , 0\right) 
+626220000\,\left(1 , 0\right)\,\left(1 , 1\right)\,\left(7 , 0\right) 
\nonumber\\ 
&
-812357985600\,\left(6 , 4\right) 
+88084752000\,\left(1 , 0\right)\,\left(6 , 3\right) 
-127285830000\,\left(2 , 0\right)\,\left(6 , 2\right) 
\nonumber\\ 
&
+43731360000\,\left(1 , 1\right)\,\left(6 , 2\right) 
-2261880000\,\left(1 , 0\right)^2\,\left(6 , 2\right) 
-28186032000\,\left(3 , 0\right)\,\left(6 , 1\right) 
\nonumber\\ 
&
+20725200000\,\left(2 , 1\right)\,\left(6 , 1\right) 
+4241970000\,\left(1 , 0\right)\,\left(2 , 0\right)\,\left(6 , 1\right) 
-1587600000\,\left(1 , 0\right)\,\left(1 , 1\right)\,\left(6 , 1\right) 
\nonumber\\ 
&
+16200340800\,\left(4 , 0\right)\,\left(6 , 0\right) 
-183256752000\,\left(3 , 1\right)\,\left(6 , 0\right) 
-845824000\,\left(1 , 0\right)\,\left(3 , 0\right)\,\left(6 , 0\right) 
\nonumber\\ 
&
+185496318000\,\left(2 , 2\right)\,\left(6 , 0\right) 
+3750990000\,\left(1 , 0\right)\,\left(2 , 1\right)\,\left(6 , 0\right) 
+3410316000\,\left(2 , 0\right)^2\,\left(6 , 0\right) 
\nonumber\\ 
&
-8324160000\,\left(1 , 1\right)\,\left(2 , 0\right)\,\left(6 , 0\right) 
+1587600000\,\left(1 , 1\right)^2\,\left(6 , 0\right) 
+376838832384\,\left(5 , 5\right) 
\nonumber\\ 
&
-61431088320\,\left(1 , 0\right)\,\left(5 , 4\right) 
+147728931840\,\left(2 , 0\right)\,\left(5 , 3\right) 
-134957995200\,\left(1 , 1\right)\,\left(5 , 3\right) 
\nonumber\\ 
&
+8564612000\,\left(1 , 0\right)^2\,\left(5 , 3\right) 
+77209342560\,\left(3 , 0\right)\,\left(5 , 2\right) 
+84796480800\,\left(2 , 1\right)\,\left(5 , 2\right) 
\nonumber\\ 
&
-227982000\,\left(1 , 0\right)\,\left(2 , 0\right)\,\left(5 , 2\right) 
+1505952000\,\left(1 , 0\right)\,\left(1 , 1\right)\,\left(5 , 2\right) 
+271824000\,\left(1 , 0\right)^3\,\left(5 , 2\right) 
\nonumber\\ 
&
-105001606080\,\left(4 , 0\right)\,\left(5 , 1\right) 
+377966635200\,\left(3 , 1\right)\,\left(5 , 1\right) 
-3152072000\,\left(1 , 0\right)\,\left(3 , 0\right)\,\left(5 , 1\right) 
\nonumber\\ 
&
-413747200800\,\left(2 , 2\right)\,\left(5 , 1\right) 
-12619152000\,\left(1 , 0\right)\,\left(2 , 1\right)\,\left(5 , 1\right) 
-11114538000\,\left(2 , 0\right)^2\,\left(5 , 1\right) 
\nonumber\\ 
&
+15876000000\,\left(1 , 1\right)\,\left(2 , 0\right)\,\left(5 , 1\right) 
-271824000\,\left(1 , 0\right)^2\,\left(2 , 0\right)\,\left(5 , 1\right) 
-24356973312\,\left(5 , 0\right)^2 
\nonumber\\ 
&
+275616936000\,\left(4 , 1\right)\,\left(5 , 0\right) 
+1347091200\,\left(1 , 0\right)\,\left(4 , 0\right)\,\left(5 , 0\right) 
-243228636000\,\left(3 , 2\right)\,\left(5 , 0\right) 
\nonumber\\ 
&
-5864633600\,\left(1 , 0\right)\,\left(3 , 1\right)\,\left(5 , 0\right) 
-5484931200\,\left(2 , 0\right)\,\left(3 , 0\right)\,\left(5 , 0\right) 
+9836573600\,\left(1 , 1\right)\,\left(3 , 0\right)\,\left(5 , 0\right) 
\nonumber\\ 
&
-1691955600\,\left(1 , 0\right)\,\left(2 , 2\right)\,\left(5 , 0\right) 
+7107435600\,\left(2 , 0\right)\,\left(2 , 1\right)\,\left(5 , 0\right) 
-4762800000\,\left(1 , 1\right)\,\left(2 , 1\right)\,\left(5 , 0\right) 
\nonumber\\ 
&
-271824000\,\left(1 , 0\right)^2\,\left(2 , 1\right)\,\left(5 , 0\right) 
+271824000\,\left(1 , 0\right)\,\left(1 , 1\right)\,\left(2 , 0\right)\,\left(5 , 0\right) 
-64729980000\,\left(2 , 0\right)\,\left(4 , 4\right) 
\nonumber\\ 
&
+107398030950\,\left(1 , 1\right)\,\left(4 , 4\right) 
-6161648325\,\left(1 , 0\right)^2\,\left(4 , 4\right) 
-60799200000\,\left(3 , 0\right)\,\left(4 , 3\right) 
\nonumber\\ 
&
-90863551800\,\left(2 , 1\right)\,\left(4 , 3\right) 
-2914240700\,\left(1 , 0\right)\,\left(2 , 0\right)\,\left(4 , 3\right) 
+148176000\,\left(1 , 0\right)\,\left(1 , 1\right)\,\left(4 , 3\right) 
\nonumber\\ 
&
-222264000\,\left(1 , 0\right)^3\,\left(4 , 3\right) 
-73995075000\,\left(4 , 0\right)\,\left(4 , 2\right) 
-18131623200\,\left(3 , 1\right)\,\left(4 , 2\right) 
\nonumber\\ 
&
-6136822800\,\left(1 , 0\right)\,\left(3 , 0\right)\,\left(4 , 2\right) 
+2222640000\,\left(2 , 2\right)\,\left(4 , 2\right) 
+5882635500\,\left(1 , 0\right)\,\left(2 , 1\right)\,\left(4 , 2\right) 
\nonumber\\ 
&
-14804370000\,\left(2 , 0\right)^2\,\left(4 , 2\right) 
+19024540500\,\left(1 , 1\right)\,\left(2 , 0\right)\,\left(4 , 2\right) 
-1210524000\,\left(1 , 0\right)^2\,\left(2 , 0\right)\,\left(4 , 2\right) 
\nonumber\\ 
&
-6667920000\,\left(1 , 1\right)^2\,\left(4 , 2\right) 
+864360000\,\left(1 , 0\right)^2\,\left(1 , 1\right)\,\left(4 , 2\right) 
-359389840950\,\left(4 , 1\right)^2 
\end{align*}
\begin{align*}
&
-760732350\,\left(1 , 0\right)\,\left(4 , 0\right)\,\left(4 , 1\right) 
+706144455000\,\left(3 , 2\right)\,\left(4 , 1\right) 
+21707784000\,\left(1 , 0\right)\,\left(3 , 1\right)\,\left(4 , 1\right) 
\nonumber\\ 
&
+42183687500\,\left(2 , 0\right)\,\left(3 , 0\right)\,\left(4 , 1\right) 
-33498360000\,\left(1 , 1\right)\,\left(3 , 0\right)\,\left(4 , 1\right) 
+222264000\,\left(1 , 0\right)^2\,\left(3 , 0\right)\,\left(4 , 1\right) 
\nonumber\\ 
&
+3007924500\,\left(1 , 0\right)\,\left(2 , 2\right)\,\left(4 , 1\right) 
-19580200500\,\left(2 , 0\right)\,\left(2 , 1\right)\,\left(4 , 1\right) 
+6667920000\,\left(1 , 1\right)\,\left(2 , 1\right)\,\left(4 , 1\right) 
\nonumber\\ 
&
+1210524000\,\left(1 , 0\right)\,\left(2 , 0\right)^2\,\left(4 , 1\right) 
-864360000\,\left(1 , 0\right)\,\left(1 , 1\right)\,\left(2 , 0\right)\,\left(4 , 1\right) 
-2084197500\,\left(2 , 0\right)\,\left(4 , 0\right)^2 
\nonumber\\ 
&
-903164325\,\left(1 , 1\right)\,\left(4 , 0\right)^2 
+158101440000\,\left(3 , 3\right)\,\left(4 , 0\right) 
+7923867500\,\left(1 , 0\right)\,\left(3 , 2\right)\,\left(4 , 0\right) 
\nonumber\\ 
&
-10236358800\,\left(2 , 0\right)\,\left(3 , 1\right)\,\left(4 , 0\right) 
+11642400000\,\left(1 , 1\right)\,\left(3 , 1\right)\,\left(4 , 0\right) 
+222264000\,\left(1 , 0\right)^2\,\left(3 , 1\right)\,\left(4 , 0\right) 
\nonumber\\ 
&
+4033680000\,\left(3 , 0\right)^2\,\left(4 , 0\right) 
-12051172700\,\left(2 , 1\right)\,\left(3 , 0\right)\,\left(4 , 0\right) 
-222264000\,\left(1 , 0\right)\,\left(1 , 1\right)\,\left(3 , 0\right)\,\left(4 , 0\right) 
\nonumber\\ 
&
+11840850000\,\left(2 , 0\right)\,\left(2 , 2\right)\,\left(4 , 0\right) 
-15368524500\,\left(1 , 1\right)\,\left(2 , 2\right)\,\left(4 , 0\right) 
+312900000\,\left(1 , 0\right)^2\,\left(2 , 2\right)\,\left(4 , 0\right) 
\nonumber\\ 
&
+7033624500\,\left(2 , 1\right)^2\,\left(4 , 0\right) 
+584724000\,\left(1 , 0\right)\,\left(2 , 0\right)\,\left(2 , 1\right)\,\left(4 , 0\right) 
-864360000\,\left(1 , 0\right)\,\left(1 , 1\right)\,\left(2 , 1\right)\,\left(4 , 0\right) 
\nonumber\\ 
&
-897624000\,\left(1 , 1\right)\,\left(2 , 0\right)^2\,\left(4 , 0\right) 
+864360000\,\left(1 , 1\right)^2\,\left(2 , 0\right)\,\left(4 , 0\right) 
-320060160000\,\left(3 , 1\right)\,\left(3 , 3\right) 
\nonumber\\ 
&
+3841600000\,\left(1 , 0\right)\,\left(3 , 0\right)\,\left(3 , 3\right) 
+354653434800\,\left(2 , 2\right)\,\left(3 , 3\right) 
+5276614000\,\left(1 , 0\right)\,\left(2 , 1\right)\,\left(3 , 3\right) 
\nonumber\\ 
&
+26245762200\,\left(2 , 0\right)^2\,\left(3 , 3\right) 
-33726406000\,\left(1 , 1\right)\,\left(2 , 0\right)\,\left(3 , 3\right) 
+1185408000\,\left(1 , 0\right)^2\,\left(2 , 0\right)\,\left(3 , 3\right) 
\nonumber\\ 
&
+7408800000\,\left(1 , 1\right)^2\,\left(3 , 3\right) 
-740880000\,\left(1 , 0\right)^2\,\left(1 , 1\right)\,\left(3 , 3\right) 
-338354074800\,\left(3 , 2\right)^2 
\nonumber\\ 
&
-28984774000\,\left(1 , 0\right)\,\left(3 , 1\right)\,\left(3 , 2\right) 
-41558252400\,\left(2 , 0\right)\,\left(3 , 0\right)\,\left(3 , 2\right) 
+35591206000\,\left(1 , 1\right)\,\left(3 , 0\right)\,\left(3 , 2\right) 
\nonumber\\ 
&
-297808000\,\left(1 , 0\right)^2\,\left(3 , 0\right)\,\left(3 , 2\right) 
+5927040000\,\left(1 , 0\right)\,\left(2 , 2\right)\,\left(3 , 2\right) 
+16728768000\,\left(2 , 0\right)\,\left(2 , 1\right)\,\left(3 , 2\right) 
\nonumber\\ 
&
-8890560000\,\left(1 , 1\right)\,\left(2 , 1\right)\,\left(3 , 2\right) 
+211680000\,\left(1 , 0\right)^2\,\left(2 , 1\right)\,\left(3 , 2\right) 
-538272000\,\left(1 , 0\right)\,\left(2 , 0\right)^2\,\left(3 , 2\right) 
\nonumber\\ 
&
+282240000\,\left(1 , 0\right)\,\left(1 , 1\right)\,\left(2 , 0\right)\,\left(3 , 2\right) 
-7021994000\,\left(2 , 0\right)\,\left(3 , 1\right)^2 
-7408800000\,\left(1 , 1\right)\,\left(3 , 1\right)^2 
\nonumber\\ 
&
-16134720000\,\left(3 , 0\right)^2\,\left(3 , 1\right) 
+28865354000\,\left(2 , 1\right)\,\left(3 , 0\right)\,\left(3 , 1\right) 
-2073008000\,\left(1 , 0\right)\,\left(2 , 0\right)\,\left(3 , 0\right)\,\left(3 , 1\right) 
\nonumber\\ 
&
+1481760000\,\left(1 , 0\right)\,\left(1 , 1\right)\,\left(3 , 0\right)\,\left(3 , 1\right) 
+6979392000\,\left(2 , 0\right)\,\left(2 , 2\right)\,\left(3 , 1\right) 
+8890560000\,\left(1 , 1\right)\,\left(2 , 2\right)\,\left(3 , 1\right) 
\nonumber\\ 
&
-870240000\,\left(1 , 0\right)^2\,\left(2 , 2\right)\,\left(3 , 1\right) 
+1528800000\,\left(1 , 0\right)\,\left(2 , 0\right)\,\left(2 , 1\right)\,\left(3 , 1\right) 
+538272000\,\left(2 , 0\right)^3\,\left(3 , 1\right) 
\nonumber\\ 
&
-1152480000\,\left(1 , 1\right)\,\left(2 , 0\right)^2\,\left(3 , 1\right) 
+18276010200\,\left(2 , 2\right)\,\left(3 , 0\right)^2 
+297808000\,\left(1 , 0\right)\,\left(2 , 1\right)\,\left(3 , 0\right)^2 
\nonumber\\ 
&
+887600000\,\left(1 , 1\right)\,\left(2 , 0\right)\,\left(3 , 0\right)^2 
-740880000\,\left(1 , 1\right)^2\,\left(3 , 0\right)^2 
-29635200000\,\left(2 , 1\right)\,\left(2 , 2\right)\,\left(3 , 0\right) 
\nonumber\\ 
&
+538272000\,\left(1 , 0\right)\,\left(2 , 0\right)\,\left(2 , 2\right)\,\left(3 , 0\right) 
-282240000\,\left(1 , 0\right)\,\left(1 , 1\right)\,\left(2 , 2\right)\,\left(3 , 0\right) 
-211680000\,\left(1 , 0\right)\,\left(2 , 1\right)^2\,\left(3 , 0\right) 
\nonumber\\ 
&
-538272000\,\left(2 , 0\right)^2\,\left(2 , 1\right)\,\left(3 , 0\right) 
+493920000\,\left(1 , 1\right)\,\left(2 , 0\right)\,\left(2 , 1\right)\,\left(3 , 0\right) 
+493920000\,\left(1 , 0\right)^2\,\left(2 , 2\right)^2 
\nonumber\\ 
&
-987840000\,\left(1 , 0\right)\,\left(2 , 0\right)\,\left(2 , 1\right)\,\left(2 , 2\right) 
+493920000\,\left(1 , 1\right)\,\left(2 , 0\right)^2\,\left(2 , 2\right) 
\end{align*}
}%

\printbibliography[heading=bibintoc]

\end{document}